\documentclass[apj,twocolumn,twocolappendix,numberedappendix]{openjournal}
\usepackage{amsmath}
\usepackage{booktabs}
\usepackage{multirow}
\usepackage{color}
\usepackage{soul}
\usepackage{threeparttable}
\usepackage{float}
\usepackage{graphicx}
\usepackage{CJK}
\usepackage{xspace}
\usepackage{afterpage}
\usepackage{placeins}
\usepackage{amssymb}
\usepackage[breaklinks,colorlinks,citecolor=blue,urlcolor=blue,linkcolor=blue,filecolor=blue]{hyperref}

\makeatletter 
  \patchcmd{\NAT@citex}
    {\@citea\NAT@hyper@{%
      \NAT@nmfmt{\NAT@nm}%
      \hyper@natlinkbreak{\NAT@aysep\NAT@spacechar}{\@citeb\@extra@b@citeb}%
      \NAT@date}}
    {\@citea\NAT@nmfmt{\NAT@nm}%
    \NAT@aysep\NAT@spacechar\NAT@hyper@{\NAT@date}}{}{}

  \patchcmd{\NAT@citex}
    {\@citea\NAT@hyper@{%
      \NAT@nmfmt{\NAT@nm}%
      \hyper@natlinkbreak{\NAT@spacechar\NAT@@open\if*#1*\else#1\NAT@spacechar\fi}%
        {\@citeb\@extra@b@citeb}%
      \NAT@date}}
    {\@citea\NAT@nmfmt{\NAT@nm}%
    \NAT@spacechar\NAT@@open\if*#1*\else#1\NAT@spacechar\fi\NAT@hyper@{\NAT@date}}
    {}{}
\makeatother

\shorttitle{BH*s Across the Universe}
\shortauthors{Weibel, Naidu et al.}

\usepackage[export]{adjustbox}

\newcommand{\orcidauthor}[3]{\author{\href{http://orcid.org/#1}{#2$^{#3}$}}}

\begin{document}
\begin{CJK*}{UTF8}{gbsn}

\title{\vspace{-1cm} Black Hole Stars Across the Universe:\\Identifying Central Engine Dominated Little Red Dots at $\MakeLowercase{z} \sim1.5-9.5$ \vspace{-1.75cm}}

\orcidauthor{0000-0001-8928-4465}{Andrea Weibel}{1, *}
\orcidauthor{0000-0003-3997-5705}{Rohan P. Naidu}{2, 3, *,\dagger}
\orcidauthor{0000-0001-5851-6649}{Pascal A. Oesch}{1, 4, 5}
\orcidauthor{0000-0002-2380-9801}{Anna de Graaff}{6, 7,\ddagger}
\orcidauthor{0000-0002-4684-9005}{Raphael E. Hviding}{7}
\orcidauthor{0009-0002-8965-1303}{Zhaoran Liu}{2}
\orcidauthor{0000-0003-2871-127X}{Jorryt Matthee}{8}
\orcidauthor{0000-0003-2919-7495}{Christina C. Williams}{9, 10}
\orcidauthor{0000-0003-2680-005X}{Gabriel Brammer}{4, 5}
\orcidauthor{0000-0002-9672-3005}{Alba Covelo Paz}{1}
\orcidauthor{0000-0002-5612-3427}{Jenny E. Greene}{11}
\orcidauthor{0000-0002-8896-6496}{Christian Kragh Jespersen}{11}
\orcidauthor{0000-0001-7673-2257}{Zhiyuan Ji}{10}
\orcidauthor{0000-0003-0695-4414}{Michael V. Maseda}{12}
\orcidauthor{0000-0003-4075-7393}{David J. Setton}{11}
\orcidauthor{0009-0007-3791-7890}{Wendy Q. Sun}{2}
\orcidauthor{0000-0001-5586-6950}{Alberto Torralba}{8}
\orcidauthor{0000-0002-1369-6452}{Callum Witten}{1}
\orcidauthor{0000-0003-1207-5344}{Mengyuan Xiao}{1}

\affiliation{$^1$ Department of Astronomy, University of Geneva, Chemin Pegasi 51, 1290 Versoix, Switzerland}
\affiliation{$^2$ MIT Kavli Institute for Astrophysics and Space Research, 70 Vassar Street, Cambridge, MA 02139, USA}
\affiliation{$^3$ Institute for Astronomy, University of Hawai'i, 2680 Woodlawn Drive, Honolulu, HI 96822, USA}
\affiliation{$^4$ Cosmic Dawn Center (DAWN), Copenhagen, Denmark}
\affiliation{$^5$ Niels Bohr Institute, University of Copenhagen, Jagtvej 128, K{\o}benhavn N, DK-2200, Denmark}
\affiliation{$^6$ Center for Astrophysics, Harvard \& Smithsonian, 60 Garden St, Cambridge, MA 02138, USA}
\affiliation{$^7$ Max-Planck-Institut f\"ur Astronomie, K\"onigstuhl 17, D-69117 Heidelberg, Germany}
\affiliation{$^8$ Institute of Science and Technology Austria (ISTA), Am Campus 1, A-3400 Klosterneuburg, Austria}
\affiliation{$^9$ NSF NOIRLab, 950 N. Cherry Ave., Tucson, AZ 85719, USA}
\affiliation{$^{10}$ Steward Observatory, University of Arizona, 933 North Cherry Avenue, Tucson, AZ 85721, USA}
\affiliation{$^{11}$ Department of Astrophysical Sciences, Princeton University, 4 Ivy Lane, Princeton, NJ 08544, USA}
\affiliation{$^{12}$ Department of Astronomy, University of Wisconsin-Madison, Madison, WI 53706, USA}

\thanks{$^*$E-mail: \href{mailto:andrea.weibel@unige.ch}{andrea.weibel@unige.ch}, \href{mailto:rnaidu@hawaii.edu}{rnaidu@hawaii.edu}}
\thanks{$\dagger$ NASA Hubble Fellow, Pappalardo Fellow}
\thanks{$\ddagger$ Clay Fellow}

\begin{abstract}
Photometric selections of Little Red Dots (LRDs) largely rely on identifying their ``V-shaped'' spectral energy distribution (SED). Recent work suggests this V-shape stems from a combination of a central engine -- also referred to as a Black Hole Star (BH*) -- and a star-forming host galaxy. We present a new and highly complementary photometric selection that is based on incorporating BH* templates in the \texttt{eazy} redshift fitting code. Selecting compact sources where a BH* template contributes $>80$\% to the best fitting SED in the rest-optical, we compile a sample of 241 BH*-dominated candidates from $\sim1000\,{\rm arcmin}^2$ of legacy and pure parallel JWST imaging. Our selection does not require a blue UV-component, and it successfully identifies objects that resemble the paradigmatic sources ``MoM-BH*-1'' and ``The Cliff''. We find that BH*-dominated sources exist across a wide range of redshifts ($z\sim1.7-9.3$) and optical luminosities (log$(L_{5100}/{\rm erg}\,{\rm s}^{-1})\sim42-44.5$), and we measure a median Balmer break strength of $\sim3$, with some breaks reaching values $>10$. We estimate bolometric luminosities in the range log$(L_{\rm bol}/{\rm erg}\,{\rm s}^{-1})\sim42-45$, which, assuming accretion at the Eddington-limit, would translate to black hole masses of $M_{\rm BH}\sim10^4-10^7{\rm M_\odot}$, spanning the intermediate mass black hole to the quasar regime. The number density of BH*-dominated candidates peaks at $z\sim5-6$ ($\sim10^{-5}\,{\rm Mpc}^{-3}$) and it declines by an order of magnitude down to $z\sim2$. Tentatively, comparing to V-shaped LRD samples suggests that the fraction of BH*-dominated sources among the broader LRD population does not decrease towards lower redshift. Crucially, our work demonstrates that BH*-dominated sources are not merely an early-Universe phenomenon but rather persist at least until cosmic noon.\footnote{A machine-readable catalog of our sample is available at: \url{https://doi.org/10.5281/zenodo.20611334}}

\end{abstract} 

\section{Introduction}
\label{sec:intro}

Every massive galaxy in the local Universe hosts a supermassive black hole \citep[SMBH;][]{KormendyHo13}. It is well established that feedback from SMBHs has a significant impact on galaxy evolution throughout the history of the Universe \citep[e.g.,][]{Weinberger18}, in particular through regulating star formation and quenching \citep[e.g.,][]{DeLucia25}. However, the formation and growth of SMBHs in the early Universe is still poorly understood \citep[e.g.,][]{Volonteri21}. Thanks to its powerful near-infrared (NIR) capabilities, the James Webb Space Telescope (JWST) is revealing the early phases of SMBH evolution. Using grism data from the FRESCO \citep{Oesch23} and EIGER \citep{Kashino23} surveys, \citet{Matthee24} identified a surprisingly large number of broad-line H$\alpha$ emitters at redshifts of $z\sim4-6$, which they interpreted as being powered by active galactic nuclei (AGN). Motivated by the finding that the broad-line emitters showed red NIRCam colors and point-like morphologies they named them Little Red Dots (LRDs). LRDs have since become one of the most hotly debated discoveries of JWST. Apart from their broadened Balmer lines, they show various features that are not typically associated with AGN: weak X-ray \citep{Ananna24,Yue24} and radio emission \citep{Latif25}; a flattening or turnover in the NIR spectral energy distribution (SED), as probed by MIRI \citep[e.g.,][]{PerezGonzalez24,Williams24} as well as weak far-infrared (FIR) emission \citep[e.g.,][]{Xiao25,Setton25}, inconsistent with expectations from an AGN dust torus; a lack of variability \citep[e.g.,][]{Kokubo25,Liu26}; Balmer absorption on top of the line emission \citep[e.g.,][]{Matthee26}; high Balmer decrements \citep[e.g.,][]{Nikopoulos25,Torralba26IFU, Sun26}; strong Balmer breaks \citep[e.g.,][]{Wang24evolved,Naidu25BHstar,degraaff25}; and ``V-shaped'' SEDs \citep[e.g.,][]{PerezGonzalez24,Kocevski25stats} that inflect around the Balmer limit \citep[${\rm H}_\infty$,][]{Setton25inflection}. 

The difficulty of modeling the full SED of LRDs with standard AGN models \citep[e.g.,][]{Ma25} has led to the proposal of a wide variety of alternative scenarios. A broader class of models interprets LRDs as a central ionizing source (typically an accreting black hole) that is enshrouded by a dense gas envelope or cocoon \citep[e.g.,][]{Naidu25BHstar,Kido25,Rusakov26,Umeda25,Liu26LRDatmosphere}. Re-processing of the light in this gas envelope may account for both the Balmer break \citep{IM25,Naidu25BHstar,Ji25BT}, as well as the broadening of the Balmer lines \citep[e.g.,][]{Chang26,Naidu25BHstar,Sneppen26}. This scenario bears similarities with so-called quasi-stars \citep{Begelman08}, as explored in e.g., \citet{Begelman26,Santarelli26}. Other models and scenarios include super-Eddington accretion \citep[e.g.,][]{Lupi24LRD,Liu26BB,Secunda26,Madau26,MadauMaiolino26}, self-interacting dark matter \citep{GrantRoberts25,Jiang26}, primordial black holes \citep{Huang24,Zhang25PMBHs,Zhang26}, direct collapse black holes \citep{Pacucci26}, supermassive stars \citep{Nandal26}, and globular clusters in formation \citep{Chisholm26}. Alternatively, \citet{MadauMaiolino26LBDs} propose a unification scheme in which LRDs are the counterparts of blue AGN \citep[see also][]{Brazzini26}, viewed from a high-inclination angle, and with a specific gas and dust geometry. Most of these scenarios share the characteristic that LRDs are a combination of a compact, non-stellar engine and a surrounding host galaxy.

Photometric selections of LRDs largely focus on identifying point sources with V-shaped SEDs, either based on NIRCam-colors \citep[e.g.,][]{Labbe25LRD,Greene24,Kokorev24,PerezGonzalez24,Rinaldi26}, or by fitting power-laws to the rest-optical and the rest-UV \citep[e.g.,][]{Kocevski25stats}. Using a spectroscopic sample of LRDs from RUBIES \citep{degraaff25rubies}, \citet{Hviding25} showed that the vast majority of rest-optical point sources with V-shaped SEDs do show broad Balmer lines, suggesting that these three properties are linked, and that photometric LRD-selections successfully identify broad-line AGN. However, these selections crucially rely on significant detections in the rest-frame UV to measure a blue color or slope. Furthermore, there is nothing inherently physical about a V-shape. \citet{Billand26} recently defined an ``LRDness'' that quantifies how compact and how strongly V-shaped the SED of a given source is. They found a continuum of ``LRDness'' across the galaxy population, with no distinct locus for LRDs. 

Two recently discovered objects may contribute a key piece to the puzzle of understanding LRDs: ``The Cliff'' at $z_{\rm spec}=3.55$ \citep{degraaff25}, and the ``MoM-BH*-1'' at $z_{\rm spec}=7.76$ \citep{Naidu25BHstar}. They show red rest-optical SEDs that can be well-approximated by a blackbody with a temperature of $\sim5000\,{\rm K}$, broad Balmer lines, extremely strong Balmer breaks ($\sim7-8$), and very little UV-emission. \citet{Naidu25BHstar} combined the spectrum of the MoM-BH*-1 with that of a star-forming galaxy at the same redshift to show that this reproduces the typical V-shaped SED of LRDs, inspiring a picture where sources similar to The Cliff or the MoM-BH*-1 lie at the heart of every LRD. This scenario has been further explored by, e.g., \citet{degraaff25pop}, \citet{Barro26}, and \citet{Sun26}, showing that the diversity of LRDs can be explained as a diversity of host galaxies and central engines, and their relative contribution to the total SED \citep[see also][]{Pan26,Merida26,Cloonan2026}. However, this insight has yet to be incorporated in photometric LRD selections. Specifically, objects like The Cliff or the MoM-BH*-1 do not show a strong V-shape and are missing from current photometric samples because of their faintness in the rest-frame UV \citep{Hviding25}. Identifying such objects is crucial because they let us probe the physics of the central engine without being impeded by the host galaxy. Following the terminology established in \citet{Naidu25BHstar}, \citet{degraaff25pop}, and \citet{Sun26}, we refer to the LRD engine as a black hole star (BH*) noting that the analysis in this paper is independent of the physics of the engine, as long as the decomposition of LRDs into an engine that resembles sources like The Cliff and the MoM-BH*-1, and a host galaxy is valid.

In this work, we present a new way of photometrically selecting LRDs as a composite of a BH* and a host galaxy. This method is naturally more sensitive to BH*-dominated sources, and does not require detections in the rest-UV. The goal of this selection method is to enlarge our sample of BH*-dominated sources akin to objects like The Cliff and the MoM-BH*-1. As such, it is complementary to existing photometric LRD samples, and does not itself deliver a complete sample of LRDs. It is essential to photometrically identify BH*-dominated sources as prime candidates for spectroscopic follow-up. Furthermore, understanding the redshift evolution in the number density of both BH*-dominated sources and V-shaped LRDs may inform us about how the formation and evolution of the LRD engines and their hosts are connected, and whether the engines are increasingly outshone by their host galaxies at later cosmic times.

The paper is structured as follows: In Section \ref{sec:data}, we present the imaging data and photometric catalogs that form the basis of the subsequent analysis. We outline our sample selection in detail in Section \ref{sec:sample_sel}, followed by a spectroscopic validation of the sample, and a comparison to V-shape selections in the literature. Section \ref{sec:overview} provides an overview of the sample, highlighting the variety of identified sources, as well as outstanding candidates. We turn to sample properties in Section \ref{sec:sample_props} where we investigate optical and bolometric luminosities, Balmer break strengths, as well as the number density evolution of BH*-dominated candidates.
We discuss our findings, along with limitations and caveats of the sample selection in Section \ref{sec:discussion}, and end with a summary and conclusions in Section \ref{sec:summary_conclusions}.

Whenever relevant, we assume a $\Lambda$CDM cosmology with parameters from the \citet{Planck20} and we specify magnitudes in the AB-system \citep{Oke83}.

\section{Imaging and Catalogs}
\label{sec:data}

This paper is based on the imaging data and photometric catalog presented in \citet{Weibel26}. Below, we briefly describe the data and catalog production, and refer the reader to \citet{Weibel26} and references therein for further details.

\newpage

\subsection{Imaging}
\label{sec:data_imaging}

We retrieve imaging mosaics across legacy fields from the DAWN JWST Archive (DJA),\footnote{\url{https://dawn-cph.github.io/dja/imaging/v7/}} using version 7.0 or higher. The reduction of these mosaics starts from level 2 calibrated data products from the Mikulski Archive for Space Telescopes (MAST) using the software package \texttt{grizli} \citep{grizli} as described in e.g., \citet{Valentino23}. Specifically, these legacy fields are the EGS, UDS, and COSMOS fields, the GOODS fields (North and South), and the Abell-2744 cluster field. A complete list of all JWST programs that contribute imaging data to these mosaics is provided in Section 2.1 of \citet{Weibel26}. JWST imaging is complemented with HST/ACS and HST/WFC3 imaging where available, most importantly from CANDELS \citep{Grogin11,Koekemoer11}. Beyond the legacy data from the DJA, we include pure parallel imaging from PANORAMIC (GO-2514, PIs Williams \& Oesch, \citealt{Williams25}), using the first data release.\footnote{\url{https://panoramic-jwst.github.io/}} PANORAMIC adds $\sim250\,{\rm arcmin}^2$ of NIRCam imaging along 28 independent lines of sight, as well as additional depth and area to various legacy fields, so that we obtain a total area of $\sim1000\,{\rm arcmin}^2$ in six or more NIRCam filters. To remove the brightest cluster galaxies from the Abell-2744 science mosaics before catalog construction, we subtract a running median filter with a box-size of $101\times101$ pixels \citep[$4.04$\arcsec $\times$ 4.04\arcsec; see][]{Naidu24}.

\subsection{Photometry}
\label{sec:data_catalog}

We create a photometric master catalog that contains sources from all the fields mentioned above following the methods outlined in \citet{Weibel24} for the legacy fields, and \citet{Williams25} for PANORAMIC. In short, we run \texttt{SourceExtractor} in dual-mode using an inverse-variance weighted stack of the F277W, F356W, and F444W images as the detection image. We then measure fluxes through circular apertures ($r=0.16$\arcsec) on convolved versions of each available JWST and HST band, so as to match the PSF resolution in F444W. We scale these fluxes to total based on the Kron correction in the detection image, and an additional correction to account for flux beyond the Kron aperture based on the F444W PSF. The full master catalog contains 1,380,654 sources and forms the basis of the subsequent sample selection.

\section{Sample Selection and Validation}
\label{sec:sample_sel}

In the following, we outline the selection of a robust sample of BH*-dominated sources from our photometric catalogs, and validate the sample with publicly available spectra.

\subsection{Pre-Selection}
\label{sec:sample_sel_presel}

We start by pre-selecting compact sources from our photometric catalogs as ${\rm c(F444W)} < 1.7$, where ${\rm c(F444W)}=f_{\rm F444W}(r=0.2\arcsec)/f_{\rm F444W}(r=0.1\arcsec)$ is the ratio of the fluxes measured through circular apertures with radii of 0.2\arcsec\ and 0.1\arcsec\ in the F444W image. This broadly removes sources that are extended, following the definition of compactness for LRDs originally proposed in \citet{Labbe25LRD}. Since compactness can only be robustly measured for sufficiently bright sources, we complement this with a magnitude cut as ${\rm mag(F444W)}<27$ \citep[see, e.g.,][]{Hviding25}. We further require (1) ${\rm SNR(F444W)}>10$, (2) at least two additional filters with ${\rm SNR}>3$, and (3) data in at least the six NIRCam filters F115W, F150W, F200W, F277W, F356W, and F444W, concurrent with the minimal PANORAMIC setup (see \citealt{Williams25}). This ensures a robust detection at the red end of the NIRCam wavelength range, and continuous photometric coverage from $1-5\,\mu{\rm m}$. These cuts reduce the initial master catalog to 11,529 sources. We cross-match this reduced catalog with objects in the GAIA DR3 \citep{GAIA2016, GAIA2023} that have a proper motion measured at ${\rm SNR}>3$, as well as a ${\rm SNR}>3$ photometric detection in the G-band, and that are neither flagged as a galaxy, nor as a QSO in the GAIA catalog. This yields 113 matches with secure bright stars which we remove from our catalog, leaving us with 11,416 compact objects. 

\subsection{BH* Templates}
\label{sec:sample_sel_templates}

The key idea of our sample selection is to incorporate BH* templates in the \texttt{eazy} redshift fitting code \citep{Brammer08}, and to identify objects whose best-fitting SED is dominated by those templates. 

We start from the \texttt{blue\_sfhz} template set (\texttt{sfhz} hereafter)\footnote{\url{https://github.com/gbrammer/eazy-photoz/tree/master/templates/sfhz}} which contains 13 templates generated with the Flexible Stellar Population Synthesis (\texttt{FSPS}) code \citep{Conroy2009b, Conroy2010} with a broad range of redshift-dependent properties, and an additional template with extremely strong emission lines, based on a JWST/NIRSpec spectrum of a $z=8.5$ galaxy \citep{Carnall23}. We then complement this template set with one empirical BH* template at a time, re-running \texttt{eazy} for six different BH* templates that are described in the following. 

\citet{Sun26} decomposed LRD spectra into a host galaxy and a BH* component by subtracting the spectra of star-forming galaxies at the same redshift from LRD spectra, after matching them in [O\,{\sc iii}] luminosity. This assumes that the [O\,{\sc iii}] doublet originates from HII regions associated with star formation, and is backed up by the fact that the [O\,{\sc iii}] line widths are usually much narrower than the Balmer lines (see \citealt{Sun26} for details). They then stacked both the host galaxies and the BH*s to investigate their average properties. Here, we use three BH*-stacks that were created by binning the individual BH* components in luminosity as log($L_{5500}/{\rm erg}\,{\rm s}^{-1})<43.2$, $43.2<{\rm log}(L_{5500}/{\rm erg}\,{\rm s}^{-1}) < 44.2$ and log($L_{5500}/{\rm erg}\,{\rm s}^{-1})>44.2$. Interestingly, these stacks also show different effective temperatures, as derived by parameterizing their continuum shape with a blackbody, such that there is a positive correlation between luminosity and temperature ($T_{\rm eff}=3505^{+182}_{-149}$, $4409^{+124}_{-170}$, and $5122^{+32}_{-416}\,$K for the three luminosity bins respectively; see also \citealt{degraaff25pop}). We extend the three stacks that cover $\lambda_{\rm rest}\lesssim1\,\mu$m to longer wavelengths assuming a single blackbody of the respective temperature to obtain our first three templates that represent average BH* properties across a range of temperatures and luminosities.

Next, we include templates based on two paradigmatic BH*s, the MoM-BH*-1 \citep{Naidu25BHstar} and The Cliff \citep{degraaff25}. Due to the high redshift of the MoM-BH*-1 ($z_{\rm spec}=7.76$), the NIRSpec PRISM spectrum only covers $\lambda_{\rm rest}\lesssim0.6\,\mu{\rm m}$. We therefore use the best-fitting \texttt{CLOUDY} \citep{Ferland17} model from \citet{Naidu25BHstar} for this source. Briefly, this model is selected to match the steep Balmer break, Balmer emission lines, and the overall SED shape from the UV to the NIR, providing a self-consistent template across the rest-wavelengths of interest for our search.

For The Cliff, we retrieve the public spectrum from the DJA which covers $\lambda_{\rm rest}\lesssim1.15\,\mu{\rm m}$. To extend the template to longer wavelengths, we fit a blackbody to the spectrum at $\lambda_{\rm rest}>4000$\AA. Masking the H$\alpha$ as well as the [O\,{\sc iii}] and H$\beta$ emission lines, we find a best-fitting $T_{\rm eff}=4635\,{\rm K}$ and we stitch the spectrum and the best-fitting blackbody together at $\lambda_{\rm rest}=1.15\,\mu{\rm m}$ to create our fifth template. 

Finally, we wish to include a template that represents the most extreme BH*s at the hot end of the temperature distribution. To this end, we add another empirical template that is based on a source tagged GN-9771, initially published as part of the NIRCam/grism sample in \citet{Matthee24}, and followed up with NIRSpec/IFU spectroscopy as presented in \citet{Torralba26IFU}. It is one of the most luminous LRDs known to date at $z_{\rm spec}=5.5$ and its spectrum closely resembles that of A2744-45924 \citep[][]{Labbe24} at $z_{\rm spec}=4.47$, another extremely luminous LRD ($L_V\approx10^{45}{\rm erg}\,{\rm s}^{-1}$). Specifically, GN-9771 is characterized by strong Balmer emission lines, a forest of optical [Fe\,{\sc ii}] lines, and a sharp Balmer break. We use the host-subtracted spectrum of GN-9771 from \citet{Sun26} which - compared to the other templates constructed so far - displays the brightest UV-emission with strong Fe\,{\sc ii} lines as well as He\,{\sc ii}$\lambda3203\,$\AA, Mg\,{\sc ii}$\lambda2799\,$\AA, and C\,{\sc iii}]$\lambda1909\,$\AA, and a downturn at $\lambda_{\rm rest}\lesssim2000\,$\AA. This template thus accounts for the possibility that some UV-emission originates from the BH* as recently discussed in e.g., \citet{Ando26}. Indeed, the \citet{Sun26} BH* templates allow for $\approx0-50\%$ contribution in the UV from the central engine, with the brightest sources showing the highest UV fraction (see their Fig. 12). Similar to the procedure applied to the spectrum of The Cliff, we fit a single blackbody to the continuum at $0.4\,\mu{\rm m}<\lambda_{\rm rest}<0.82\,\mu{\rm m}$, masking prominent emission lines. We obtain a temperature of $T_{\rm eff}=5410\,{\rm K}$, and use this fit to extend the template to $\lambda_{\rm rest}>1\,\mu$m. We discuss the inclusion of an additional template representing the cold end of the temperature distribution in Section \ref{sec:discussion_caveats_intrinsic_variation}.

\begin{figure*}
     \centering \includegraphics[width=1.8\columnwidth]{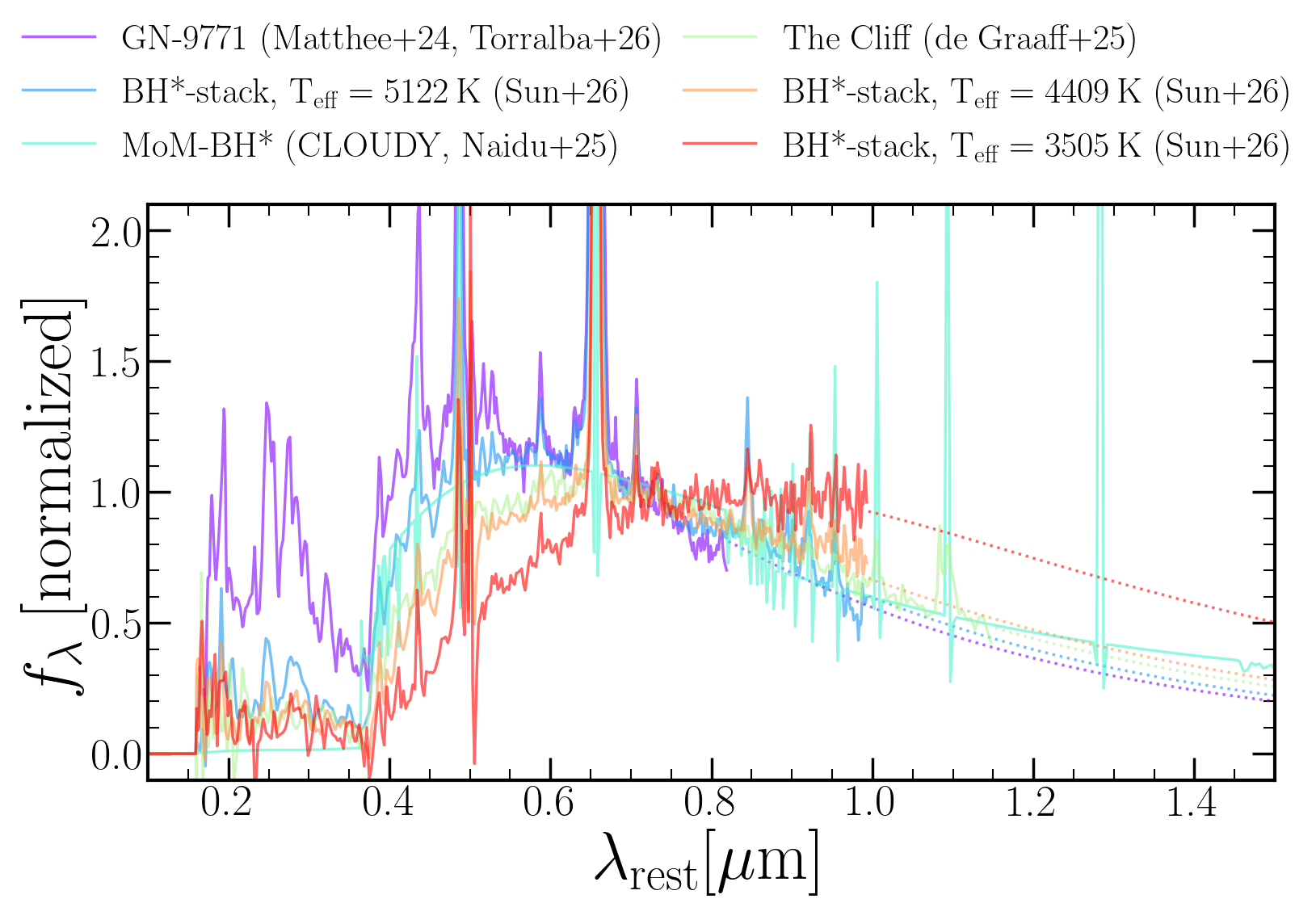}
     \caption{BH* templates used to select BH*-dominated candidates with \texttt{eazy}, normalized to the median in the range $0.7\,\mu{\rm m}<\lambda_{\rm rest}<0.8\,\mu{\rm m}$. The extrapolated part of the templates is shown as a dotted line respectively. This illustrates that our templates cover a broad range of continuum shapes, with the SED peaking at longer wavelengths for lower blackbody temperatures. As a consequence, the Balmer break is stronger and sharper for the hotter sources where there is more continuum to absorb at $\sim0.4\,\mu{\rm m}$.}
     \label{fig:templates}
 \end{figure*}

Extending our template spectra using single blackbodies may not accurately represent the true continuum shape of BH*-dominated sources. For example, the source A2744-45924, with a similar spectrum to GN-9771 that we use as a template, shows a significant excess in MIRI relative to the expectation of a pure blackbody continuum \citep{Setton25}. However, this only affects our sample selection at the low redshift end because at $z\gtrsim4$, the NIRCam data used in this work do not probe the extrapolated part of the templates. Further, the rest-optical continuum shapes of LRDs have been shown to be well described by blackbody continuum at least out to $\sim1\,\mu{\rm m}$ \citep{degraaff25pop,Umeda25}. To avoid any impact of the noisy far UV part of our empirical templates on the selection, we further set all fluxes to 0 at $\lambda_{\rm rest}<0.16\,\mu{\rm m}$. We show our BH* templates in Figure \ref{fig:templates}.

\subsection{Template Fitting with \texttt{eazy}}
\label{sec:sample_sel_eazy}

We run \texttt{eazy} six times on the pre-selected catalog, complementing the \texttt{sfhz} template set with a different BH* template each time. We do not include multiple BH* templates simultaneously to avoid potentially unrealistic combinations of different BH* templates and to limit the number of degrees of freedom in each fit. We allow the redshift to vary freely in the range $z\in(0.01,20)$ and apply a noise floor at 5\%\ of the flux in each filter to allow for additional flexibility. One concern in selecting BH*s from photometry is contamination by Milky Way stars, and in particular brown dwarfs (BDs), which can show strong breaks in their spectra and are known to contaminate photometric LRD samples \citep[e.g.,][]{Greene24}. We use the internal \texttt{eazy} function \texttt{fit\_phoenix\_stars} to fit our candidates against a large grid of stellar templates that is composed of (1) the grid of PHOENIX templates available directly through \texttt{eazy}, (2) a grid of low-temperature, low-metallicity BDs from \citet{Meisner21}, (3) a set of 22 empirical BD templates based on NIRSpec fixed-slit PRISM spectra from GO-2302 (PI Cushing, \citealt{Beiler24}), and (4) the Sonora Elf Owl grid \citep{Mukherjee24} consisting of 43,251 templates spanning a wide range of BD and giant exoplanet properties.

\subsection{Selection Cuts}
\label{sec:sample_sel_cuts}

To obtain our sample of BH*-dominated candidates, we apply the following selection cuts to each of the six \texttt{eazy} runs corresponding to the six different BH* templates:

\begin{itemize}
    \item The integrated BH* template contribution to the best-fitting (lowest $\chi^2$) \texttt{eazy}-SED is $>80\,$\% between $\lambda_{\rm rest,\,min}=0.4\,\mu{\rm m}$ and $\lambda_{\rm rest,\,max}={\rm min}(1\mu{\rm m}\, ,5\mu{\rm m}/(1 +z))$, i.e., red-wards of the Balmer break, and out to either $1\,\mu{\rm m}$ rest-frame (the range covered by all our empirical templates), or to the maximum rest-frame wavelength that is covered by NIRCam at the best-fitting redshift, if this is $<1\mu{\rm m}$.
    \item $1.5 < z_{\rm phot,\,eazy} < 9.5$
    \item The source is nominally better fit with the \texttt{sfhz}+BH* template set, than with the large grid of stellar templates, $\Delta\chi^2({\rm stars})>0$, as well as than with the \texttt{sfhz} template set only: $\Delta\chi^2(\texttt{sfhz})>0$.
    \item $\chi^2_{\rm eazy}/{\rm N_{filters}}<5$ where ${\rm N_{filters}}$ is the number of available NIRCam and HST filters for each object.
    
\end{itemize}

For sources that satisfy all these criteria in more than one \texttt{eazy} run (i.e., with more than one of our BH* templates), we choose the fit with the lowest $\chi^2$. Of the final sample, 44.8, 22.2, 12.3, 5.6, and 0.4\%\ of all sources are selected with 2, 3, 4, 5, and all six templates. 

The first and most important cut requires that the best-fitting SED from \texttt{eazy} is dominated by a BH* template. We discuss the implications of choosing a somewhat arbitrary cut of a template contribution of $>80$\%\ in Section \ref{sec:discussion_completeness}. This cut alone yields a sample of 1,434 objects. We then remove sources fit at $z<1.5$ and at $z>9.5$. At the low redshift end, the distinction between robust BH*-dominated candidates and Milky Way stars becomes increasingly difficult, and we lack deep and uniform photometric coverage blue-wards of the Balmer break. On the other hand, at $z\sim9.5$, the NIRCam-coverage is limited to $\lambda_{\rm rest}\lesssim0.48\,\mu{\rm m}$, meaning that we lose sufficient access to the rest-optical continuum at even higher redshifts. The redshift cuts reduce the sample size to 297 objects. The third cut makes sure that the BH* candidates are better fit with our \texttt{sfhz}+BH* template set than with the large stellar grid described above, as well as than with the \texttt{sfhz} template set only. For five sources, \texttt{eazy} does not find any fit with the \texttt{sfhz} templates and returns an arbitrary value of $\chi^2(\texttt{sfhz})=-1$. However, it does find a decent fit with a BH* template, so while these would nominally fail our $\Delta\chi^2(\texttt{sfhz})$ cut, we consider them to be promising candidates and include them in our sample. The $\Delta\chi^2$-based cuts are complemented by the fourth cut which requires that the best fit with a BH* template has a reasonably low $\chi^2/{\rm N_{filters}}$ to remove cases where either fit is bad. The third and fourth cuts combined remove 45 objects, so that we are left with 252. 
We visually inspect imaging cutouts and SED-fits, and remove eight sources that we identify as spurious detections or hot pixels, or whose photometry is affected by data quality issues, diffraction spikes, or a bright foreground source. We further remove a source in the Abell-2744 field that corresponds to a secondary lensed image of the triply imaged LRD A2744-QSO1 \citep{Furtak23QSO1}, and only keep the brightest to avoid double-counting the source, leaving us with 243 candidates. We identify two of those as contaminants in Section \ref{sec:overview_contaminants}, yielding our final sample size of 241 BH*-dominated candidates shown from Section \ref{sec:sample_props}.

Finally, we correct all fluxes in the Abell-2744 cluster field for lensing magnification using the UNCOVER DR4 magnification catalog which is based on the lens modeling from \citet{Furtak23}, building on photometric redshifts from \citet{Weaver24} and \citet{Suess24} and updated with spectroscopic redshifts in \citet{Price25}.

\begin{figure}
     \centering
     \includegraphics[width=0.47\textwidth]{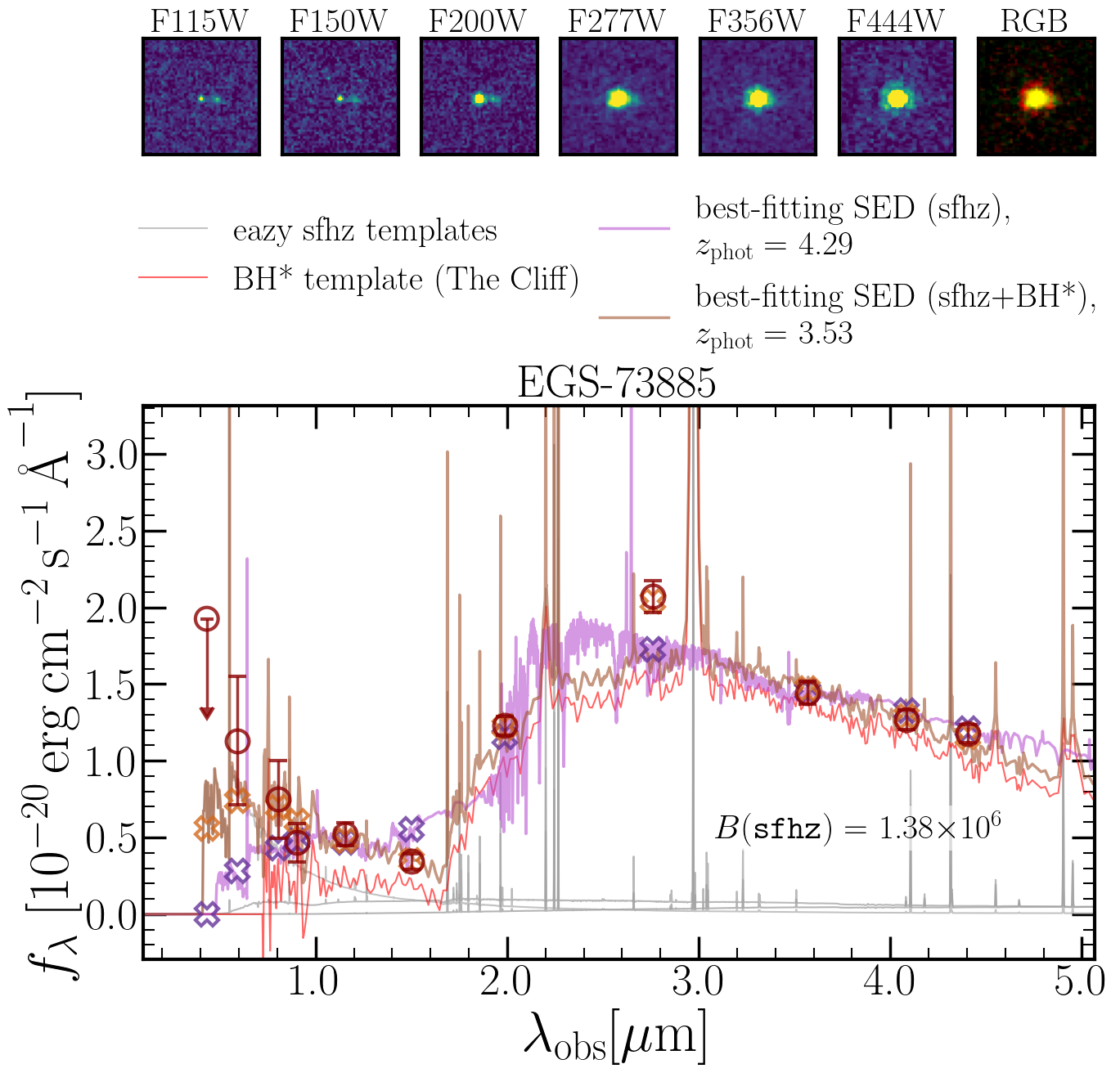}
     \caption{Example of a BH*-dominated candidate selected with the template based on The Cliff \citep{degraaff25}. The stamps show $2.04$\arcsec\ $\times$ $2.04$\arcsec imaging cutouts in JWST filters, as well as an RGB image which is composed of the F444W (red), F200W (green) and F115W (blue) filters. With the \texttt{sfhz}+Cliff template set, the photometry of EGS-73885 is fit at $z_{\rm phot}=3.53$, with the fit being dominated by The Cliff template whose contribution to the total SED is shown in red. Contributions of \texttt{sfhz} templates to the best-fitting SED are displayed in gray. The best-fitting SED with the \texttt{sfhz} template set only (purple line) corresponds to a QG at $z_{\rm phot}=4.29$ but does not reproduce the flux in F277W, the strong Balmer break, and the detected rest-UV flux at $\lambda_{\rm obs}<1\,\mu{\rm m}$, as can be seen from the model photometry (crosses). This results in a Bayes factor $B(\texttt{sfhz})=1.38\times10^6$ suggesting strong evidence in favor of the BH*-dominated solution.}
     \label{fig:worked_example}
 \end{figure}

Figure \ref{fig:worked_example} shows an example of a galaxy in the EGS field with imaging from CEERS that is selected in the \texttt{eazy} run with The Cliff template which produces a good fit at $z_{\rm phot}=3.53$. The best-fitting SED without The Cliff template (\texttt{sfhz} only) is that of a quiescent galaxy (QG) at $z_{\rm phot}=4.29$, and it does not match the flux in the F277W filter which is likely boosted by H$\alpha$ emission. This illustrates that H$\alpha$ emission is an important distinguishing feature between BH*-dominated sources and contaminants such as stars or QGs, since LRDs show typical H$\alpha$ equivalent widths of $\sim300-1500\,$\AA\ \citep[see][]{degraaff25pop}. However, depending on the redshift of the source, and the photometric filter coverage, different solutions still remain degenerate. Further, the QG solution for EGS-73885 does not reproduce the strength of the Balmer break as probed by deep imaging in F150W, and the detected rest-UV flux at $\lambda_{\rm obs}<1\mu{\rm m}$. To quantify how significantly the fit with the BH* template is favored, we compute the Bayesian information criterion (BIC) as ${\rm BIC} = k\,{\rm log}({\rm N_{filters}}) - 2\,{\rm log}(L_{\rm max})$ where $k$ is the number of degrees of freedom of the fit, ${\rm N_{filters}}$ is the number of available photometric filters, and $L_{\rm max}$ is the maximum likelihood, or equivalently the minimum $\chi^2$ of the fit, $L_{\rm max}\propto e^{-\chi_{\rm min}^2/2}$. We can then approximate the Bayes factor as $B(\texttt{sfhz})\approx e^{-\Delta{\rm BIC}/2} = e^{(\Delta\chi^2(\texttt{sfhz}) + {\rm log}({\rm N_{filters}}))/2}$. A high Bayes factor suggests strong evidence in favor of the fit with the BH* template (e.g., $B>100$ is a common threshold for decisive evidence following \citealt{Kass95}). In principle, the BIC is a good approximation only for well-constrained models (i.e., for ${\rm N_{filters}}\gg k$), which is not true in our case. Further, a  high $\Delta\chi^2$ (and thus a low BIC and high Bayes factor) does not guarantee that the underlying model -- in our case a combination of galaxy templates and a BH* template -- is correct. Because we do not claim to have access to the ``true'' model for each spectrum, the exact values of the resulting Bayes factors should be interpreted cautiously. Nevertheless, this formalism takes into account that different sources are covered by different numbers of photometric filters, ranging from ${\rm N_{filters}}=6$ for most PANORAMIC pointings to ${\rm N_{filters}}=23$ in some parts of the Abell-2744 cluster field, and is therefore better suited than only comparing $\chi^2_{\rm min}$ values. For EGS-73885 shown in Figure \ref{fig:worked_example}, we find $B(\texttt{sfhz})=1.38\times10^6$. This example illustrates how our sample selection successfully identifies promising BH*-dominated candidates. We note that EGS-73885 has not been observed by any (public or proprietary) spectroscopic program to date (see Section \ref{sec:sample_sel_spec}). The example also shows a potential degeneracy between QG and BH* solutions in cases where the H$\alpha$-excess is less prominent or not well probed by the available photometry. We discuss this further in Section \ref{sec:overview_contaminants}. 

\subsection{Spectroscopic Validation}
\label{sec:sample_sel_spec}

To validate our sample, we leverage the large and growing archive of spectroscopic data from JWST. We start by cross-matching our sample with all public spectra available on the DJA as of April 29, 2026, finding 55 spectra with grade 3 redshifts (i.e., robust redshifts as determined by visual inspection). Of those, 51 have at least one robust redshift from a PRISM spectrum, the other four sources have robust redshifts from G235M and G395M medium resolution grating spectra (two each). A full list of programs contributing public spectra used in this work is provided in Appendix \ref{sec:appendix_list_of_programs}. We additionally include six PRISM spectra with grade 3 redshifts from two recently observed programs, GO-7511 (PI Covelo-Paz, four spectra), and GO-8915 (PI Weibel, two spectra), increasing the total number of sources with robust spectroscopic redshifts to 61, i.e., $\sim25\%$ of the sample. 

This spectroscopic sub-sample includes various well-known LRDs. Besides The Cliff and the MoM-BH*-1 which are selected by design (i.e., because they are used as templates), we select A2744-45924 \citep{Labbe24}, A2744-QSO1 \citep{Furtak23QSO1}, UNCOVER-20466 \citep{Kokorev23}, CAPERS-LRD-z9 \citep{Taylor25}, and RUBIES-BLAGN-1 \citep{Wang25Outflow}. 

To estimate the possible contamination rate of our photometric sample, we focus on the sub-set of 57 sources with grade 3 PRISM spectra, as a more in-depth analysis of the medium resolution spectra is beyond the scope of this work. First, we identify one source as a BD and return to this contaminant in Section \ref{sec:overview_contaminants}. Then, we cross-match with the spectroscopic LRD-sample from \citet{degraaff25pop}, and find 34 matches. Their selection method builds on \citet{Hviding25} where LRDs are defined through a combination of a (spectroscopic) V-shape, compactness, and the detection of broad lines. The latter requires a medium grating spectrum with a sufficiently high SNR. However, \citet{degraaff25pop} show that sources that satisfy the V-shape and compactness criteria turn out to have broad lines in $98$\%\ of the cases when such data are available. Thus, we consider candidates that are part of their sample to be robust LRDs (additionally selected here to have a dominant BH*-component). To further investigate the remaining 22 sources with PRISM spectra, we apply the V-shape classification from \citet{Hviding25}. The details of this analysis are described in Appendix \ref{sec:appendix_vshape_classification}. Here, we directly move to estimating contamination rates. In doing so, we exclude the MoM-BH*-1 and The Cliff, because they were used as templates in the sample selection. Only considering sources with confirmed V-shapes as per \citet{Hviding25} to be robust LRDs, we find a very conservative upper limit on the contamination rate of 14/55 ($25.5\%$). However, except for the confirmed BD contaminant, we cannot rule out a V-shaped SED for any of these candidates due to insufficient SNR in the measured slopes (see Appendix \ref{sec:appendix_vshape_classification}). If we additionally consider sources that show breaks or turnovers around ${\rm H}_\infty$, along with strong H$\alpha$-emission and/or tentative broadened lines to be confirmed LRDs, the potential contamination rate decreases to 7/55 ($12.7\%$). In addition to the BD contaminant, we only identify one likely contaminant (UDS-27591, showing a blue UV-slope and a tentatively blue optical slope, along with strong narrow emission lines, see Appendix \ref{sec:appendix_vshape_classification}), resulting in a lower limit on the contamination rate of 2/55 ($3.6\%$).

\begin{figure*}
     \centering
     \includegraphics[width=1.8\columnwidth]{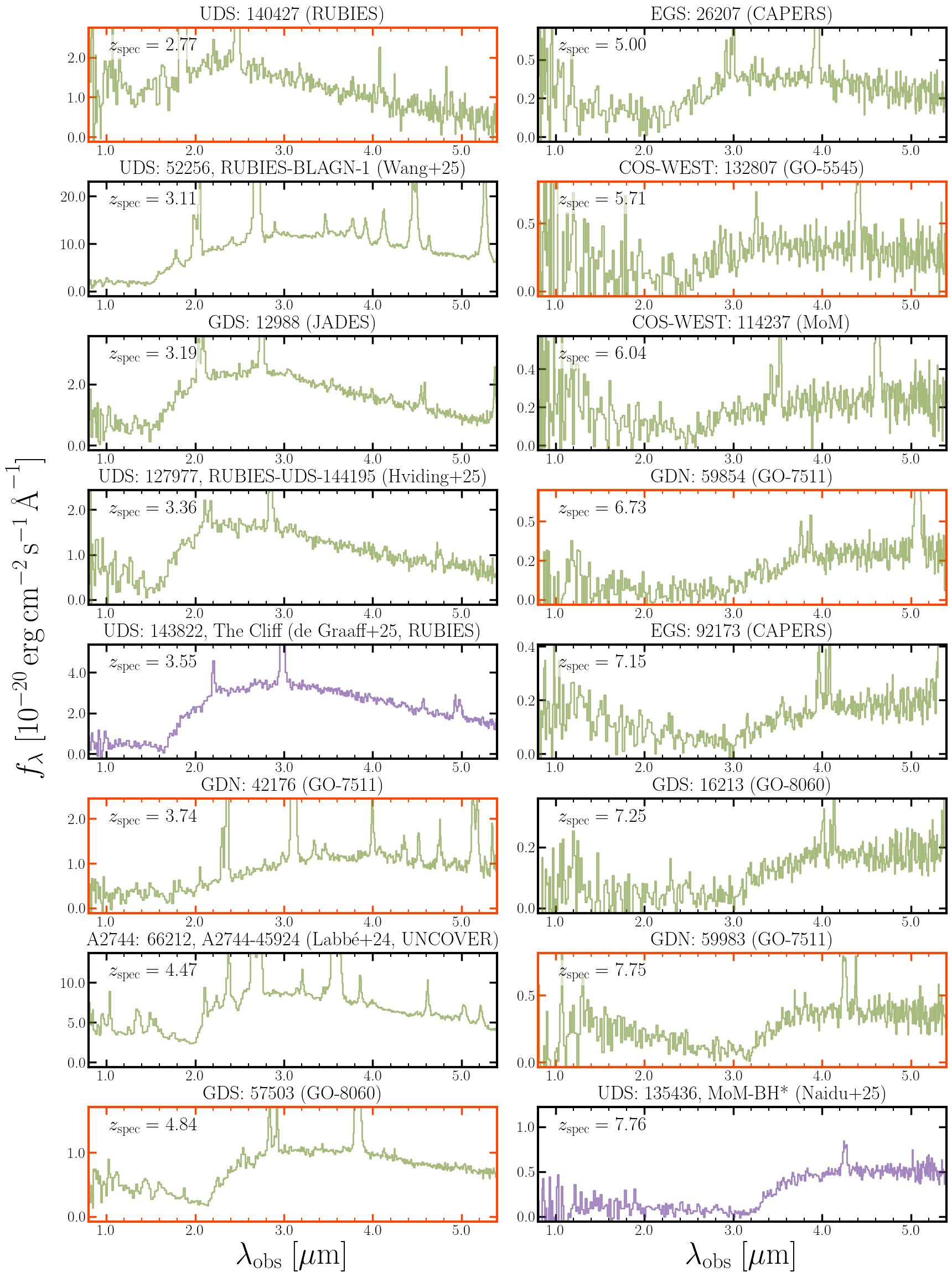}
     \caption{Example PRISM spectra of sources selected in this work, ordered by redshift. Spectra of well-known LRDs/BH*s (single object papers) are labeled accordingly, and spectra presented for the first time here are highlighted with red frames. This illustrates that our selection successfully identifies objects that resemble the paradigmatic sources The Cliff \citep{degraaff25} and the MoM-BH*-1 \citep{Naidu25BHstar} over a wide redshift range. These two sources are plotted in purple to indicate that they were used as templates in the sample selection. Most of the objects shown here are not picked up by prevalent photometric LRD selection methods.}
     \label{fig:gallery_spectra}
 \end{figure*}

In Figure \ref{fig:gallery_spectra}, we show a gallery of 16 PRISM spectra of sources selected as BH*-dominated candidates in this work. This includes known sources from the literature which are labeled accordingly, as well as spectra presented here for the first time, which are highlighted with red frames. GDN-59854 and GDN-59983 from GO-7511 (PI Covelo-Paz) are new BH*-dominated sources at $z_{\rm spec}>6.5$ (see Witten et al. in preparation). The spectra of The Cliff and the MoM-BH*-1 are plotted in purple to mark them as templates used for the sample selection (Section \ref{sec:sample_sel_templates}). The gallery highlights that the selection presented here successfully identifies sources that resemble these paradigmatic BH*s across a wide range of redshifts. Importantly, many of the objects shown in Figure \ref{fig:gallery_spectra} are not picked up by prevalent photometric LRD searches. We cross-match the 16 sources with the public catalogs from \citet{Kokorev24} and \citet{Kocevski25stats}, noting that both works do not include the GOODS-N field, and \citet{Kokorev24} also do not include the Abell-2744 cluster field. We find that \citet{Kokorev24} only select three of the 12 sources in overlapping fields, and \citet{Kocevski25stats} select five out of 13; the only object selected by both is GDS-16213. Most notably, both selections miss The Cliff as already discussed in \citet{Hviding25}. \citet{Kokorev24} also miss the MoM-BH*-1 \citep{Naidu25BHstar} as well as the RUBIES-BLAGN-1 \citep{Wang25Outflow}, while \citet{Kocevski25stats} do not include A2744-45924 \citep{Labbe24} and our GDS-57503. We further discuss how our sample compares to photometric LRD samples in the literature in Section \ref{sec:sample_sel_v-shapes}.

Next, we cross-match our sample in the Abell-2744 cluster field with the DR1 catalog from the All the Little Things (ALT) survey (GO-3516, PIs Naidu \& Matthee, \citealt{Naidu24}). We find three matches, two of which already have a NIRSpec-based spectroscopic redshift. The ALT catalog thus adds one grism redshift to our spectroscopic sample. Interestingly, the corresponding source at $z_{\rm grism}=2.49$ (ID 49263, see Section \ref{sec:overview_lowz}) was highlighted in Figure 19 of the ALT survey paper \citep{Naidu24} to show a broadened Pa-$\gamma$ line as well as HeI in absorption close to the systemic redshift. The authors therefore hypothesized that it may be a lower redshift version of an LRD.

We further cross-match our sample in the GOODS-N and GOODS-S fields with the public H$\alpha$-emitter catalog from \citet{CoveloPaz25} that is based on the FRESCO survey \citep{Oesch23}, finding two matches with additional spectroscopic redshifts that are consistent with our $z_{\rm phot}$ estimates. One of these sources is a confirmed broad line H$\alpha$ emitter from \citet[][their GOODS-N-13733]{Matthee23}, the other is too low SNR to determine whether the H$\alpha$ line in the grism spectrum is broadened.

Another of our BH*-dominated candidates has been observed as a backup target of program 116.294D (PI Matthee) with VLT/X-Shooter, and it was then followed-up by the DDT program ID 116.2AQ0 (PI Matthee). These observations confirmed its redshift of $z_{\rm spec}=1.73$ through the detection of broad H$\alpha$ and H$\beta$ lines. The source with ID 1115 is highlighted again in Section \ref{sec:overview_lowz}; it has been named PAN-BH*-1 and is discussed in detail in \citet{Torralba26}.

\begin{figure}
     \centering
     \includegraphics[width=0.47\textwidth]{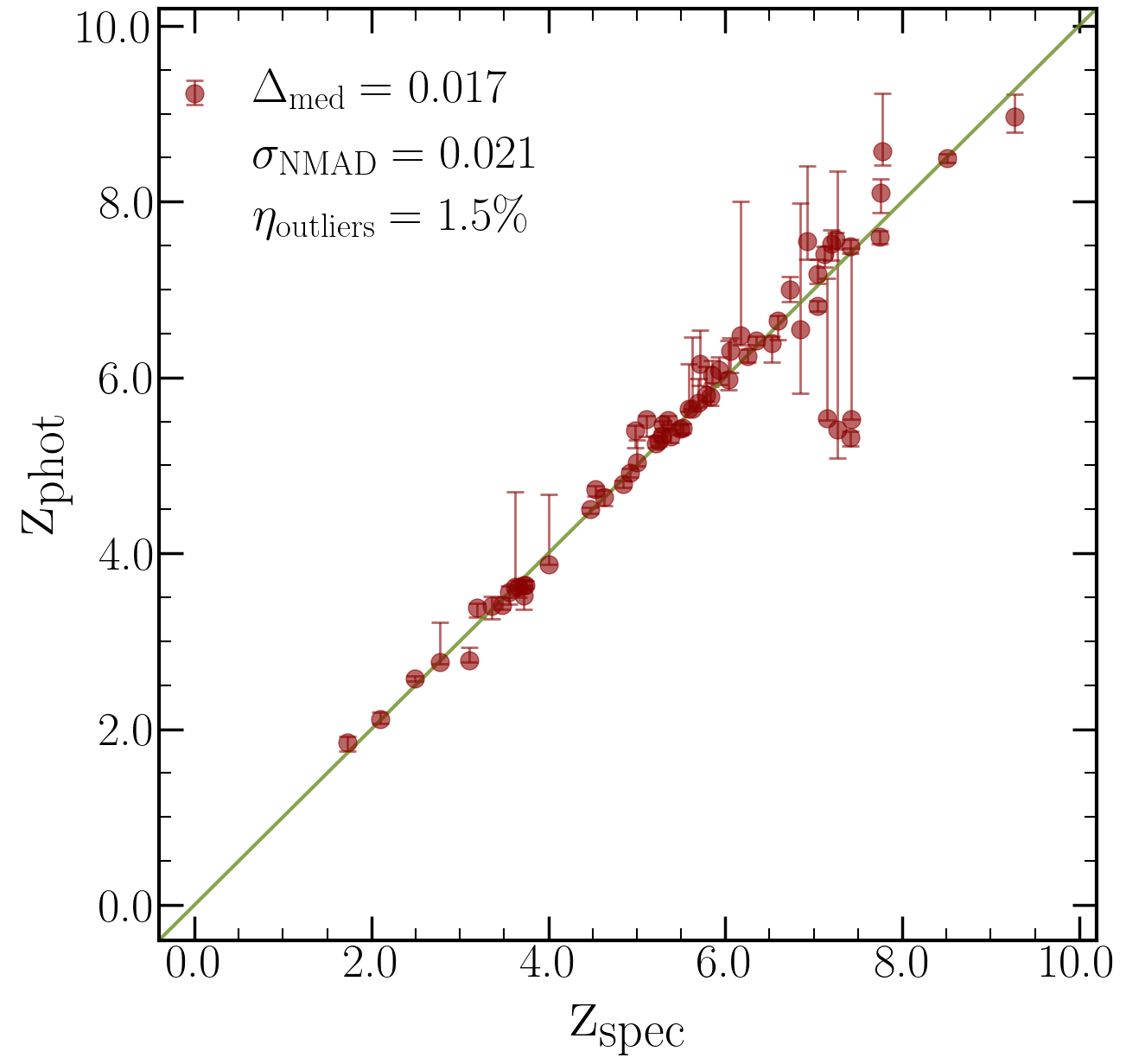}
     \caption{Photometric redshifts from \texttt{eazy} against spectroscopic redshifts for 65 of our BH*-dominated candidates that have robust redshifts from NIRSpec spectra (61), public grism catalogs (3), and the source named PAN-BH*-1 \citep{Torralba26} with a $z_{\rm spec}=1.73$ from VLT/X-Shooter.}
     \label{fig:zphot_zspec}
 \end{figure}

Overall, 65 of our 243 initial candidates (including the contaminants discussed in Section \ref{sec:overview_contaminants}) have robust spectroscopic redshifts, corresponding to $26.7$\%\ of the sample. We plot them against our photometric redshifts in Figure \ref{fig:zphot_zspec}. The comparison shows remarkably good agreement between $z_{\rm phot}$ and $z_{\rm spec}$ with no significant systematic offset. Defining the photometric redshift deviation as $\Delta = | z_{\rm phot} - z_{\rm spec}|/(1 + z_{\rm spec})$, we find a median deviation of $\Delta_{\rm med}=0.017$ and a scatter of $\sigma_{\rm NMAD}=0.021$. We then measure the outlier fraction, $\eta_{\rm outliers}$, as the number of sources satisfying $\Delta>0.1$, and find $\eta_{\rm outliers} = 1.5\,$\%. We compare these numbers to \citet{Hviding25} who derived photometric redshifts for their sample of spectroscopic LRDs using \texttt{eazy} with the \texttt{blue\_sfhz} template set and found much larger values of $\Delta_{\rm med}=0.045$, $\sigma_{\rm NMAD}=0.127$ and $\eta_{\rm outliers}=0.44$. They identified photometric redshift failures, represented by the high outlier fraction, as a major reason for incompleteness of photometric LRD samples. This comparison shows that the inclusion of BH* templates vastly improves photometric redshift estimates for (BH*-dominated) LRDs, confirming the suitability of our templates and the corresponding composite ${\rm BH*}+{\rm host}$ galaxy fits. We note that the four sources at $z_{\rm phot}\sim5-6$ and $z_{\rm spec}\sim7-8$ correspond to confusion between H$\alpha$ and [O{\sc iii}]/H$\beta$ boosting the F444W filter, as is reflected by the error bars on $z_{\rm phot}$ in three cases. The remaining outlier at $z_{\rm spec}=0$ is the BD which we discuss in Section \ref{sec:overview_contaminants}.

We finally use the DJA to identify sources for which NIRSpec spectra have been taken but where extractions are not available, usually because the data are still proprietary, and we find 13 such sources. This means that 169 ($69.5$\%) of our BH*-dominated candidates do not yet have any NIRSpec spectroscopy.

\subsection{Selection Completeness}
\label{sec:sample_sel_completeness}

As is illustrated in Figure \ref{fig:gallery_spectra}, the selection presented here is sensitive to BH*-dominated LRDs, i.e., LRDs with relatively weak UV-emission, due to our requirement of a sub-dominant ($<20\%$) host galaxy contribution in the rest-optical. This causes us to miss a substantial fraction of the total LRD population, in particular strongly V-shaped, UV-blue LRDs, meaning that the sample compiled here is an incomplete subset of the total LRD population. We explore this by comparing to various photometric LRD selection methods in the next Section. Here, we wish to roughly quantify the incompleteness of our sample due to photometric noise, taking into account the varying imaging depth across the different fields used in this work. 

For simplicity, we only calculate the incompleteness caused by our requirement of ${\rm SNR(F444W)}>10$, and ${\rm mag(F444W)}<27$. This part of the selection is crucial because the F444W filter probes different rest-frame wavelengths as a function of redshift, introducing redshift-dependent completeness at, e.g., fixed optical luminosity. While the \texttt{eazy}-based requirement of a $>80\%$ BH* template contribution introduces substantial incompleteness relative to the full LRD population, it is less sensitive to photometric noise, in particular given the stringent SNR-cut in F444W. We therefore start from the six BH* templates presented in Section \ref{sec:sample_sel_templates} and scale each template to different values of redshift and monochromatic luminosity at $5100$\AA\ ($L_{5100}$). We compute synthetic photometric fluxes in F444W and then loop through the different fields in the master catalog to add noise to the photometric fluxes according to the respective median depth. We create 1000 realizations of the photometry in F444W for each BH* template, redshift, luminosity, and field, and then apply the F444W-based pre-selection cuts. The fraction of the 1000 realizations that passes these cuts provides a rough estimate of the photometric completeness as a function of redshift, $L_{5100}$, imaging depth, and for the different BH* templates. We return to these estimates in Section \ref{sec:sample_props_Lopt_BB}.

\subsection{Comparison to ``Classic'' V-Shape Selections}
\label{sec:sample_sel_v-shapes}

In the following, we compare our sample selection to photometric LRD selections from the literature. Building on the color selection cuts from \citet{Labbe25LRD}, \citet{Greene24} defined two complementary selections named \textit{red1} and \textit{red2}, and combined them with a compactness cut to select LRDs at lower and higher redshifts, corresponding to $z\lesssim6$ and $z\gtrsim6$ \citep{Kokorev24}. Specifically, their color cuts are

\begin{gather}
    {\it red1} = {\rm F115W - F150W < 0.8}\,\land\nonumber\\[1ex]
    {\rm F200W - F277W > 0.7}\,\land\nonumber\\[1ex]
    {\rm F200W - F356W > 1.0}\nonumber\\[1ex]
    \text{and}\nonumber\\[1ex]
    {\it red2} = {\rm F150W - F200W < 0.8}\,\land\nonumber\\[1ex]
    {\rm F277W - F356W > 0.7}\,\land\nonumber\\[1ex]
    {\rm F277W - F444W > 1.0}\nonumber\\[1ex]
    \label{eq:greene_lrd_sel}
\end{gather}

with all quantities in AB-magnitudes. This selects for a V-shaped SED, requiring a red color at the red end of the NIRCam-coverage (rest-optical), and a moderately blue color at the blue end (rest-UV). \citet{Kokorev24} modified the \textit{red2} cut to be somewhat less stringent, requiring ${\rm F277W - F356W > 0.6}$ and ${\rm F277W - F444W > 0.7}$. Simplifying the selection, \citet{PerezGonzalez24} only required ${\rm F277W - F444W > 1}$, combined with a slightly more stringent cut at the blue end, ${\rm F150W - F200W < 0.5}$. \citet{Barro24} and \citet{Akins25} instead opted for a selection more targeted at extremely red sources, without imposing cuts on the rest-UV, as ${\rm F277W - F444W > 1.5}$. Slightly changing the approach, \citet{Kocevski25stats} defined a V-shape based on $\beta_{\rm opt}$ and $\beta_{\rm UV}$, the slopes obtained from fitting a power law to the photometry red-wards and blue-wards of ${\rm H}_\infty$. More recently, \citet{Rinaldi26} defined more relaxed color cuts as ${\rm F277W - F444W > 0.5}$ and ${\rm F150W - F200W < 1}$ with the goal of obtaining a more complete sample of LRDs. Some color-based selections add further cuts to remove BDs by requiring, e.g., ${\rm F115W - F200W > -0.5}$ \citep{Greene24}.

\begin{figure}
     \centering
     \includegraphics[width=0.47\textwidth]{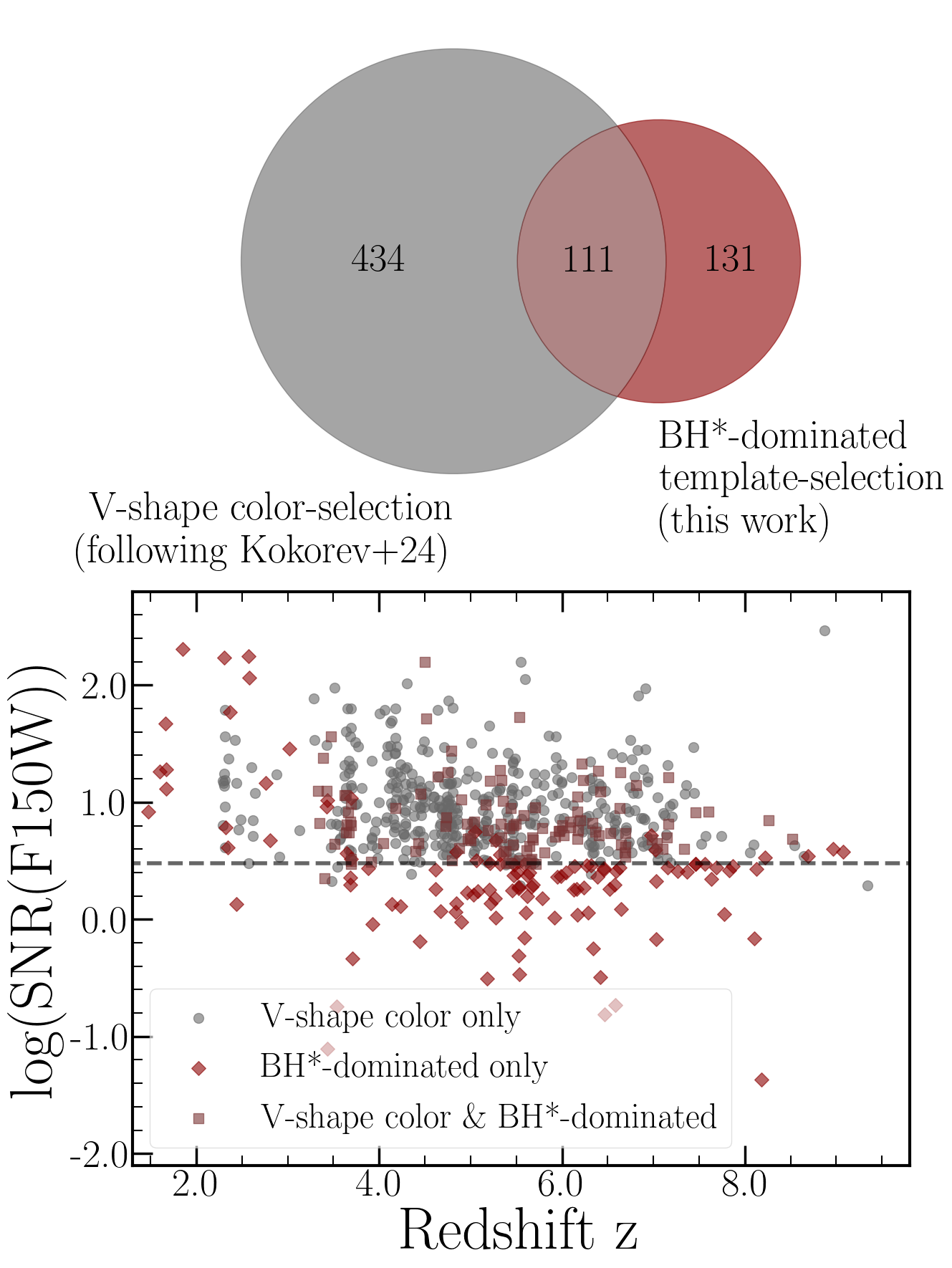}
     \caption{Top: Euler diagram showing the overlap between a color-based V-shape LRD selection following \citet{Kokorev24}, and the templated-based selection of BH*-dominated sources applied here. This highlights how these selections are highly complementary. Bottom: SNR in the F150W filter against redshift for our sample galaxies, as well as the color-selected V-shaped LRDs, illustrating the new parameter space probed in this work: low SNR in the rest-UV (as probed by F150W for $z\gtrsim3$) and the lowest redshifts ($z\lesssim2.5)$. Sources picked up by both selections are plotted as squares.}
     \label{fig:color_sel}
\end{figure}

To directly compare our template-based selection to a color-based V-shape selection, we apply the selection from \citet{Kokorev24} to our photometric catalog. For consistency, we adopt the same SNR and magnitude cuts as used in this work (${\rm SNR(F444W)}>10$ and ${\rm mag(F444W)}<27$), similar but not identical to the cuts applied in \citet[][${\rm SNR(F444W)}>14$ and ${\rm mag(F444W)}<27.7$]{Kokorev24}. Conveniently, their compactness cut is identical to the cut used in this work. We then apply their color selection including the BD removal cut from \citet{Greene24}. We require $3\sigma$ detections in all bands needed to measure colors, or replace non-detections ($<3\sigma$) by $2\sigma$ upper limits where appropriate to constrain the respective colors, consistent with \citet{Kokorev24}. We visually inspect all selected sources and remove spurious detections to obtain a sample of 545 V-shaped LRDs. In Figure \ref{fig:color_sel}, we show an Euler diagram illustrating the overlap between this color-based V-shape selection and our template-based search for BH*-dominated sources, showing that the two selections are highly complementary. Only $20.4\%$ of the V-shaped LRDs are also selected as BH*-dominated candidates. Conversely, only $45.9\%$ of our BH*-dominated candidates are picked up by the color-based V-shape selection. The main reason for this is that many BH*-dominated candidates are faint in the rest-UV, making it impossible to constrain the blue part of the V-shape. We illustrate this in the bottom of Figure \ref{fig:color_sel} where we plot the SNR in F150W against redshift for our sample of BH*-dominated candidates, and the V-shaped LRD sample. The vast majority of our BH*-dominated candidates that are missed by the V-shape selection lie at ${\rm SNR(F150W)}<3$ (dashed line). The importance of rest-UV SNR-cuts for the selection of UV-faint LRDs has already been discussed in \citet{Hviding25}.

Irrespective of the SNR in F150W, most of our sample galaxies do show red colors in ${\rm F277W - F444W}$. Specifically, $71.1\%$ of our sample galaxies satisfy ${\rm F277W - F444W}>1$. For sources with $z_{\rm phot}>3,$ 4, and 5 this fraction increases to $77.1$, $89.1$, and $100\%$. Many of the lower redshift candidates that are missed by a ${\rm F277W - F444W}$-based selection (\textit{red2} in the \citealt{Kokorev24} selection, see Equation \ref{eq:greene_lrd_sel}) instead show a red color in ${\rm F200W - F356W}$ (\textit{red1}). E.g., $85.5\%$ of all our sample galaxies satisfy ${\rm F200W - F356W}>1$, though even this color selection fails at $z\lesssim2.5$ (see the examples in Section \ref{sec:overview_lowz}).

Towards the low redshift end of our sample ($z\lesssim3$), the SNR in the F150W filter generally increases as it starts probing rest-optical wavelengths. While the \textit{red1} selection relies on a detection in F115W instead of F150W, we note that $\sim90\%$ of the sources undetected in F150W are also undetected in F115W. We note that there is no overlap between BH*-dominated candidates and V-shaped LRDs at $z\lesssim3$, and there are no V-shaped LRDs at $z<2$, reflecting that color-based selections are inherently redshift-dependent, while our template-based search is, in principle, redshift-independent. 

For further comparison, we apply various other color cuts from the literature to our sample. The selections from \citet{Greene24}, \citet{PerezGonzalez24}, and \citet{Rinaldi26} recover $47.1\%$, $34.7\%$, and $42.6\%$ of our candidates, similar to the $45.9\%$ inferred for the \citet{Kokorev24} selection above. Interestingly, the stringent single color cut from \citet{Akins25} selects $58.7\,$\% of our sample because it does not rely on rest-UV fluxes. By design, it is however biased towards the reddest, most extreme, and typically high redshift sources ($z\gtrsim5$). We compare these findings to \citet{Hviding25} who inferred the completeness of the \citet{Kocevski25stats}, \citet{Kokorev24} and the single-color ${\rm F277W}-{\rm F444W}>1.5$ \citep[e.g.,][]{Akins25} samples based on their spectroscopic sample of RUBIES LRDs, and found values of $61.8\%$, $50.0\%$, and $35.29\%$. This means that the selection based on $\beta_{\rm UV}$ and $\beta_{\rm opt}$ achieves the highest completeness for LRDs in general, and also picks up close to half of our BH*-dominated candidates. A single color cut such as used in \citet{Akins25} is relatively successful at selecting the most extreme LRDs, including BH*-dominated sources, but is highly incomplete for the general population of LRDs. Typical color-based selections pick up roughly half of the population of LRDs, and an even smaller fraction of BH*-dominated objects. We emphasize that none of the photometric selection methods discussed above picks up any of our BH*-dominated candidates at $z_{\rm phot}<2.5$ where the NIRCam wide filters at $\lambda_{\rm obs}\gtrsim2\mu{\rm m}$ probe the declining part of the rest-frame optical to NIR SED (see Section \ref{sec:overview_lowz}).

The findings of this Section illustrate that the sample presented here is a subset of the LRD population that consists of objects with a dominant BH* contribution and that are relatively faint in the rest-UV, due to the faint, sub-dominant host galaxy. As such, our sample is highly complementary to previous photometric selections, and sensitive to a different part of the LRD parameter space.

\section{Sample Overview}
\label{sec:overview}

\begin{figure*}
     \centering
     \includegraphics[width=1.9\columnwidth]{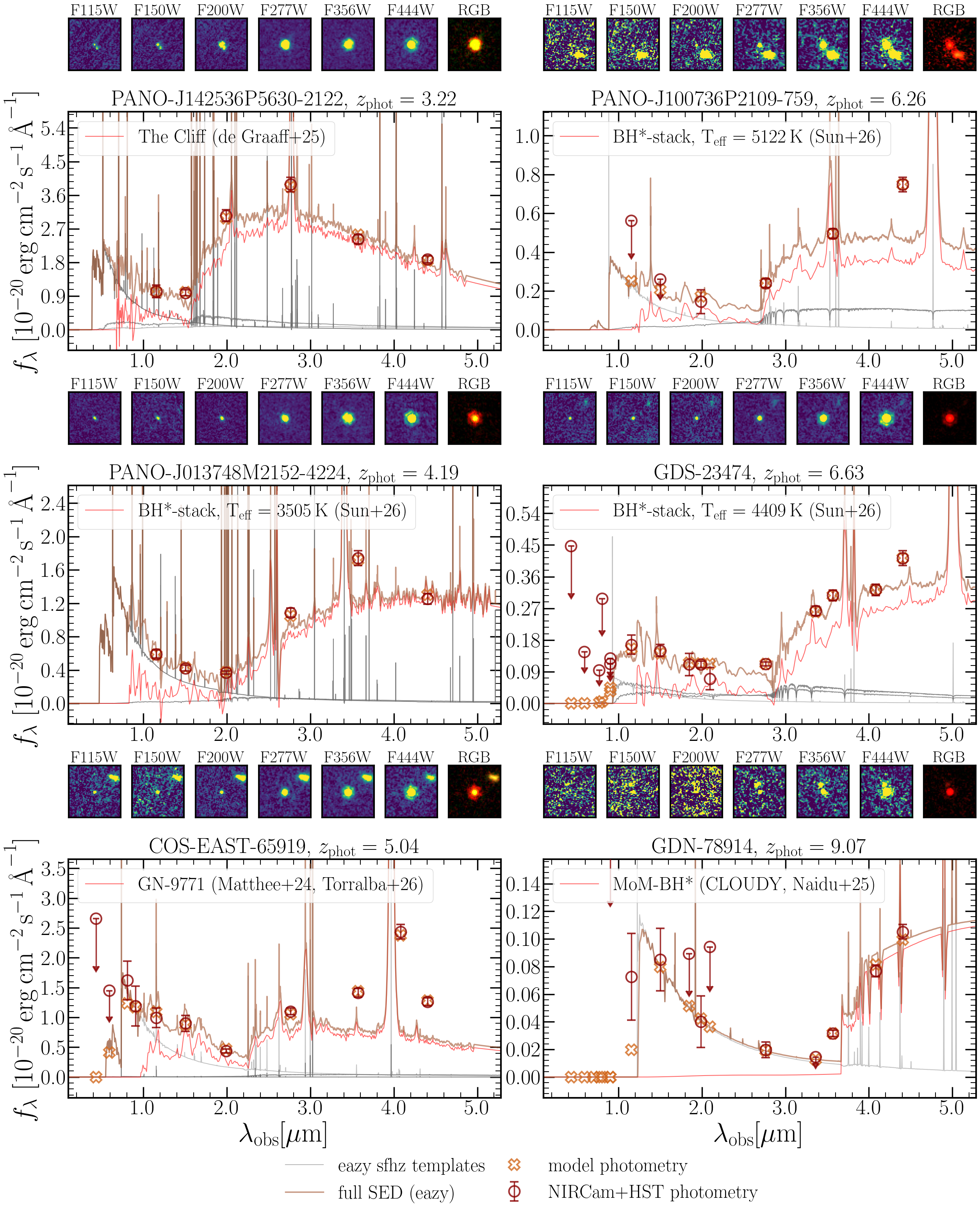}
     \caption{Overview of the sample selection. We show one BH*-dominated candidate selected by each of the six BH* templates used in this work (Section \ref{sec:sample_sel_templates}). The displayed candidates span a wide range in redshift ($z\sim3-9$), and they highlight various levels of rest-frame UV emission, as well as different photometric constraints depending on redshift and wavelength coverage. The RGB stamp shown for each source is constructed from F444W (red), F277W (green) and F115W (blue). Gray lines show the contribution of different \texttt{eazy} templates to the best-fitting SED (brown line). The contribution of the respective BH* template that dominates the fit is shown in red.}
     \label{fig:overview}
 \end{figure*}

To provide a first overview of our photometric sample, we show six BH*-dominated candidates that do not have public spectra on the DJA in Figure \ref{fig:overview}. We choose one source selected by each of the six BH* templates, and a wide range of photometric redshifts ($z\sim3-9$). Lower redshift candidates are discussed separately in Section \ref{sec:overview_lowz}. The displayed candidates illustrate the varying filter coverage across fields ranging from six NIRCam wide filters in PANORAMIC to a wealth of HST and NIRCam wide and medium bands in legacy fields such as the GDS. We emphasize that BH*-dominated candidates at different redshifts can be successfully selected from pure parallel imaging in six NIRCam filters. The candidates in Figure \ref{fig:overview} further show different levels of rest-frame UV flux that can be reconciled with a $>80\%$ contribution of the BH* template in the rest-optical. Subsequently, we discuss the lowest and highest redshift candidates in  our sample in more detail.

\subsection{Low Redshift BH*s ($z<2.5$)}
\label{sec:overview_lowz}

BH*-dominated sources at low redshift are particularly valuable for various reasons: First, their apparent brightness is relatively high at a fixed intrinsic luminosity, enabling high-resolution and high-SNR JWST spectroscopy in relatively short integration times. For the brightest objects at $z\sim1.5-2.5$, even spectroscopy from the ground becomes viable \citep[see][]{Torralba26}. Second, at lower redshift, JWST/NIRSpec has access to different spectral features such as Paschen lines, HeI at $1.08\mu{\rm m}$, and even potential water absorption features at $\sim1.4\mu{\rm m}$ \citep[e.g.,][]{Wang26} which may provide crucial insights into the physical conditions that prevail in BH*-dominated sources. Third, the rapid decline in the number density of LRDs at $z\lesssim4$ is poorly understood \citep[e.g.,][]{Ma26} and it may in part be related to selection effects inherent to photometric color selections that are mitigated by the selection strategy applied here.

\begin{figure*}
     \centering
     \includegraphics[width=1.9\columnwidth]{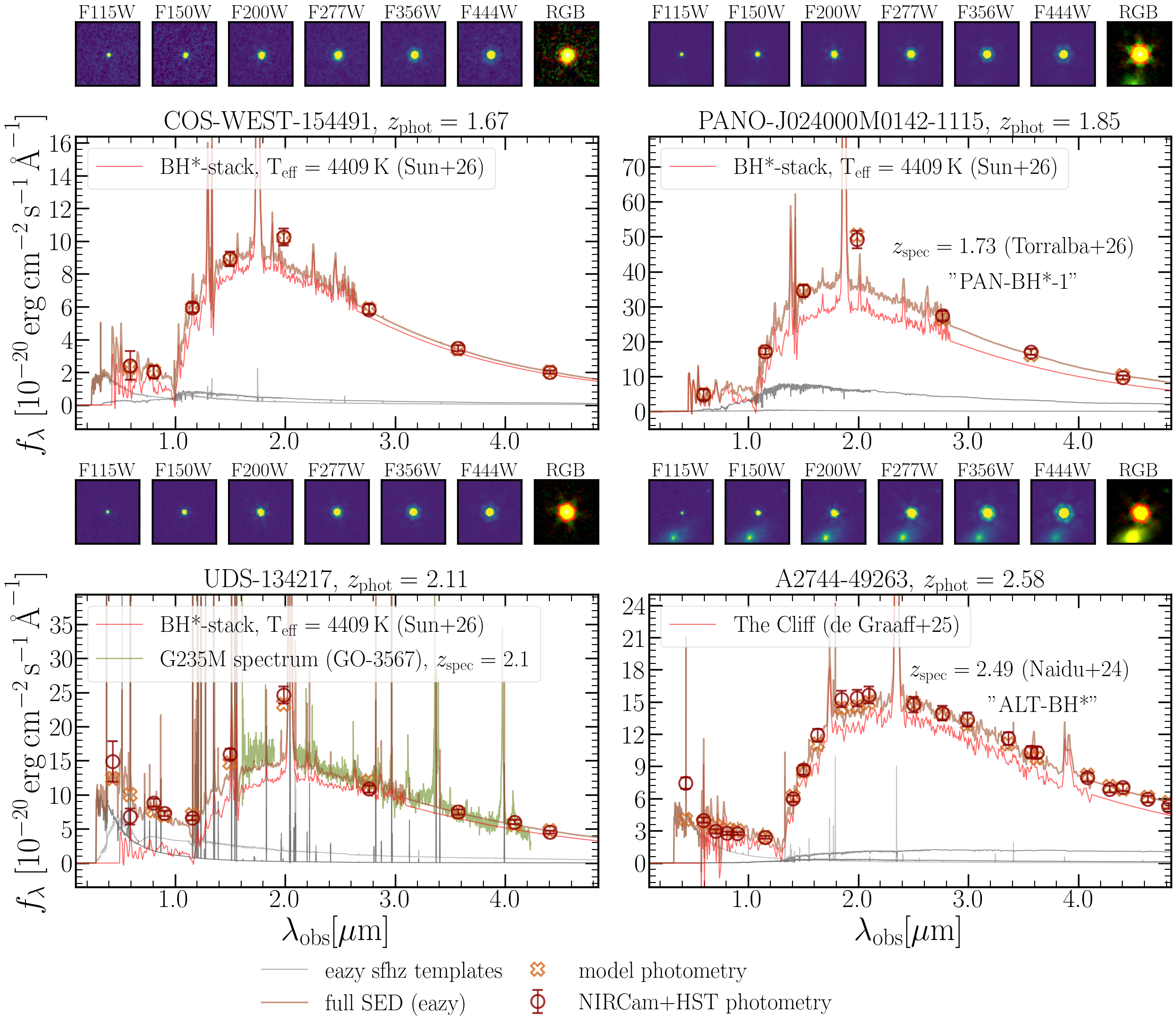}
     \caption{Most promising BH*-dominated candidates at low redshift ($1.5<z<2.5$), in analogy to Figure \ref{fig:overview}. Top left: Pure photometric candidate at $z_{\rm phot}=1.67$ where the photometry closely follows the BH* stack at $T_{\rm eff}=4409\,{\rm K}$, including the Balmer break, and the extension to $\lambda_{\rm rest}>1.15\,\mu{\rm m}$ assuming a pure blackbody. Top right: PAN-BH*-1 spectroscopically confirmed in \citet{Torralba26} with a spectrum from VLT/X-Shooter showing remarkably strong absorption on top of a broad H$\alpha$ line at $z_{\rm spec}=1.73$. Bottom left: A source with a NIRSpec/G235M spectrum from GO-3567 (PI Valentino), confirming its redshift as $z_{\rm spec}=2.1$ through broadened Paschen-lines and HeI/$1.08\,\mu{\rm m}$ in emission and absorption. Bottom right: BH*-dominated candidate in the Abell-2744 cluster field that was highlighted in the ALT survey paper \citep[][thus we name it ``ALT-BH*'']{Naidu24} with a broad Pa$\gamma$ line as well as HeI in absorption at $z_{\rm spec}=2.49$.}
     \label{fig:lowz}
 \end{figure*}

In Figure \ref{fig:lowz}, we present our four most promising $z<2.5$ BH*-dominated candidates. Sorted by increasing redshift, COS-WEST-154491 at $z_{\rm phot}=1.67$ in the top left shows a continuum shape that is remarkably similar to the BH* stack at $T_{\rm eff}=4409\,{\rm K}$ \citep{Sun26}, including its extension to longer wavelengths as a pure blackbody. PANO-J024000M0142-1115 is the source tagged PAN-BH*-1 in \citet{Torralba26} with a spectroscopic redshift of $z_{\rm spec}=1.73$ from VLT/X-Shooter. UDS-134217 in the bottom left has a G235M spectrum from GO-3567 (PI Valentino), confirming its photometric redshift as $z_{\rm spec}=2.1$ through various prominent emission lines, in particular broadened Pa$\beta$ and Pa$\gamma$ as well as HeI at $1.08\,\mu{\rm m}$ in both emission and absorption. Finally, A2744-49263 in the bottom right has a NIRCam/grism spectrum from the ALT survey ($z_{\rm grism}=2.49$), as discussed in Section \ref{sec:sample_sel_spec}. In our photometric sample, the source is fit with The Cliff template which yields $z_{\rm phot}=2.55$, close to the spectroscopic redshift. We refer to this source as the ALT-BH* hereafter.

\subsection{High Redshift BH*s ($z>7$)}
\label{sec:overview_highz}

\begin{figure*}
     \centering
     \includegraphics[width=1.9\columnwidth]{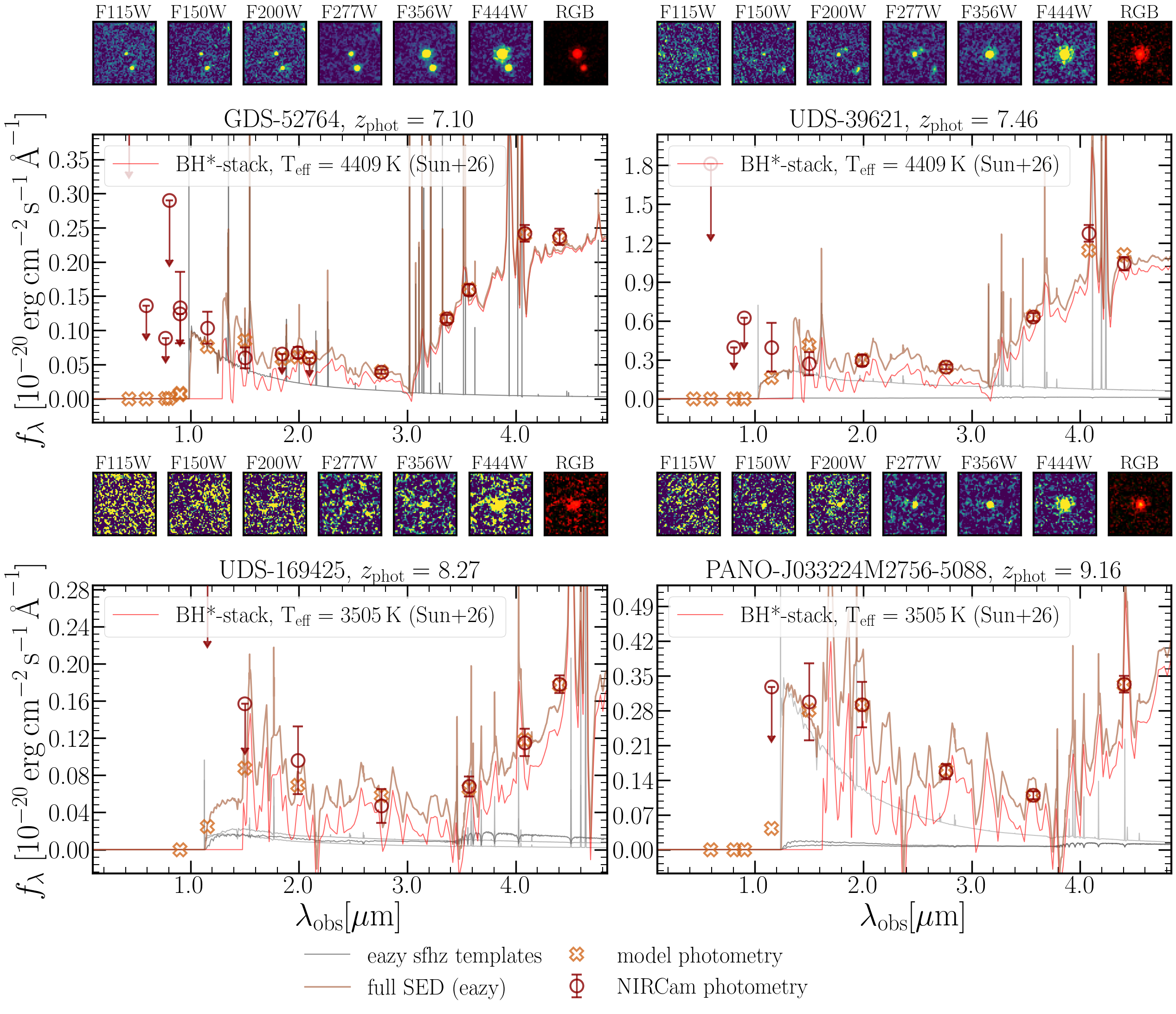}
     \caption{Most promising BH*-dominated candidates at high redshift ($z>7$), in analogy to Figure \ref{fig:overview}. Top left: Candidate at $z_{\rm phot}=7.10$ with deep photometry from JWST and HST in the GOODS-S field. Top right: Slightly higher redshift ($z_{\rm phot}=7.46$), brighter candidate in the UDS field. Bottom left: Fainter candidate at $z_{\rm phot}=8.27$ with a smoothly rising continuum from $\lambda_{\rm obs}\sim3.5-4.5\mu{\rm m}$ that is fit with the coldest BH*-stack ($T_{\rm eff}=3505\,{\rm K}$). Bottom right: A candidate at $z_{\rm phot}=9.16$ identified in pure parallel imaging from PANORAMIC adjacent to legacy imaging in the GOODS-S field.}
     \label{fig:highz}
 \end{figure*}

Next, we turn to the other end of the redshift distribution, and highlight four of our highest redshift ($z>7$) BH*-dominated candidates in Figure \ref{fig:highz}. None of these sources has a public spectrum on the DJA. GDS-52764 at $z_{\rm phot}=7.10$ shown in the top left has deep HST and JWST photometry, including medium bands that constrain the Balmer break/turnover as well as the rest-frame UV. UDS-39621 in the top right has sparser wavelength coverage and shallower imaging from PRIMER, but is significantly brighter than GDS-52764, and fit at a slightly higher redshift ($z_{\rm phot}=7.46$). UDS-169425 in the bottom left is again fainter at $z_{\rm phot}=8.27$ and thus the least robust of the displayed candidates. Finally, PANO-J033224M2756-5088 in the bottom right is selected from PANORAMIC pure parallel imaging adjacent to legacy imaging in the GOODS-S field. At $z_{\rm phot}=9.16$ it is nominally the source with the highest bolometric luminosity in the entire sample as we discuss in Section \ref{sec:sample_props_lowhighLbol}.

\subsection{Contaminants}
\label{sec:overview_contaminants}

\begin{figure}
     \centering
     \includegraphics[width=0.47\textwidth]{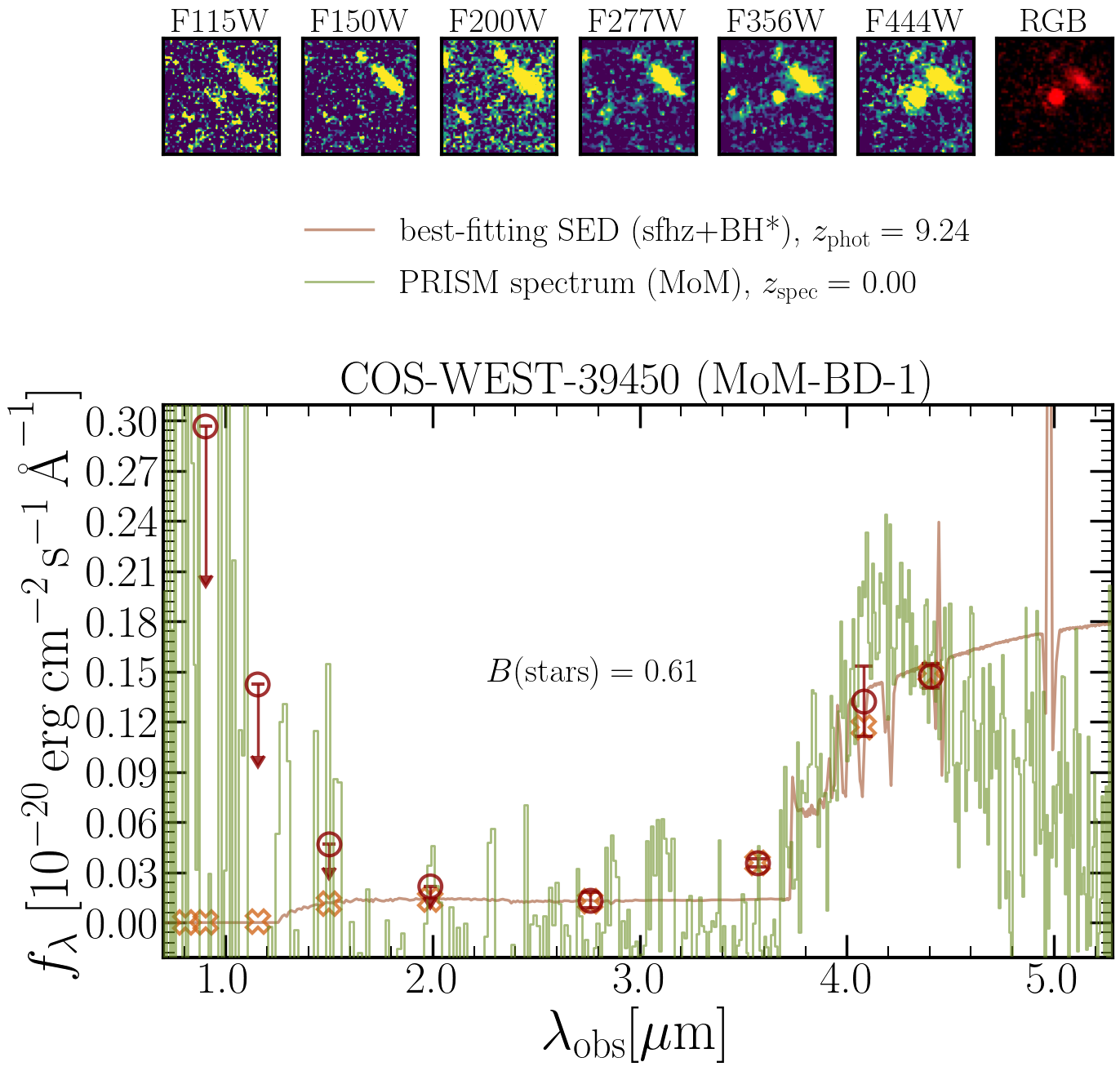}
     \caption{BH*-dominated candidate at $z_{\rm phot}=9.24$, spectroscopically identified as a cold BD through the MoM program. We refer to this source as the MoM-BD-1 hereafter. The best-fitting \texttt{eazy} SED and the photometry are displayed in analogy to Figure \ref{fig:overview}. The PRISM spectrum from MoM is shown in green. The Bayes factor $B({\rm stars})=0.61$ indicates no evidencde in favor of the fit with the BH* template, reflecting the degeneracy between solutions when only NIRCam photometry is considered.}
     \label{fig:mom_bd}
 \end{figure}

In Section \ref{sec:sample_sel_spec}, we estimate a contamination rate of $3.6-25.5\%$ for our photometric sample based on the subset of our candidates with grade 3 PRISM spectra. The only unambiguously identified contaminant is shown in Figure \ref{fig:mom_bd}. It is a candidate BH*-dominated source at $z_{\rm phot}=9.24$, selected with the template based on a CLOUDY model of the MoM-BH*-1 \citep{Naidu25BHstar}. It shows a huge break in the photometry of a factor $>5$ between F444W and F356W, and even of $>20$ between F444W and F277W, the strongest break measured in our entire sample. The source was observed by MoM (GO-5224, PIs Oesch \& Naidu), and the PRISM spectrum reveals that it is a cold BD, with a steeply rising continuum that peaks at $\sim4.1\,\mu{\rm m}$ (see Hainline et al. in preparation).

Importantly, this type of contaminant only affects the high redshift end of the sample, where the supposed Balmer break of the BH* shifts to the red end of the NIRCam wavelength coverage, so that only one or two filters (F444W and F410M) are substantially boosted with respect to the other filters. As opposed to the BH* SED, the BD spectrum drops again rapidly at  $\lambda_{\rm obs}\gtrsim4.1\,\mu{\rm m}$, meaning that it can be easily distinguished with sufficiently deep MIRI data.

The photometry of COS-WEST-39450, MoM-BD-1 hereafter, resembles that of the object tagged \textit{Capotauro} in \citet{Gandolfi26}. The latter is roughly $3\times$ fainter in F444W which may explain why it is not detected in any filters blue-wards of F410M. \citet{Gandolfi26} discuss various physical origins of the source, including the possibility that it is a $z\sim30$ galaxy, a BH*, as well as a BD. In the latter case, a temperature of $\lesssim400\,{\rm K}$ is required to produce a sufficiently strong break between F444W and F356W, as well as to not produce emission at shorter wavelengths that would be detected by JWST. MoM-BD-1 illustrates that such sources indeed exist and produce NIRCam photometry that is largely degenerate with BH* solutions at $z\gtrsim7$. This is reflected in the measured small $\Delta\chi^2({\rm stars})=1.3$ between the \texttt{sfhz}+BH* templates and the large grid of stellar templates (Section \ref{sec:sample_sel_eazy}), resulting in a Bayes factor $B({\rm stars})=0.6$, suggesting no evidence in favor of the BH*-dominated solution \citep{Kass95}. We note that our sample selection cut based on $\Delta\chi^2({\rm stars})$ (Section \ref{sec:sample_sel_cuts}) likely removes most stellar contaminants from our sample, so that we are left with this extreme and potentially rare type of BD that is not entirely captured by our stellar templates. Future work may include this source as another stellar template to further refine searches for BH*-dominated sources at high redshift. 

We further identify a source, COS-WEST-29852, that is fit at $z_{\rm phot}=5.61$ with the BH* template based on GN-9771. If true, its magnitude of ${\rm mag(F444W)}=19.81$ would make it more than an order of magnitude brighter than any other BH*-dominated candidate at similar redshifts. We show imaging cutouts, the measured photometry, as well as the best-fitting SED of this source in Appendix \ref{sec:appendix_contaminant}. The photometry does not show distinctive features such as a break or filters boosted by emission lines, but suggests a smoothly declining red SED over the wavelength range probed by NIRCam. A more detailed investigation of COS-WEST-29852 is beyond the scope of this work, although we argue that it is unlikely to be a BH*-dominated source at $z_{\rm phot}=5.61$. As noted in Section \ref{sec:sample_sel_cuts}, we therefore remove this source, as well as the MoM-BD-1 from our sample, reducing our total sample size to 241.

For many of our BH*-dominated candidates, in particular at the low redshift end ($z\lesssim3.5$), the best fitting SED with the \texttt{sfhz} galaxy templates is that of a QG. We show and discuss a corresponding example in Figure \ref{fig:worked_example} where the fit with the BH* template is clearly preferred due to a strong Balmer break and H$\alpha$ emission boosting a photometric filter. Some objects in our sample indeed have Balmer breaks that are too strong to be reproduced by a stellar population (see the discussion regarding The Cliff in \citet{degraaff25}, and Figure \ref{fig:Lopt_zphot_BB}), but not all BH*-dominated candidates have strong Balmer breaks. The SEDs of the colder sources peak at longer wavelengths and their SEDs decline towards the rest-UV. At the Balmer limit there is thus not much continuum to break (see Figure \ref{fig:templates}). Further, in some cases, the limited depth of the photometric bands blue-wards of the Balmer break does not allow us to probe its full strength. On the other hand, the detection of an H$\alpha$-excess relative to the continuum depends on the redshift of the source, and the available photometric filters, causing the photometry of some candidates to be more degenerate with QG solutions, even if many photometric bands are available. For example, our source GDS-57503, spectroscopically confirmed to be a BH*-dominated LRD through GO-8060 (PI Egami) as shown in Figure \ref{fig:gallery_spectra}, has been confidently modeled to be a QG based on 22 photometric bands including 7 NIRcam and 8 MIRI filters in \citet{Williams24}. Crucially, any degeneracy between BH*-dominated and QG solutions is captured by the Bayes factor $B(\texttt{sfhz})$ which we use to define sample tiers in the following.

\subsection{Gold and Silver Sample}
\label{sec:overview_gold_sample}

We use the Bayes factors $B(\texttt{sfhz})$ and $B({\rm stars})$ to define a gold sample of BH*-dominated candidates that we consider the most robust because the fit with the BH* template is clearly favored over fits with standard galaxy templates as well as over fits with stellar templates. Specifically, we require $B(\texttt{sfhz})>100$ and $B({\rm stars})>100$, following Bayes factor thresholds for decisive evidence as provided in \citet{Kass95}. We consider the five sources for which \texttt{eazy} does not find any fit with the \texttt{sfhz} templates (see Section \ref{sec:sample_sel_cuts}) to have high Bayes factors. This yields a sample of 127 BH*-dominated sources that we refer to as the gold sample hereafter; the remaining 114 sources are referred to as the silver sample. We show the photometry, best-fitting SEDs from \texttt{eazy}, as well as RGB cutouts of all our gold sample candidates in Appendix \ref{sec:appendix_gallery}.

\section{Sample Properties}
\label{sec:sample_props}

We now turn to exploring sample properties of both our gold and silver samples of BH*-dominated candidates. We start with purely empirical quantities, specifically $L_{5100}$ and the Balmer break strength, before moving on to estimating bolometric luminosities. 

\subsection{Optical Luminosity and Balmer break strength}
\label{sec:sample_props_Lopt_BB}

\begin{figure*}
     \centering
     \includegraphics[width=1.75\columnwidth]{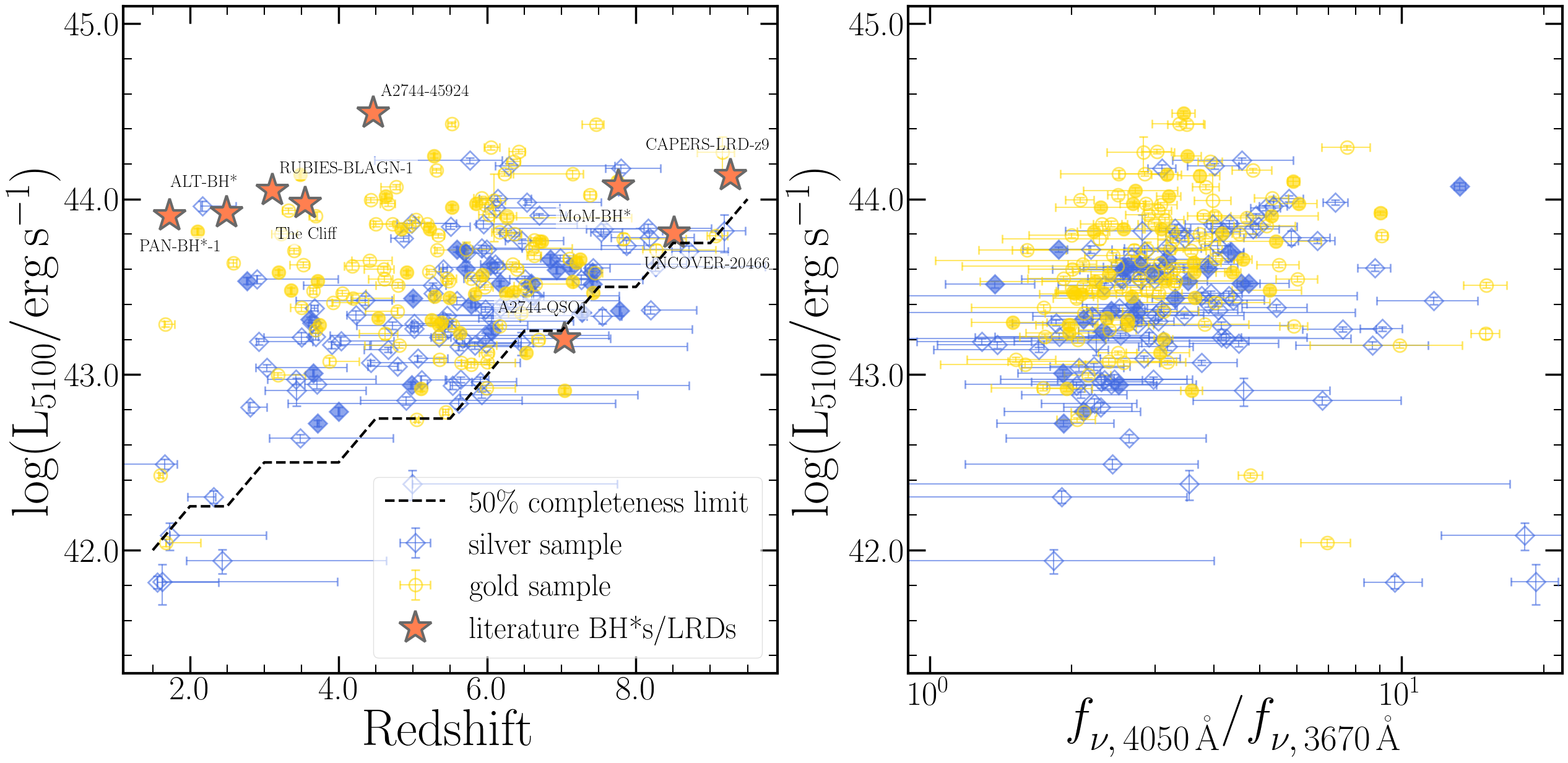}
     \caption{Monochromatic optical luminosity $L_{5100}$ against redshift ($z_{\rm phot}$, replaced by $z_{\rm spec}$ where available) on the left and against Balmer break strength ($f_{\nu,\,4050\mathrm{\mathring{A}}}/f_{\nu,\,3670\mathrm{\mathring{A}}}$) on the right. The gold and silver sample are displayed with different markers and colors, and sources with spectroscopic redshifts are plotted as filled markers. In the left panel, we further plot the rough $50\%$ completeness limit of our sample selection based on The Cliff template at the median depth across all fields, illustrating that the lower right edge of the distribution in the $z$-$L_{5100}$ diagram reflects a selection effect. These plots show that we identify BH*-dominated candidates across a wide range of redshifts ($z\sim1.5-9.5$) and luminosities $L_{5100}\sim10^{42}-10^{44.5}\,{\rm erg}\,{\rm s}^{-1}$, with the bulk of the sample being concentrated around $z\sim3-7$ and $L_{5100}\sim10^{43}-10^{44}\,{\rm erg}\,{\rm s}^{-1}$. The sample displays a broad range of Balmer break strengths of $\sim1.9-8.7$ (5th and 95th percentiles) with a high median of $3.05$, and some of the strongest Balmer breaks measured in the Universe to date. There is a weak apparent correlation between Balmer break strength and $L_{5100}$, consistent with the trend seen for LRDs in \citet{degraaff25pop}.}
     \label{fig:Lopt_zphot_BB}
 \end{figure*}

We use the best-fitting \texttt{eazy} SED to infer the monochromatic luminosity $L_{5100}$, as well as the Balmer break strength, where the latter is given by the ratio between the median fluxes in the ranges $\lambda_{\rm rest}\in[4060,4140]\,$\AA\ and $\lambda_{\rm rest}\in[3560,3640]\,$\AA, measured in $f_\nu$. We plot $L_{5100}$ against redshift (photometric or spectroscopic where available) and the derived Balmer break strength in Figure \ref{fig:Lopt_zphot_BB}. Uncertainties in $z_{\rm phot}$ are computed by excluding any probability density at $z>10$ because solutions at $z>10$ largely correspond to un-physical combinations of galaxy and BH* templates. We use uncertainties in rest-frame fluxes from \texttt{eazy} to compute uncertainties for $L_{5100}$ and the Balmer break strength.

The Figure highlights that we identify BH*-dominated sources across the entire redshift range we select for ($1.5<z<9.5$), and over almost three orders of magnitude in $L_{5100}$. Our sample is mostly concentrated around intermediate redshifts of $z\sim3-7$, and luminosities of $L_{5100}\sim10^{43}-10^{44}\,{\rm erg}\,{\rm s}^{-1}$ with the fraction of objects belonging to the silver sample slightly increasing towards the upper right (bright, high redshift) as well as the lower left (faint, low redshift) parts of the $z$-$L_{5100}$ diagram. The lower right edge of the distribution reflects a selection effect where we only detect intrinsically brighter objects at higher redshifts. This is illustrated by the $50\%$ completeness limit at the median imaging depth in F444W across all fields, computed according to Section \ref{sec:sample_sel_completeness} for the BH* template based on The Cliff. There are only a handful of candidates at $L_{5100}\lesssim10^{42.5}$, only two of which are part of the gold sample, and we return to these candidates in Section \ref{sec:sample_props_lowhighLbol}. At the other end of the distribution, we identify relatively few objects at $L_{5100}\gtrsim10^{44}$, with the source A2744-45924 \citep{Labbe24} standing out as the most optically luminous source. 
Our BH*-dominated candidates also show a broad range of Balmer break strengths, with most sources showing strong breaks. Specifically, $95\%$ of all candidates show Balmer breaks $>1.85$, and the median break strength is $3.05$, close to the strongest breaks observed in high redshift QGs \citep[$\sim3.1$;][]{degraaff25}. A few objects show some of the strongest Balmer breaks observed in the Universe to date (e.g., 21 objects have break strengths $>6$). We highlight various spectroscopically confirmed objects that are part of our sample: PAN-BH*-1 \citep{Torralba26}, the ALT-BH* \citep{Naidu24}, RUBIES-BLAGN-1 \citep{Wang25Outflow}, The Cliff \citep{degraaff25}, A2744-45924 \citep{Labbe24}, A2744-QSO1 \citep{Furtak23QSO1}, the MoM-BH*-1 \citep{Naidu25BHstar}, UNCOVER-20466 \citep{Kokorev23}, and CAPERS-LRD-z9 \citep{Taylor25}. All these sources lie at high optical luminosities relative to the sample at their respective redshift, except for A2744-QSO1 which is strongly lensed \citep{Furtak23QSO1}, illustrating that well-known literature sources tend to probe an extreme part of the parameter space. We finally note a weak correlation between $L_{5100}$ and Balmer break strength, consistent with the findings of \citet{degraaff25pop}.

\subsection{Bolometric Luminosities}
\label{sec:sample_props_Lbol}

\begin{figure}
     \centering
     \includegraphics[width=0.47\textwidth]{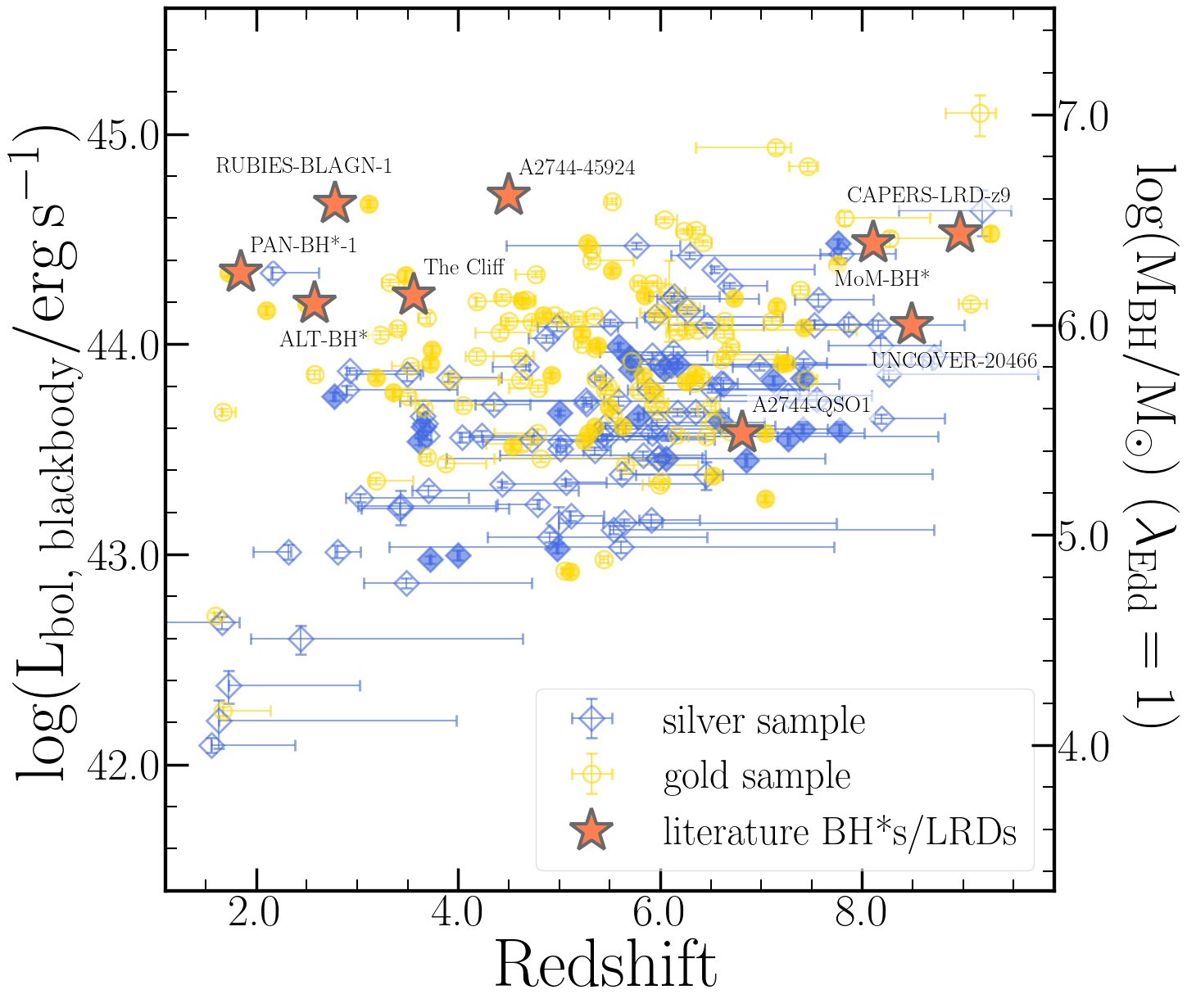}
     \caption{Bolometric luminosities derived by integrating the BH* template contribution to the best-fitting SED, plotted against redshifts ($z_{\rm phot}$, replaced by $z_{\rm spec}$ where available). Filled markers indicate sources with spectroscopic redshifts and various objects known from the literature are highlighted as stars, and labeled accordingly \citep{Torralba26,Naidu24,Wang25Outflow,degraaff25,Labbe24,Naidu25BHstar,Furtak23QSO1,Taylor25,Kokorev23}. The secondary y-axis specifies black hole masses derived from $L_{\rm bol}$ using Equation \ref{eq:bhmass}, and assuming $\lambda_{\rm Edd}=1$. This again highlights the broad range of redshifts and luminosities spanned by our sample, as well as the rapid decline in the number of candidates towards the highest ($L_{\rm bol}\gtrsim10^{44.5}\,{\rm erg}\,{\rm s}^{-1}$) and lowest ($L_{\rm bol}\lesssim10^{43.5}\,{\rm erg}\,{\rm s}^{-1}$) luminosities.}
     \label{fig:zphot_Lbol}
 \end{figure}
 
We calculate bolometric luminosities ($L_{\rm bol}$) of all our BH*-candidates by integrating over the BH* template contribution to the best-fitting \texttt{eazy} SED. For sources with spectroscopic redshifts, we instead use the best-fitting SED at $z=z_{\rm spec}$. The resulting $L_{\rm bol}$ is meant to quantify the total energy output of the BH* itself, and to not include any host galaxy contribution. We note that, by construction, the \texttt{eazy} fits are dominated by the BH* template in the rest-frame optical. The templates then fall off rapidly towards longer wavelengths due to their blackbody extensions (Section \ref{sec:sample_sel_templates}). Based on rest-frame FIR constraints on two of the most luminous LRDs, \citet{Setton25} showed that the IR-luminosity of LRDs is very limited. \citet{Greene26} specifically studied implications for $L_{\rm bol}$ estimates and found that more than half of $L_{\rm bol}$ is emitted in the rest-frame optical, justifying the approach followed here. Figure \ref{fig:zphot_Lbol} shows the derived bolometric luminosities plotted against our photometric redshifts from \texttt{eazy}. 

$L_{\rm bol}$ is directly related to the black hole mass as

\begin{equation}
\label{eq:bhmass}
M_{\rm BH} \approx 0.81\times10^5\,{\rm M_\odot}\lambda_{\rm Edd}^{-1}\left(\frac{L_{\rm bol}}{10^{43}\,{\rm erg}\,{\rm s}^{-1}}\right)
\end{equation}

where $\lambda_{\rm Edd}$ is the Eddington ratio \citep[see, e.g.,][]{Umeda25}. Assuming $\lambda_{\rm Edd}=1$, i.e., accretion at the Eddington limit, then yields a black hole mass that, in principle, constitutes a lower limit. However, recent studies proposed that LRDs may accrete at or even above the Eddington limit \citep[e.g.,][]{Lambrides26,Liu26BB, MadauMaiolino26}, suggesting $\lambda_{\rm Edd}=1$ may be a reasonable assumption, or even over-estimate the true black hole masses. We assume $\lambda_{\rm Edd}=1$, noting that the resulting $M_{\rm BH}$ represents a rough reference and has to be interpreted cautiously which is why we only show $M_{\rm BH}$ as a secondary y-axis on the right of Figure \ref{fig:zphot_Lbol}. 

In rough analogy to the results for $L_{5100}$, we detect BH*-dominated candidates over three orders of magnitude in $L_{\rm bol}$. The full sample, and especially the gold sample, is mostly concentrated around intermediate to high luminosities ($L_{\rm bol}\sim10^{44}\,{\rm erg}\,{\rm s}^{-1}$), while the number of candidates declines rapidly towards higher ($L_{\rm bol}\gtrsim10^{44.5}\,{\rm erg}\,{\rm s}^{-1}$) and lower ($L_{\rm bol}\lesssim10^{43.5}\,{\rm erg}\,{\rm s}^{-1}$) luminosities. The implied black hole masses (for $\lambda_{\rm Edd}=1$) span $M_{\rm BH}\sim10^4-10^7\,{\rm M_\odot}$.

\subsection{Intrinsically Brightest and Faintest BH*s Candidates}
\label{sec:sample_props_lowhighLbol}

The objects at the extremes of the luminosity distribution are of particular interest because they may inform us about the physics driving the engines of the LRDs. On the one hand, the intrinsically brightest sources may host the most massive black holes and/or have the highest accretion rates and they are the easiest and most obvious targets for spectroscopic follow-up. On the other hand, intrinsically faint LRDs remain relatively unexplored, as the best-studied objects so far are also among the brightest (see Figure \ref{fig:Lopt_zphot_BB}). Further, if confirmed, our faintest candidates would imply that BH*-dominated sources exist over three orders of magnitude in $L_{5100}$ (and $L_{\rm bol}$), and may push us into the intermediate mass black hole regime \citep[e.g.,][]{Greene20}. 

In Figure \ref{fig:Lopt_zphot_BB}, the source A2744-45924 \citep{Labbe24} stands out as the most optically luminous source. In terms of $L_{\rm bol}$, it is slightly less extreme because it is fit with the BH* template corresponding to the hottest stack \citep[$T=5122\,{\rm K}$;][]{Sun26} which peaks around $\lambda_{\rm rest}\approx5000\,\mathrm{\mathring{A}}$ and falls off rapidly towards longer wavelengths. Interestingly, we measure even higher $L_{\rm bol}$ for three candidates at $z_{\rm phot}>7$, with the highest luminosity (log(L$_{\rm bol}/{\rm erg}\,{\rm s}^{-1})=45.1$) measured for the source PANO-J033224M2756-5088 which is shown in the bottom right of Figure \ref{fig:highz}. However, its high redshift of $z_{\rm phot}=9.16$ implies that only the F444W filter probes wavelengths red-wards of the Balmer break, and it is limited to $\lambda_{\rm rest}\lesssim4850\,\mathrm{\mathring{A}}$ at the red end. Nominally, the best-fitting BH* template is the coldest stack \citep[$T=3505\,{\rm K}$;][]{Sun26}, which peaks at $\lambda_{\rm rest}\sim0.8\mu{\rm m}$, meaning that most of the bolometric luminosity is emitted in a wavelength range that is not directly constrained by NIRCam. Further, a cold temperature for the most luminous source is in contrast to the trend that more luminous LRDs are associated with higher temperatures \citep[i.e., shorter peak wavelengths,][]{degraaff25pop}. Indeed, the fits with other BH* templates are almost equally good, pointing to a degeneracy between templates at high redshift, and suggesting that the true $L_{\rm bol}$ may be lower. To test this, we infer $L_{\rm bol}$ again from the best-fitting \texttt{eazy} SED with the hottest stack ($T=5122\,{\rm K}$) and find log(L$_{\rm bol}/{\rm erg}\,{\rm s}^{-1})=44.6$ which still places the source among the most luminous in the sample, but below, e.g., A2744-45924. We further elaborate on the degeneracy between BH* templates at high redshift in Section \ref{sec:discussion_caveats_highz} and subsequently focus on $L_{5100}$ as a more directly constrained quantity. 

\begin{figure*}
     \centering
     \includegraphics[width=1.9\columnwidth]{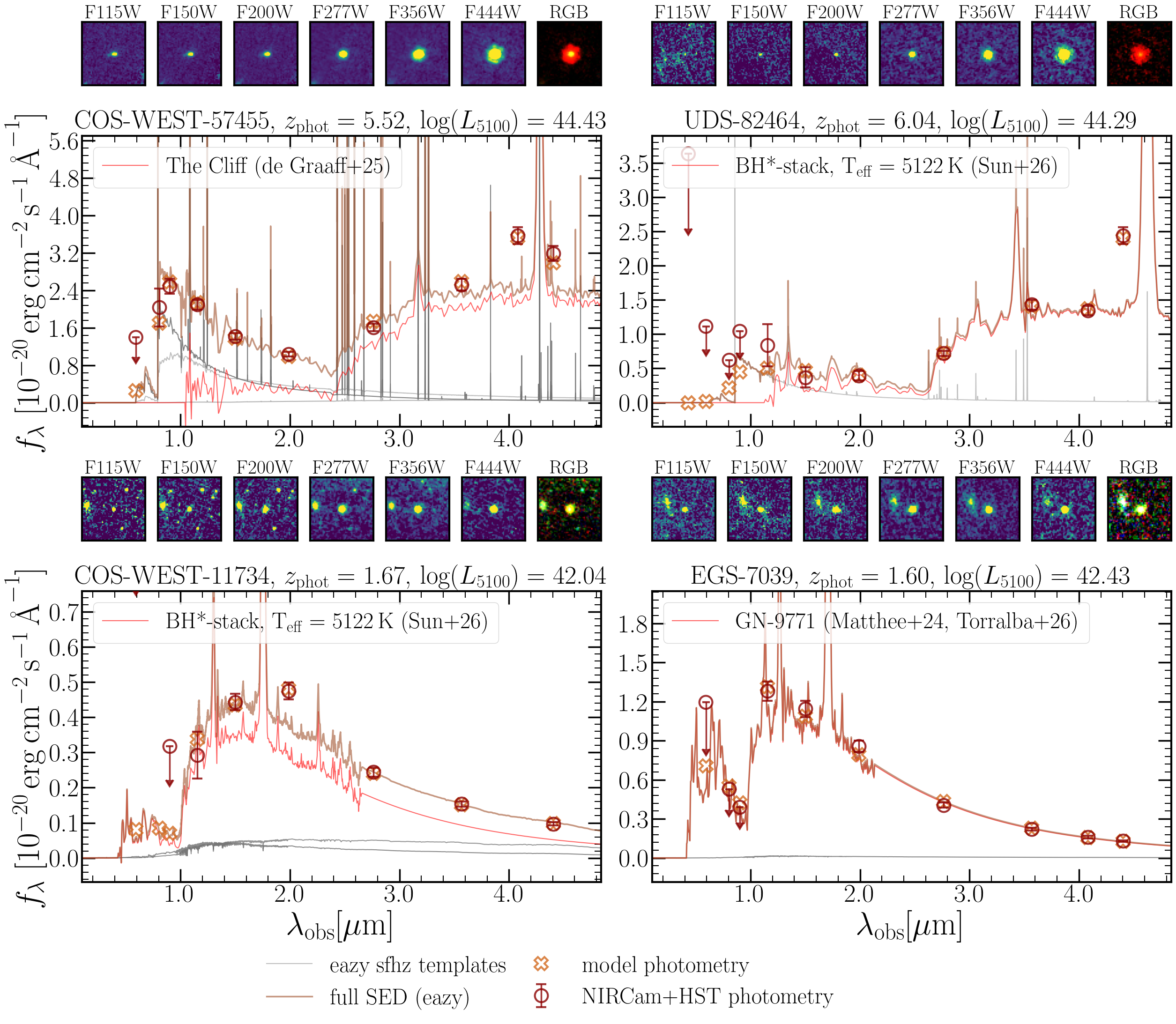}
     \caption{BH*-dominated candidates at the highest (top panels) and lowest (bottom panels) optical luminosities. The former constitute some of the most outstanding candidates and are ideal targets for spectroscopic follow-up. The latter probe a relatively unexplored part of the LRD parameter space given most well studied objects are among the intrinsically brightest. However, we can only detect the faintest sources at the lowest redshifts where we often lack strong constraints blue-wards of the Balmer break. Top left: $z_{\rm phot}=5.52$ candidate in the COSMOS field and the second most luminous source in the sample. Top right: Fourth most luminous candidate at $z_{\rm phot}=6.04$, identified in the UDS field. Bottom left: Intrinsically faintest candidate in the gold sample at $z_{\rm phot}=1.67$ showing a good fit with the BH*-stack at $T_{\rm eff}=5122\,{\rm K}$, but lacking tight constraints blue-wards of the Balmer break. Bottom right: Slightly brighter candidate at a similar redshift ($z_{\rm phot}=1.60$) in the EGS field with tighter constraints on the Balmer break from NIRCam/F090W and HST imaging.}
     \label{fig:brightest_faintest}
 \end{figure*}

Despite the larger area ($\sim0.3\,{\rm deg}^2$) searched here compared to previous JWST-based searches, A2744-45924 remains the most optically luminous source in our sample. This is in line with the sharp cutoff found at the bright end of the optical luminosity function (LF) in \citet{Ma25cutoff}, although we lack sufficient area to put tighter constraints on the LF (see Section \ref{sec:sample_props_numden}). In Figure \ref{fig:brightest_faintest}, we show the second and fourth most optically luminous BH*-dominated candidates (apart from A2744-49524 and the third most luminous source, UDS-39621 shown in Figure \ref{fig:highz}), as well as the two faintest sources that are part of the gold sample. The most luminous sources are outstanding BH*-dominated candidates that warrant spectroscopic follow-up, in particular because they appear to be rare. 

Moving to the intrinsically faintest sources, COS-WEST-11734 in the bottom left is well fit with the hottest BH*-stack ($T=5122\,{\rm K}$), though it lacks sufficiently deep rest-UV constraints to fully probe the Balmer break. On the other hand, EGS-7039 in the bottom right shows a stronger break of $\sim3$ based on the photometry (using a $2\sigma$ upper limit on the flux) and a good fit with the GN-9771 template, including its extension to longer wavelengths as a pure blackbody. We note that these are the only two sources in the gold sample with log($L_{5100}/{\rm erg}\,{\rm s}^{-1})<42.5$. All other faint candidates show $B(\texttt{sfhz})<100$, i.e., the photometry is more degenerate with galaxy template fits. They are further preferentially found at low redshift where they appear brighter, but where we often lack strong constraints blue-wards of the Balmer break as is illustrated by COS-WEST-11734. We also note that even if we consider the full sample, we only find nine candidates at log($L_{5100}/{\rm erg}\,{\rm s}^{-1})<42.5$, and only three silver candidates at log($L_{5100}/{\rm erg}\,{\rm s}^{-1})<42$, in line with the findings of \citet{degraaff25pop} who discussed a possible physical origin of the lack of LRDs at low luminosities but could not rule out a spectroscopic selection effect. The $50\%$ completeness limit shown in Figure \ref{fig:Lopt_zphot_BB} suggests that the faint cut-off observed here is indeed a selection effect, and that current photometric data sets are insufficient to robustly detect fainter BH*-dominated sources, if they exist.

\subsection{Number Densities}
\label{sec:sample_props_numden}

The mere fact that BH*-dominated sources exist over such a wide range of redshifts, and at least down to $z=1.73$ \citep{Torralba26} implies that they are not merely an early Universe phenomenon requiring primordial conditions or primordial chemistry \citep[e.g.,][]{Dayal26}. Note that lower redshift LRDs such as J1025+1402 \citep[``The Egg'';][]{Lin26} would likely not qualify as BH*-dominated due to a stronger host galaxy contribution. 

\begin{figure}
     \centering
     \includegraphics[width=0.47\textwidth]{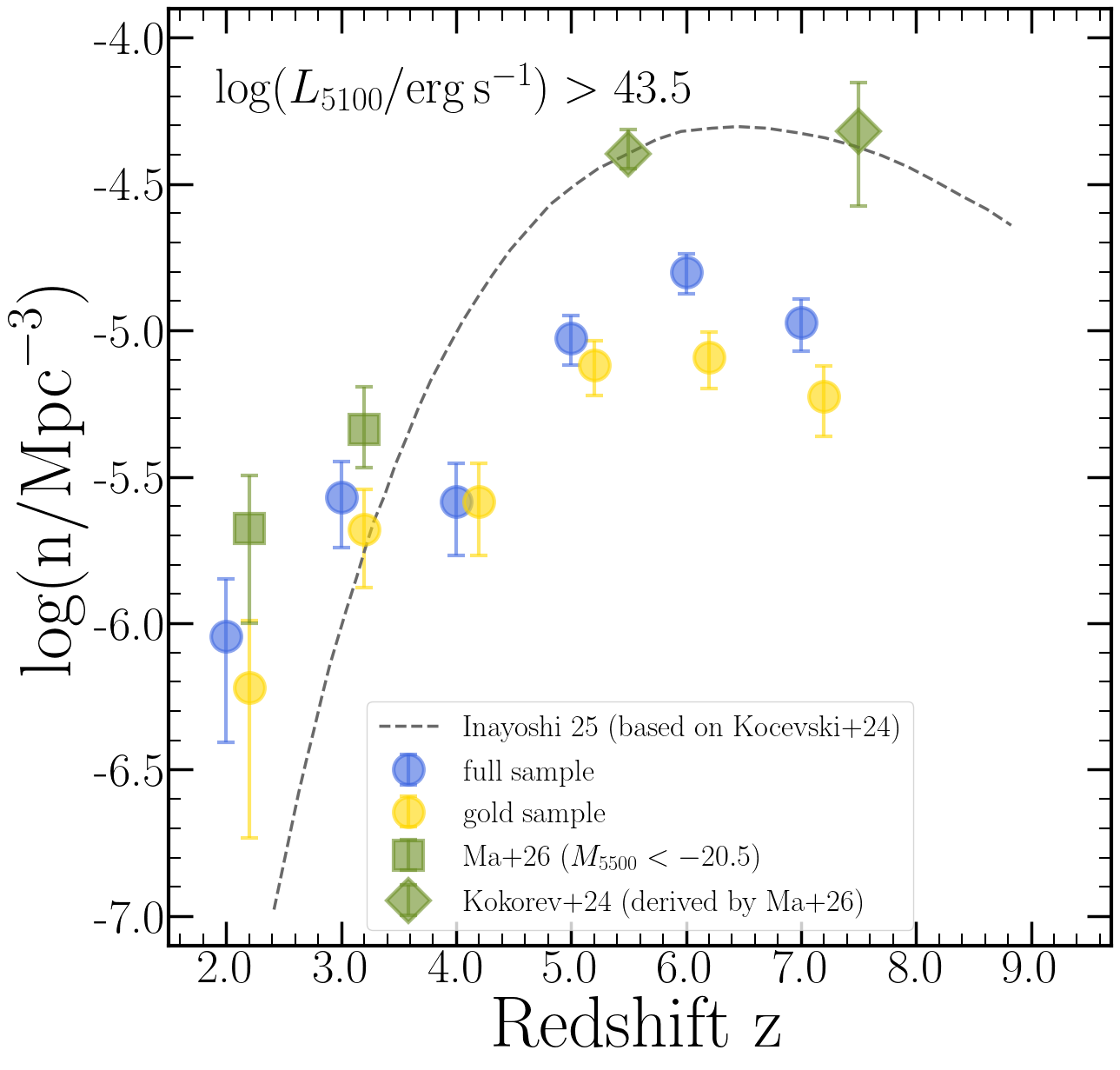}
     \caption{Rough number density of our BH*-dominated candidates as a function of redshift for both the full and the gold sample. These are derived from dividing the number of sources in redshift bins centered at $z=2$, 3, 4, 5, 6, and 7 by the respective co-moving volume. The error bars represent Poisson uncertainties, and the markers for the gold sample are displaced by 0.2 on the x-axis for better visual separation. For comparison, we show number densities at $z<4$ from \citet{Ma26}, as well as their numbers provided for the sample from \citet{Kokorev24}, and the log-normal relation from \citet{Inayoshi25} based on the sample from \citet{Kocevski25stats}.}
     \label{fig:number_density}
 \end{figure}

To provide a rough estimate of the number density evolution of BH*-dominated sources as a function of redshift, we build on the completeness calculation described in Section \ref{sec:sample_sel_completeness}. Calculating $80\%$ completeness limits in terms of $L_{5100}$ at the median F444W imaging depth in analogy to the $50\%$ completeness limits shown in Figure \ref{fig:Lopt_zphot_BB}, we find that our sample is $>80\%$ complete for log($L_{5100}/{\rm erg}\,{\rm s}^{-1})>43.5$ at $z<7.5$ for all six BH* templates. We thus define six redshift bins centered at $z=2$, 3, 4, 5, 6, and 7 with a width of $\Delta z=1$. In each bin, we count the number of BH*-dominated candidates with optical luminosity ${\rm log}(L_{5100}/{\rm erg}\,{\rm s}^{-1})>43.5$ in fields whose F444W depth is sufficient to detect a BH*-dominated source at the minimum optical luminosity and at the central redshift of the bin. We calculate the survey area across the sufficiently deep fields from the number of un-flagged pixels in all six NIRCam wide filters (F115W, F150W, F200W, F277W, F356W, and F444W) that are required in the sample selection (Section \ref{sec:sample_sel}), and convert this to a co-moving volume in each redshift bin by which we divide the respective number count. The resulting number densities are shown for both the full sample and the gold sample in Figure \ref{fig:number_density}. We compare to number densities inferred through a ground-based search at $z<4$ from \citet{Ma26}. These are derived for sources below a treshold in the absolute magnitude at $5500$\AA\, $M_{5500}<-20.5$, which corresponds to log($L_{5500}/{\rm erg}\,{\rm s}^{-1})=43.57$, close to our threshold of log($L_{5100}/{\rm erg}\,{\rm s}^{-1})>43.5$. \citet{Ma26} also provide equivalent number density estimates at higher redshifts based on the luminosity functions from \citet{Kokorev24} which we include in the comparison. Finally, we plot the log-normal relation from \citet{Inayoshi25} which is inferred from the LRD-sample of \citet{Kocevski25stats}.

Our measurements suggest that the number density of BH*-dominated sources peaks at $z\sim5-6$ and declines by about an order of magnitude from $z\sim6$ to $z\sim2$. We also identify a weak decline from $z\sim6$ to $z\sim7$ which is, however, not significant within uncertainties. At even higher redshifts, the number of candidates in our sample drops rapidly, though this is affected by a selection effect where we only detect the most luminous sources at the highest redshifts. Those appear to be rare at all redshifts as can be seen in Figure \ref{fig:Lopt_zphot_BB}. Moreover, the limited coverage of the rest-frame optical by NIRCam at those high redshifts, makes it harder to robustly identify BH*-dominated sources using the method outlined here, which may further decrease our number densities. We argue that larger samples, and/or the inclusion of longer wavelength data, in particular from MIRI is required to shed further light on the number density evolution at $z\gtrsim7$. 

 Intriguingly, our number density estimates for the full (gold) sample are a factor of 2.5 to 4.5 (5 to 8) lower than those based on the LRD-samples from \citet{Kokorev24} and \citet{Kocevski25stats} at $z\sim5-8$. Moving to lower redshifts, our number densities are only factors of 2.3 (3.5) and 1.7 (2.2) below those from \citet{Ma26} at $z\sim2$ and $z\sim3$, and lie slightly above the log-normal curve from \citet{Inayoshi25}, albeit with large uncertainties. Taken at face value, and assuming that BH*-dominated LRDs are a subset of the full LRD population, this indicates that while the total LRD number density declines towards lower redshift, the fraction of BH*-dominated sources among them increases. However, there are a few important caveats to consider.
 
As we have discussed in Section \ref{sec:sample_sel_v-shapes}, photometric V-shape selections tend to miss objects with a dominant BH*-component because they are too faint in the rest-UV to measure a blue UV-color. Specifically, applying the selection criteria from \citet{Kokorev24} to our photometric catalog in Section \ref{sec:sample_sel_v-shapes} yields a sample of 545 LRDs, compared to our 241 BH*-dominated candidates. Moreover, the V-shaped LRD sample only contains about half of our sources (see Figure \ref{fig:color_sel}). Similarly, \citet{Ma26} selected LRDs at $z\sim2-4$ defining a V-shape criterion in the $r$, $z$, $J$, and $K_s$ bands which likely also misses BH*-dominated sources. Different number densities in Figure \ref{fig:number_density} therefore trace partially disjoint samples, complicating their interpretation. Nevertheless, our detection of robust BH*-dominated sources around $z\sim2$, including some with spectroscopic confirmation (see Figure \ref{fig:lowz}) suggests that the number density of such objects remains surprisingly high around cosmic noon, relative to expectations from LRD number densities. While this may point to a significant incompleteness in existing photometric samples of (V-shaped) LRDs at $z\sim2-3$, \citet{Ma26} also point out that their sample likely contains a significant fraction of contaminants, and they argue their number densities should be considered as upper limits. We further note that our extension of the BH* templates to longer wavelengths assuming single blackbodies more strongly affects the selection at low redshift where NIRCam probes rest-frame wavelengths in the extended part of the templates. While we do find candidates whose SED shape closely follows the extended template (e.g., COS-WEST-154491 in the top left of Figure \ref{fig:lowz}), this may not be true for all BH*-dominated sources, and could thus be a source of incompleteness of our low redshift sample. In summary, this points to a surprising abundance of BH*-dominated sources around cosmic noon which remains to be confirmed through spectroscopic follow-up and larger samples in the future.

\section{Discussion}
\label{sec:discussion}

In the preceding sections, we have developed a novel method of selecting BH*s from photometric data based on template fitting with \texttt{eazy}. Using six different empirical BH* templates covering a wide range of intrinsic luminosities and blackbody temperatures, we compiled a sample of 241 BH*-dominated candidates. We showed that our robust candidates range from $z\sim1.7$ to $z\sim9.3$ and span more than two orders of magnitude in optical luminosity (${\rm log}(L_{5100}/{\rm erg}\,{\rm s}^{-1})\sim42-44.5$). Subsequently, we discuss some caveats and limitations of our sample selection, along with possible future avenues for improvement. We end the discussion by exploring the potential of this new selection method to compile samples of LRDs more broadly.

\subsection{Caveats and Limitations}
\label{sec:discussion_caveats}

\subsubsection{Intrinsic Variation in BH* Spectra}
\label{sec:discussion_caveats_intrinsic_variation}

The completeness of our sample selection hinges upon the extent to which our BH* templates cover the intrinsic variation in BH* spectra. To this day, only few spectroscopically confirmed BH*-dominated objects are known in the literature \citep[e.g.,][]{degraaff25,Naidu25BHstar,Torralba26}. In principle, future spectra may reveal new features and add diversity to the library of observed BH*s that is not yet captured by our templates. On the other hand, if the larger population of LRDs can be reliably decomposed into a host galaxy and a BH* \citep[e.g.,][]{Barro26}, then the host-subtracted spectra and stacks from \citet{Sun26} used in this work provide a solid baseline for the identification of BH*-dominated objects over a wide range of intrinsic properties. 

Nevertheless, current samples of LRDs, both spectroscopic and photometric, may still be missing extreme and rare sources. Focusing on the extremes in blackbody temperature, the hottest sources are represented by our template based on GN-9771 \citep{Torralba26IFU}. While hot sources are characterized by strong Balmer lines and Balmer breaks \citep{degraaff25pop} that make them easy to identify in principle, they may also show significant UV-emission that is intrinsic to the BH* component \citep{Labbe24,Torralba26IFU,Ando26}. If true, this complicates their selection from photometry because they are more degenerate with LRDs that show a significant host galaxy contribution. 

We explore the cold end of the temperature distribution by adding a host-subtracted version of the source UNCOVER-20698 \citep{degraaff25pop,Wang26} to our template set. It is highlighted as one of the coldest sources ($T_{\rm eff}\sim2300\,{\rm K}$) in the sample of \citet[][their Fig. 5]{degraaff25pop}, and was shown to display a potential water absorption feature at $\lambda_{\rm rest}\sim1.4\mu$m, indicative of a cold atmosphere ($\lesssim3000\,$K) in \citet{Wang26}. First, this template only adds a small number of candidates to the sample which may indicate that extremely cold sources ($T\lesssim3000\,{\rm K}$) are rare. However, the template picks up a confirmed QG at $z_{\rm spec}=3.25$ as a BH*-dominated candidate, illustrating that the degeneracy between BH* and QG solutions is particularly pronounced for cold sources (see also Section \ref{sec:overview_contaminants}). At the same time, we do not select the source UNCOVER-20698 itself from our photometric catalog due to degeneracies with galaxy templates, also related to its host galaxy. Further, with this extremely cold template, some spectroscopically confirmed sources are fit at the wrong redshift, e.g., the ALT-BH* at $z_{\rm spec}=2.59$ (bottom right panel in Figure \ref{fig:lowz}) is fit at $z_{\rm phot}=1.68$, indicating further degeneracy with other BH* templates and making it extremely difficult to identify (potentially rare) cold BH*s from photometric data alone. Additional medium band and MIRI imaging may help to put tighter constraints on the emission line strength (colder sources tend to show weaker H$\alpha$ emission, \citealt{degraaff25pop}), as well as the continuum shape of such objects. Nevertheless, spectroscopy will be indispensable to unambiguously identify and characterize cold BH*-dominated sources, as well as potentially new and unexplored types of LRDs or BH*s.

\subsubsection{Limitations at High Redshift}
\label{sec:discussion_caveats_highz}

At the highest redshifts, our sample selection is increasingly limited by the rest-frame optical coverage of the available NIRCam filters. In particular, at $z\gtrsim9$, only F444W (and the medium bands probing similar wavelengths) trace the light red-wards of the Balmer break. This leads to a degeneracy with at least a certain type of BDs as exemplified by the MoM-BD-1 (Section \ref{sec:overview_contaminants}). Such BDs may, however, be rare and most stellar contaminants are likely captured by our large grid of stellar templates. In general, the rest-optical continuum shape of the highest redshift candidates is poorly constrained by NIRCam imaging causing degeneracies between different BH* templates. Thus, it is nearly impossible to constrain the peak wavelength of the optical continuum and the blackbody temperature. As discussed in Section \ref{sec:sample_props_lowhighLbol}, this introduces substantial uncertainty in the bolometric luminosity of the highest redshift candidates, and may specifically cause us to overestimate $L_{\rm bol}$ for some of them. We also note that we select the highest redshift spectroscopically confirmed LRD, CAPERS-LRD-z9 \citep{Taylor25} at $z_{\rm spec}=9.27$, showing that our selection successfully identifies LRDs at such high redshift.

In the high redshift regime, MIRI data will be extremely useful. First, Figure \ref{fig:mom_bd} shows how the spectrum of the MoM-BD-1 declines much more sharply at $\lambda_{\rm obs}>4\,\mu{\rm m}$ than the BH* template, so that the two can be easily distinguished based on MIRI fluxes. Besides removing this remaining type of contaminant, MIRI can probe H$\alpha$ at $z\gtrsim7$, and it provides valuable constraints on the rest-optical continuum shape of BH*-dominated candidates. A major limitation of MIRI is the depth that can be achieved, though many of our BH*-dominated candidates are sufficiently bright to be detected in existing or upcoming MIRI surveys.

\subsubsection{Limitations at Low Redshift}
\label{sec:discussion_caveats_lowz}

At the lower end of the redshift distribution, our sample selection is mainly limited by the availability of sufficiently deep imaging at rest-frame wavelengths blue-wards of the Balmer break. At $z\lesssim2.5$, this regime is mainly probed by HST/ACS and the NIRCam SW filters F070W and F090W which are not uniformly available across the fields studied here, and are often significantly shallower than the NIRCam/LW imaging. Without sufficient constraints below the Balmer break, i.e., without actually constraining the full Balmer break strength, the photometry becomes degenerate with QGs or even sources showing relatively weak stellar Balmer breaks, such as ``mini''-quenched or ``napping'' galaxies \citep[e.g.,][]{Looser24,CoveloPaz26}.
However, the few robust sources we identify at $z<2.5$ (specifically the four sources highlighted in Figure \ref{fig:lowz}) are particularly valuable as they allow us to probe redder rest-frame wavelengths with NIRSpec, and to explore lines such as Pa$\beta$, Pa$\gamma$ and HeI at $1.08\mu{\rm m}$ (see the spectrum of UDS-134217 in the top left panel of Figure \ref{fig:lowz}), or even molecular absorption features \citep{Wang26}.

\subsection{A New Way of Selecting LRDs}
\label{sec:discussion_completeness}

In Section \ref{sec:sample_sel_v-shapes}, we discussed how the selection method applied here is highly complementary to photometric LRD selections in the literature that are based on colors between NIRCam filters and/or a V-shaped SED. Specifically, we do not require any detection in the rest-frame UV which is often very faint for the sources presented here. Our requirement of a BH* template contribution of $>80$\% at $\lambda_{\rm rest}>4000\,$\AA\ leaves some room for UV flux from a host galaxy which is why we recover some ``classic'' V-shaped LRDs. However, with increasing host emission in the rest-UV, the contribution of the host galaxy also increases in the rest-optical (depending on the age of the stellar population and its dust content), so that the BH* contribution quickly drops to $<80$\%. 

In principle, we could use the search method presented here to select LRDs more broadly by lowering our threshold below $80$\%. As a simple test, we compare to the spectroscopic LRD sample of \citet{degraaff25pop}. Considering only overlapping fields, and sources that pass our pre-selection cuts (in particular ${\rm SNR(F444W)}>10$ and ${\rm mag(F444W)}<27$), we find that our sample contains 34/84 of their sources, corresponding to $40.5\%$. If we decrease our template contribution threshold to 75\% and 50\%, these numbers increase to 44/84 (52.4\%) and 70/84 (83.3\%), while the total sample size increases by a factor of 1.4 and 3.1. This indicates that while it is possible to recover most LRDs with the method presented here, this likely comes at the cost of a significantly higher contamination fraction. The extra BH* template adds a degree of freedom to the \texttt{eazy} fits, so that the fits to some galaxies can be nominally improved with a limited contribution of a BH* template which is not physically motivated but simply reflects degeneracies with other templates. Our choice of a relatively high threshold of $80$\% is motivated by our objective of finding the most extreme, BH*-dominated sources, resembling the paradigmatic sources The Cliff and the MoM-BH*-1. In other words, we wish to identify BH*-dominated sources with a small host galaxy contribution. With a threshold of 80\%, we achieve a high completeness for such sources with a limited contamination fraction. At the same time, we are increasingly incomplete for sources with a stronger host galaxy contribution, thus missing a substantial fraction of V-shaped LRDs as explored in detail in Section \ref{sec:sample_sel_v-shapes}.



\section{Summary and Conclusions}
\label{sec:summary_conclusions}

In this paper, we present a new selection method for LRDs that focuses on sources whose SED is dominated by the central engine, the BH*. This is motivated by the two first spectroscopically confirmed BH*s, The Cliff and the MoM-BH*-1, which are characterized by an optical continuum that resembles a blackbody at a temperature of $T_{\rm eff}\sim5000\,{\rm K}$, a strong Balmer break, and very weak UV emission. Our method is based on incorporating six different BH* templates in the \texttt{eazy} redshift-fitting code to then identify objects whose best-fitting SED is dominated by these templates. We construct empirical templates that cover the range of spectroscopically observed BH* properties, in particular a wide range of intrinsic luminosities and temperatures, based on individual sources and BH* stacks from \citet{Sun26}. Applying the method to a photometric catalog across $1000\,{\rm arcmin}^{2}$ of legacy and pure parallel NIRCam-imaging in six or more filters, we identify 241 BH*-dominated candidates. Cross-matching with publicly available spectroscopic data, we find that $\sim27\%$ of our candidates have spectroscopic redshifts, largely confirming our SED-modeling and photometric redshifts. The vast majority of the available spectra with sufficiently high SNR show V-shaped SEDs, blackbody shaped continua, strong Balmer breaks and relatively weak rest-UV emission, indeed resembling objects like The Cliff and the MoM-BH*-1. Among all the spectra, we only identify one secure contaminant, a BD, and one likely contaminant with strong emission lines boosting the NIRCam/LW photometry. To mitigate the possible impact of such contaminants, we calculate Bayes factors between our fits with galaxy and a BH* template, and pure galaxy fits, as well as fits with stellar templates. We use these Bayes factors to define a gold sample of 127 particularly robust BH*-candidates showing Bayes factors $>100$.

Our results can be summarized as follows:

\begin{itemize}
    \item BH*-dominated sources exist across the Universe, and at least over the range $z=1.7-9.3$, indicating they are not merely an early-Universe phenomenon but persist to cosmic noon.
    \item Comparing our sample to photometric V-shape selections in the literature, we show that our selection is highly complementary, and identifies objects that are missing from existing photometric samples, in particular because they are too faint in the rest-UV to measure a blue UV-color. Conversely, and by design, our sample misses strongly V-shaped LRDs with significant host galaxy contributions.
    \item We identify four promising candidates at $z<2.5$, three of which have been confirmed spectroscopically through NIRCam/grism, NIRSpec/G235M and ground based VLT/X-Shooter spectroscopy, but lack continuous spectroscopic coverage in the rest-frame optical. These sources provide unique opportunities to study the engines of the LRDs in great detail owing to their apparent brightness ($21.9-23.6\,{\rm mag}$ at $\sim2\,\mu{\rm m}$), and by enabling NIRSpec spectroscopy out to $\lambda_{\rm rest}\sim1.8\,\mu{\rm m}$.
    \item We measure optical luminosities spanning at least two orders of magnitude, ${\rm log}(L_{5100}/{\rm erg}\,{\rm s}^{-1})\sim42.5-44.5$, with candidates as faint as $L_{5100}\sim10^{42}\,{\rm erg}\,{\rm s}^{-1}$ at $z<2$.
    \item The Balmer break strengths of our candidates are centered around a median of $\sim3$, close to the maximum possible stellar Balmer breaks, and with a long tail out to break strengths $>10$, the strongest Balmer breaks measured in the Universe to date.
    \item We estimate bolometric luminosities by integrating over the BH* template contribution to the best-fitting SED, and find that our sample covers ${\rm log}(L_{\rm bol}/{\rm erg}\,{\rm s}^{-1})\sim42-45$. Converting the bolometric luminosities to black hole masses assuming accretion at the Eddington limit, we find $M_{\rm BH}\sim10^5-10^7\,{\rm M_\odot}$, with the faintest candidates approaching $\sim10^4{\rm M_\odot}$, suggesting that BH*-dominated sources range from the intermediate mass black hole to the quasar regime.
    \item We compute approximate number densities of BH*-dominated sources at $L_{5100}>10^{43.5}\,{\rm erg}\,{\rm s}^{-1}$ and find that they peak at $z\sim5-6$ ($\sim10^{-5}\,{\rm Mpc}^{-3}{\rm dex}^{-1}$), and decline towards higher and lower redshifts. While the weak decline towards higher redshift is not significant, not least due to possible selection effects, we measure a drop in the number density by an order of magnitude from $z\sim6$ to $z\sim2$.
    \item Our number densities at $z\sim5-7$ are a factor of $2.5-8$ below LRD number densities based on photometric V-shape selections in the literature, but they are only a factor of $1.7-3.5$ below estimates based on V-shapes identified from ground-based imaging at $z\sim2-3$. Despite some caveats regarding the comparison of these partly complementary samples, this suggests that the fraction of BH*-dominated sources among the total LRD population does not decline towards lower redshifts, or that lower redshift samples of LRDs are still highly incomplete.
\end{itemize}

The sample presented here provides ideal targets for spectroscopic follow-up, enabling unique insights into the physics driving the SEDs of LRDs through a direct view into their engines that outshines the host galaxy in the rest-frame optical. Further, it highlights the power of identifying such sources through template-based searches. The robustness of our sample selection is limited by the photometric wavelength-coverage and depth of the available NIRCam-imaging, by the extent to which our templates capture intrinsic variations in the spectra of LRD engines, and by how accurately they trace the continuum shapes, especially at $\lambda_{\rm rest}\gtrsim1\,\mu{\rm m}$. Future studies including MIRI data as well as additional medium-band imaging as is for example obtained through the cycle 4 program MINERVA (GO-7814, PI Muzzin, \citealt{Muzzin25}) will likely help improve the sample selection, and shed further light on the rest-frame optical to NIR properties of BH*-dominated objects and LRDs more generally, as well as their number density evolution. NIRSpec spectra of a larger sample of BH*-dominated sources will be crucial to better understand the diversity in their continuum shapes, and the physical mechanisms behind their peculiar and still enigmatic appearance.

\section*{Acknowledgments}

This work has received funding from the Swiss State Secretariat for Education, Research and Innovation (SERI) under contract number MB22.00072, as well as from the Swiss National Science Foundation (SNSF) through project grant 200020\_207349. RPN, WQS, and ZL acknowledge funding from {\it JWST} programs GO-3516, GO-5224, and the MIT Undergraduate Research Opportunities Program (UROP). Support for this work was provided by NASA through the NASA Hubble Fellowship grant HST-HF2-51515.001-A awarded by the Space Telescope Science Institute, which is operated by the Association of Universities for Research in Astronomy, Incorporated, under NASA contract NAS5-26555. RPN thanks Neil Pappalardo and Jane Pappalardo for their generous support of the MIT Pappalardo Fellowships in Physics, and for their enthusiasm and encouragement for pursuing the earliest galaxies and black holes. This research was supported by the International Space Science Institute (ISSI) in Bern, through ISSI International Team 25-659, ``Little Red Dots, Big Open Questions: Unraveling the Mystery of the James Webb Space Telescope's Most Debated Discovery" led by MX and RPN. AdG acknowledges support from a Clay Fellowship awarded by the Smithsonian Astrophysical Observatory. REH acknowledges support by the German Aerospace Center (DLR) and the Federal Ministry for Economic Affairs and Energy (BMWi) through program 50OR2403 `RUBIES'. JM and AT acknowledge funding from the European Union (ERC, AGENTS,  101076224). Views and opinions expressed are, however, those of the authors only and do not necessarily reflect those of the European Union or the European Research Council. Neither the European Union nor the granting authority can be held responsible for them. The data products presented herein were retrieved from the Dawn JWST Archive (DJA). DJA is an initiative of the Cosmic Dawn Center (DAWN), which is funded by the Danish National Research Foundation under grant DNRF140. This work is based on observations made with the NASA/ESA/CSA James Webb Space Telescope. The data were obtained from the Mikulski Archive for Space Telescopes at the Space Telescope Science Institute, which is operated by the Association of Universities for Research in Astronomy, Inc., under NASA contract NAS 5-03127 for JWST. Support for programs \#2514, was provided by NASA through grants from the Space Telescope Science Institute, which is operated by the Association of Universities for Research in Astronomy, Inc., under NASA contract NAS 5-03127. 

Software used in developing this work includes: \texttt{matplotlib} \citep{matplotlib}, \texttt{jupyter} \citep{jupyter}, \texttt{IPython} \citep{ipython}, \texttt{numpy} \citep{numpy}, \texttt{scipy} \citep{scipy}, and \texttt{Astropy} \citep{astropy}.

\end{CJK*}

\bibliography{MasterBiblio}

@ARTICLE{Gandolfi26,
       author = {{Gandolfi}, G. and {Rodighiero}, G. and {Castellano}, M. and {Fontana}, A. and {Santini}, P. and {Dickinson}, M. and {Finkelstein}, S. and {Catone}, M. and {Calabr{\`o}}, A. and {Merlin}, E. and {Pentericci}, L. and {Bisigello}, L. and {Grazian}, A. and {Napolitano}, L. and {Vulcani}, B. and {Taylor}, A.~J. and {Arrabal Haro}, P. and {Kirkpatrick}, A. and {Backhaus}, B.~E. and {Holwerda}, B.~W. and {Giulietti}, M. and {Bianchetti}, A. and {Cassata}, P. and {Cleri}, N.~J. and {Daddi}, E. and {Ferguson}, H.~C. and {Girardi}, G. and {Hirschmann}, M. and {Koekemoer}, A.~M. and {Lapi}, A. and {Pacucci}, F. and {P{\'e}rez-Gonz{\'a}lez}, P.~G. and {de la Vega}, A. and {Vietri}, A. and {Wilkins}, S. and {Yung}, L.~Y.~A. and {Bagley}, M. and {Bhatawdekar}, R. and {Kartaltepe}, J. and {Papovich}, C. and {Pirzkal}, N.},
        title = "{Mysteries of Capotauro: Investigating the puzzling nature of an extreme F356W-dropout}",
      journal = {\aap},
         year = 2026,
        month = feb,
       volume = {706},
          eid = {A364},
        pages = {A364},
          doi = {10.1051/0004-6361/202557061},
archivePrefix = {arXiv},
       eprint = {2509.01664},
 primaryClass = {astro-ph.GA},
       adsurl = {https://ui.adsabs.harvard.edu/abs/2026A&A...706A.364G}
}

@ARTICLE{Sun26,
       author = {{Sun}, Wendy Q. and {Naidu}, Rohan P. and {Matthee}, Jorryt and {de Graaff}, Anna and {Chisholm}, John and {Greene}, Jenny E. and {Oesch}, Pascal A. and {Torralba}, Alberto and {Hviding}, Raphael E. and {Brammer}, Gabriel and {Simcoe}, Robert A. and {Bose}, Sownak and {Bouwens}, Rychard and {Dayal}, Pratika and {Eilers}, Anna-Christina and {Fei}, Qinyue and {Furtak}, Lukas J. and {Gottumukkala}, Rashmi and {Goulding}, Andy and {Heintz}, Kasper E. and {Hirschmann}, Michaela and {Kokorev}, Vasily and {Leja}, Joel and {Liu}, Zhaoran and {Natarajan}, Priyamvada and {Santarelli}, Andrew D. and {Setton}, David J. and {Smith}, Aaron and {Tacchella}, Sandro and {Volonteri}, Marta and {Walter}, Fabian and {Weibel}, Andrea and {Williams}, Christina C.},
        title = "{Little Red Dot $-$ Host Galaxy $=$ Black Hole Star: A Gas-Enshrouded Heart at the Center of Every Little Red Dot}",
      journal = {arXiv e-prints},
         year = 2026,
        month = jan,
          eid = {arXiv:2601.20929},
        pages = {arXiv:2601.20929},
          doi = {10.48550/arXiv.2601.20929},
archivePrefix = {arXiv},
       eprint = {2601.20929},
 primaryClass = {astro-ph.GA},
       adsurl = {https://ui.adsabs.harvard.edu/abs/2026arXiv260120929S}
}

@ARTICLE{Lupi24LRD,
       author = {{Lupi}, Alessandro and {Trinca}, Alessandro and {Volonteri}, Marta and {Dotti}, Massimo and {Mazzucchelli}, Chiara},
        title = "{Size matters: are we witnessing super-Eddington accretion in high-redshift black holes from JWST?}",
      journal = {\aap},
         year = 2024,
        month = sep,
       volume = {689},
          eid = {A128},
        pages = {A128},
          doi = {10.1051/0004-6361/202451249},
archivePrefix = {arXiv},
       eprint = {2406.17847},
 primaryClass = {astro-ph.HE},
       adsurl = {https://ui.adsabs.harvard.edu/abs/2024A&A...689A.128L}
}

@ARTICLE{Volonteri21,
       author = {{Volonteri}, Marta and {Habouzit}, M{\'e}lanie and {Colpi}, Monica},
        title = "{The origins of massive black holes}",
      journal = {Nature Reviews Physics},
         year = 2021,
        month = sep,
       volume = {3},
       number = {11},
        pages = {732-743},
          doi = {10.1038/s42254-021-00364-9},
archivePrefix = {arXiv},
       eprint = {2110.10175},
 primaryClass = {astro-ph.GA},
       adsurl = {https://ui.adsabs.harvard.edu/abs/2021NatRP...3..732V}
}

@ARTICLE{Greene20,
       author = {{Greene}, Jenny E. and {Strader}, Jay and {Ho}, Luis C.},
        title = "{Intermediate-Mass Black Holes}",
      journal = {\araa},
         year = 2020,
        month = aug,
       volume = {58},
        pages = {257-312},
          doi = {10.1146/annurev-astro-032620-021835},
archivePrefix = {arXiv},
       eprint = {1911.09678},
 primaryClass = {astro-ph.GA},
       adsurl = {https://ui.adsabs.harvard.edu/abs/2020ARA&A..58..257G}
}

@ARTICLE{Dayal26,
       author = {{Dayal}, Pratika and {Maiolino}, Roberto},
        title = "{The properties of primordially-seeded black holes and their hosts in the first billion years: implications for JWST}",
      journal = {\aap},
         year = 2026,
        month = feb,
       volume = {706},
          eid = {A72},
        pages = {A72},
          doi = {10.1051/0004-6361/202555959},
archivePrefix = {arXiv},
       eprint = {2506.08116},
 primaryClass = {astro-ph.GA},
       adsurl = {https://ui.adsabs.harvard.edu/abs/2026A&A...706A..72D}
}

@ARTICLE{Ma25,
       author = {{Ma}, Yilun and {Greene}, Jenny E. and {Setton}, David J. and {Volonteri}, Marta and {Leja}, Joel and {Wang}, Bingjie and {Bezanson}, Rachel and {Brammer}, Gabriel and {Cutler}, Sam E. and {Dayal}, Pratika and {van Dokkum}, Pieter and {Furtak}, Lukas J. and {Glazebrook}, Karl and {Goulding}, Andy D. and {de Graaff}, Anna and {Kokorev}, Vasily and {Labbe}, Ivo and {Pan}, Richard and {Price}, Sedona H. and {Weaver}, John R. and {Williams}, Christina C. and {Whitaker}, Katherine E. and {Zitrin}, Adi},
        title = "{UNCOVER: 404 Error{\textemdash}Models Not Found for the Triply Imaged Little Red Dot A2744-QSO1}",
      journal = {\apj},
         year = 2025,
        month = mar,
       volume = {981},
       number = {2},
          eid = {191},
        pages = {191},
          doi = {10.3847/1538-4357/ada613},
archivePrefix = {arXiv},
       eprint = {2410.06257},
 primaryClass = {astro-ph.GA},
       adsurl = {https://ui.adsabs.harvard.edu/abs/2025ApJ...981..191M}
}

@ARTICLE{Umeda25,
       author = {{Umeda}, Hiroya and {Inayoshi}, Kohei and {Harikane}, Yuichi and {Murase}, Kohta},
        title = "{A Black-Hole Envelope Interpretation for Cosmological Demographics of Little Red Dots}",
      journal = {arXiv e-prints},
         year = 2025,
        month = dec,
          eid = {arXiv:2512.04208},
        pages = {arXiv:2512.04208},
          doi = {10.48550/arXiv.2512.04208},
archivePrefix = {arXiv},
       eprint = {2512.04208},
 primaryClass = {astro-ph.GA},
       adsurl = {https://ui.adsabs.harvard.edu/abs/2025arXiv251204208U}
}

@ARTICLE{Kokorev23,
       author = {{Kokorev}, Vasily and {Fujimoto}, Seiji and {Labbe}, Ivo and {Greene}, Jenny E. and {Bezanson}, Rachel and {Dayal}, Pratika and {Nelson}, Erica J. and {Atek}, Hakim and {Brammer}, Gabriel and {Caputi}, Karina I. and {Chemerynska}, Iryna and {Cutler}, Sam E. and {Feldmann}, Robert and {Fudamoto}, Yoshinobu and {Furtak}, Lukas J. and {Goulding}, Andy D. and {de Graaff}, Anna and {Leja}, Joel and {Marchesini}, Danilo and {Miller}, Tim B. and {Nanayakkara}, Themiya and {Oesch}, Pascal A. and {Pan}, Richard and {Price}, Sedona H. and {Setton}, David J. and {Smit}, Renske and {Stefanon}, Mauro and {Wang}, Bingjie and {Weaver}, John R. and {Whitaker}, Katherine E. and {Williams}, Christina C. and {Zitrin}, Adi},
        title = "{UNCOVER: A NIRSpec Identification of a Broad-line AGN at z = 8.50}",
      journal = {\apjl},
         year = 2023,
        month = nov,
       volume = {957},
       number = {1},
          eid = {L7},
        pages = {L7},
          doi = {10.3847/2041-8213/ad037a},
archivePrefix = {arXiv},
       eprint = {2308.11610},
 primaryClass = {astro-ph.GA},
       adsurl = {https://ui.adsabs.harvard.edu/abs/2023ApJ...957L...7K}
}

@ARTICLE{Muzzin25,
       author = {{Muzzin}, Adam and {Suess}, Katherine A. and {Marchesini}, Danilo and {Robbins}, Luke and {Willott}, Chris J. and {Alberts}, Stacey and {Antwi-Danso}, Jacqueline and {Asada}, Yoshihisa and {Brammer}, Gabriel and {Cutler}, Sam E. and {Iyer}, Kartheik G. and {Labbe}, Ivo and {Martis}, Nicholas S. and {Miller}, Tim B. and {Mitsuhashi}, Ikki and {Pope}, Alexandra and {Sajina}, Anna and {Sarrouh}, Ghassan T.~E. and {Sharma}, Monu and {Stefanon}, Mauro and {Whitaker}, Katherine E. and {Abraham}, Roberto and {Atek}, Hakim and {Bradac}, Marusa and {Berek}, Samantha and {Bezanson}, Rachel and {Brown}, Westley and {Burgasser}, Adam J. and {Chicoine}, Nathalie and {Cloonan}, Aidan P. and {Cooper}, Olivia R. and {Dayal}, Pratika and {de Graaff}, Anna and {Desprez}, Guillaume and {Feldmann}, Robert and {Forrest}, Ben and {Franx}, Marijn and {Fudamoto}, Yoshinobu and {Fujimoto}, Seiji and {Furtak}, Lukas J. and {Glazebrook}, Karl and {Goovaerts}, Ilias and {Greene}, Jenny E. and {Jagga}, Naadiyah and {Jarvis}, William W.~H. and {Kriek}, Mariska and {Khullar}, Gourav and {La Torre}, Valentina and {Leja}, Joel and {Lin}, Jamie and {Lorenz}, Brian and {Lyon}, Daniel and {Markov}, Vladan and {Maseda}, Michael V. and {McConachie}, Ian and {Merchant}, Maya and {Merida}, Rosa M. and {Mowla}, Lamiya and {Myers}, Katherine and {Naidu}, Rohan P. and {Nanayakkara}, Themiya and {Nelson}, Erica J. and {Noirot}, Gael and {Oesch}, Pascal A. and {Omori}, Kiyoaki C. and {Pan}, Richard and {Porraz Barrera}, Natalia and {Price}, Sedona H. and {Ravindranath}, Swara and {Sawicki}, Marcin and {Setton}, David J. and {Smit}, Renske and {Sok}, Visal and {Speagle}, Joshua S. and {Taylor}, Edward N. and {Tan}, Vivian Yun Yan and {Tripodi}, Roberta and {van der Wel}, Arjen and {Perez Vidal}, Edgar and {Wang}, Bingjie and {Weaver}, John R. and {Williams}, Christina C. and {Withers}, Sunna and {Zaidi}, Kumail},
        title = "{MINERVA: A NIRCam Medium Band and MIRI Imaging Survey to Unlock the Hidden Gems of the Distant Universe}",
      journal = {arXiv e-prints},
         year = 2025,
        month = jul,
          eid = {arXiv:2507.19706},
        pages = {arXiv:2507.19706},
          doi = {10.48550/arXiv.2507.19706},
archivePrefix = {arXiv},
       eprint = {2507.19706},
 primaryClass = {astro-ph.GA},
       adsurl = {https://ui.adsabs.harvard.edu/abs/2025arXiv250719706M}
}

@ARTICLE{Santarelli26,
       author = {{Santarelli}, Andrew D. and {Farag}, Ebraheem and {Bellinger}, Earl P. and {Natarajan}, Priyamvada and {Naidu}, Rohan P. and {Campbell}, Claire B. and {Caplan}, Matthew E.},
        title = "{Evolutionary Tracks and Spectral Properties of Quasi-stars and Their Correlation with Little Red Dots}",
      journal = {\apjl},
         year = 2026,
        month = feb,
       volume = {998},
       number = {1},
          eid = {L4},
        pages = {L4},
          doi = {10.3847/2041-8213/ae3713},
archivePrefix = {arXiv},
       eprint = {2510.17952},
 primaryClass = {astro-ph.GA},
       adsurl = {https://ui.adsabs.harvard.edu/abs/2026ApJ...998L...4S}
}

@ARTICLE{degraaff25rubies,
       author = {{de Graaff}, Anna and {Brammer}, Gabriel and {Weibel}, Andrea and {Lewis}, Zach and {Maseda}, Michael V. and {Oesch}, Pascal A. and {Bezanson}, Rachel and {Boogaard}, Leindert A. and {Cleri}, Nikko J. and {Cooper}, Olivia R. and {Gottumukkala}, Rashmi and {Greene}, Jenny E. and {Hirschmann}, Michaela and {Hviding}, Raphael E. and {Katz}, Harley and {Labb{\'e}}, Ivo and {Leja}, Joel and {Matthee}, Jorryt and {McConachie}, Ian and {Miller}, Tim B. and {Naidu}, Rohan P. and {Price}, Sedona H. and {Rix}, Hans-Walter and {Setton}, David J. and {Suess}, Katherine A. and {Wang}, Bingjie and {Whitaker}, Katherine E. and {Williams}, Christina C.},
        title = "{RUBIES: A complete census of the bright and red distant Universe with JWST/NIRSpec}",
      journal = {\aap},
         year = 2025,
        month = may,
       volume = {697},
          eid = {A189},
        pages = {A189},
          doi = {10.1051/0004-6361/202452186},
archivePrefix = {arXiv},
       eprint = {2409.05948},
 primaryClass = {astro-ph.GA},
       adsurl = {https://ui.adsabs.harvard.edu/abs/2025A&A...697A.189D}
}

@ARTICLE{Chang26,
       author = {{Chang}, Seok-Jun and {Gronke}, Max and {Matthee}, Jorryt and {Mason}, Charlotte},
        title = "{Impact of resonance, Raman, and Thomson scattering on hydrogen line formation in Little Red Dots}",
      journal = {\mnras},
         year = 2026,
        month = feb,
       volume = {545},
       number = {4},
          eid = {staf2131},
        pages = {staf2131},
          doi = {10.1093/mnras/staf2131},
archivePrefix = {arXiv},
       eprint = {2508.08768},
 primaryClass = {astro-ph.GA},
       adsurl = {https://ui.adsabs.harvard.edu/abs/2026MNRAS.545f2131C}
}

@ARTICLE{Liu26BB,
       author = {{Liu}, Hanpu and {Jiang}, Yan-Fei and {Quataert}, Eliot and {Greene}, Jenny E. and {Ma}, Yilun},
        title = "{The Balmer Break and Optical Continuum of Little Red Dots from Super-Eddington Accretion}",
      journal = {\apj},
         year = 2025,
        month = nov,
       volume = {994},
       number = {1},
          eid = {113},
        pages = {113},
          doi = {10.3847/1538-4357/ae0c19},
archivePrefix = {arXiv},
       eprint = {2507.07190},
 primaryClass = {astro-ph.GA},
       adsurl = {https://ui.adsabs.harvard.edu/abs/2025ApJ...994..113L}
}

@ARTICLE{IM25,
       author = {{Inayoshi}, Kohei and {Maiolino}, Roberto},
        title = "{Extremely Dense Gas around Little Red Dots and High-redshift Active Galactic Nuclei: A Nonstellar Origin of the Balmer Break and Absorption Features}",
      journal = {\apjl},
         year = 2025,
        month = feb,
       volume = {980},
       number = {2},
          eid = {L27},
        pages = {L27},
          doi = {10.3847/2041-8213/adaebd},
archivePrefix = {arXiv},
       eprint = {2409.07805},
 primaryClass = {astro-ph.GA},
       adsurl = {https://ui.adsabs.harvard.edu/abs/2025ApJ...980L..27I}
}

@ARTICLE{Inayoshi25,
       author = {{Inayoshi}, Kohei},
        title = "{Little Red Dots as the Very First Activity of Black Hole Growth}",
      journal = {\apjl},
         year = 2025,
        month = jul,
       volume = {988},
       number = {1},
          eid = {L22},
        pages = {L22},
          doi = {10.3847/2041-8213/adea66},
archivePrefix = {arXiv},
       eprint = {2503.05537},
 primaryClass = {astro-ph.GA},
       adsurl = {https://ui.adsabs.harvard.edu/abs/2025ApJ...988L..22I}
}

@ARTICLE{Lin26,
       author = {{Lin}, Xiaojing and {Fan}, Xiaohui and {Cai}, Zheng and {Bian}, Fuyan and {Liu}, Hanpu and {Sun}, Fengwu and {Ma}, Yilun and {Greene}, Jenny E. and {Strauss}, Michael A. and {Green}, Richard and {Lyu}, Jianwei and {Champagne}, Jaclyn B. and {Goulding}, Andy D. and {Inayoshi}, Kohei and {Jin}, Xiangyu and {Leung}, Gene C.~K. and {Li}, Mingyu and {Liu}, Weizhe and {Liu}, Yichen and {Mao}, Junjie and {Pudoka}, Maria Anne and {Tee}, Wei Leong and {Wang}, Ben and {Wang}, Feige and {Wu}, Yunjing and {Yang}, Jinyi and {Zhang}, Haowen and {Zhu}, Yongda},
        title = "{The Discovery of Little Red Dots in the Local Universe: Signatures of Cool Gas Envelopes}",
      journal = {\apj},
         year = 2026,
        month = feb,
       volume = {997},
       number = {2},
          eid = {364},
        pages = {364},
          doi = {10.3847/1538-4357/ae2bdf},
archivePrefix = {arXiv},
       eprint = {2507.10659},
 primaryClass = {astro-ph.GA},
       adsurl = {https://ui.adsabs.harvard.edu/abs/2026ApJ...997..364L}
}

@ARTICLE{Matthee26,
       author = {{Matthee}, Jorryt and {Torralba}, Alberto and {Pezzulli}, Gabriele and {Naidu}, Rohan P. and {Chisholm}, John and {Mascia}, Sara and {Greene}, Jenny E. and {Ishikawa}, Yuzo and {Gronke}, Max and {Wuyts}, Stijn and {Bordoloi}, Rongmon and {Brammer}, Gabriel and {Chang}, Seok-Jun and {Eilers}, Anna-Christina and {de Graaff}, Anna and {Hviding}, Raphael E. and {Iani}, Edoardo and {Illingworth}, Garth and {Kashino}, Daichi and {Labbe}, Ivo and {Ma}, Yilun and {Maseda}, Michael V. and {Meyer}, Romain and {Nelson}, Erica and {Oesch}, Pascal and {Xiao}, Mengyuan},
        title = "{The Engine and its Flows: Little Red Dot spectra are shaped by the column densities of their gas envelopes}",
      journal = {arXiv e-prints},
         year = 2026,
        month = mar,
          eid = {arXiv:2603.17667},
        pages = {arXiv:2603.17667},
          doi = {10.48550/arXiv.2603.17667},
archivePrefix = {arXiv},
       eprint = {2603.17667},
 primaryClass = {astro-ph.GA},
       adsurl = {https://ui.adsabs.harvard.edu/abs/2026arXiv260317667M}
}

@ARTICLE{Latif25,
       author = {{Latif}, Muhammad A. and {Aftab}, Ammara and {Whalen}, Daniel J. and {Mezcua}, Mar},
        title = "{Radio emission from little red dots may reveal their true nature}",
      journal = {\aap},
         year = 2025,
        month = feb,
       volume = {694},
          eid = {L14},
        pages = {L14},
          doi = {10.1051/0004-6361/202453194},
archivePrefix = {arXiv},
       eprint = {2502.03742},
 primaryClass = {astro-ph.GA},
       adsurl = {https://ui.adsabs.harvard.edu/abs/2025A&A...694L..14L}
}

@ARTICLE{Kido25,
       author = {{Kido}, Daisaburo and {Ioka}, Kunihito and {Hotokezaka}, Kenta and {Inayoshi}, Kohei and {Irwin}, Christopher M.},
        title = "{Black Hole Envelopes in Little Red Dots}",
      journal = {\mnras},
         year = 2025,
        month = nov,
          doi = {10.1093/mnras/staf1898},
archivePrefix = {arXiv},
       eprint = {2505.06965},
 primaryClass = {astro-ph.HE},
       adsurl = {https://ui.adsabs.harvard.edu/abs/2025MNRAS.tmp.1794K}
}

@ARTICLE{Beiler24,
       author = {{Beiler}, Samuel A. and {Cushing}, Michael C. and {Kirkpatrick}, J. Davy and {Schneider}, Adam C. and {Mukherjee}, Sagnick and {Marley}, Mark S. and {Marocco}, Federico and {Smart}, Richard L.},
        title = "{Precise Bolometric Luminosities and Effective Temperatures of 23 Late-T and Y Dwarfs Obtained with JWST}",
      journal = {\apj},
         year = 2024,
        month = oct,
       volume = {973},
       number = {2},
          eid = {107},
        pages = {107},
          doi = {10.3847/1538-4357/ad6301},
archivePrefix = {arXiv},
       eprint = {2407.08518},
 primaryClass = {astro-ph.SR},
       adsurl = {https://ui.adsabs.harvard.edu/abs/2024ApJ...973..107B}
}

@ARTICLE{Begelman26,
       author = {{Begelman}, Mitchell C. and {Dexter}, Jason},
        title = "{Little Red Dots as Late-stage Quasi-stars}",
      journal = {\apj},
         year = 2026,
        month = jan,
       volume = {996},
       number = {1},
          eid = {48},
        pages = {48},
          doi = {10.3847/1538-4357/ae274a},
archivePrefix = {arXiv},
       eprint = {2507.09085},
 primaryClass = {astro-ph.GA},
       adsurl = {https://ui.adsabs.harvard.edu/abs/2026ApJ...996...48B}
}

@ARTICLE{Valentino23,
       author = {{Valentino}, Francesco and {Brammer}, Gabriel and {Gould}, Katriona M.~L. and {Kokorev}, Vasily and {Fujimoto}, Seiji and {Jespersen}, Christian Kragh and {Vijayan}, Aswin P. and {Weaver}, John R. and {Ito}, Kei and {Tanaka}, Masayuki and {Ilbert}, Olivier and {Magdis}, Georgios E. and {Whitaker}, Katherine E. and {Faisst}, Andreas L. and {Gallazzi}, Anna and {Gillman}, Steven and {Gim{\'e}nez-Arteaga}, Clara and {G{\'o}mez-Guijarro}, Carlos and {Kubo}, Mariko and {Heintz}, Kasper E. and {Hirschmann}, Michaela and {Oesch}, Pascal and {Onodera}, Masato and {Rizzo}, Francesca and {Lee}, Minju and {Strait}, Victoria and {Toft}, Sune},
        title = "{An Atlas of Color-selected Quiescent Galaxies at z > 3 in Public JWST Fields}",
      journal = {\apj},
         year = 2023,
        month = apr,
       volume = {947},
       number = {1},
          eid = {20},
        pages = {20},
          doi = {10.3847/1538-4357/acbefa},
archivePrefix = {arXiv},
       eprint = {2302.10936},
 primaryClass = {astro-ph.GA},
       adsurl = {https://ui.adsabs.harvard.edu/abs/2023ApJ...947...20V}
}

@ARTICLE{Huang24,
       author = {{Huang}, Hai-Long and {Jiang}, Jun-Qian and {He}, Jibin and {Wang}, Yu-Tong and {Piao}, Yun-Song},
        title = "{Sub-Eddington accreting supermassive primordial black holes explain Little Red Dots}",
      journal = {arXiv e-prints},
         year = 2024,
        month = oct,
          eid = {arXiv:2410.20663},
        pages = {arXiv:2410.20663},
          doi = {10.48550/arXiv.2410.20663},
archivePrefix = {arXiv},
       eprint = {2410.20663},
 primaryClass = {astro-ph.GA},
       adsurl = {https://ui.adsabs.harvard.edu/abs/2024arXiv241020663H}
}

@ARTICLE{Looser24,
       author = {{Looser}, Tobias J. and {D'Eugenio}, Francesco and {Maiolino}, Roberto and {Witstok}, Joris and {Sandles}, Lester and {Curtis-Lake}, Emma and {Chevallard}, Jacopo and {Tacchella}, Sandro and {Johnson}, Benjamin D. and {Baker}, William M. and {Suess}, Katherine A. and {Carniani}, Stefano and {Ferruit}, Pierre and {Arribas}, Santiago and {Bonaventura}, Nina and {Bunker}, Andrew J. and {Cameron}, Alex J. and {Charlot}, Stephane and {Curti}, Mirko and {de Graaff}, Anna and {Maseda}, Michael V. and {Rawle}, Tim and {Rix}, Hans-Walter and {Del Pino}, Bruno Rodr{\'\i}guez and {Smit}, Renske and {{\"U}bler}, Hannah and {Willott}, Chris and {Alberts}, Stacey and {Egami}, Eiichi and {Eisenstein}, Daniel J. and {Endsley}, Ryan and {Hausen}, Ryan and {Rieke}, Marcia and {Robertson}, Brant and {Shivaei}, Irene and {Williams}, Christina C. and {Boyett}, Kristan and {Chen}, Zuyi and {Ji}, Zhiyuan and {Jones}, Gareth C. and {Kumari}, Nimisha and {Nelson}, Erica and {Perna}, Michele and {Saxena}, Aayush and {Scholtz}, Jan},
        title = "{A recently quenched galaxy 700 million years after the Big Bang}",
      journal = {\nat},
         year = 2024,
        month = may,
       volume = {629},
       number = {8010},
        pages = {53-57},
          doi = {10.1038/s41586-024-07227-0},
archivePrefix = {arXiv},
       eprint = {2302.14155},
 primaryClass = {astro-ph.GA},
       adsurl = {https://ui.adsabs.harvard.edu/abs/2024Natur.629...53L}
}

@ARTICLE{Matthee24,
       author = {{Matthee}, Jorryt and {Naidu}, Rohan P. and {Brammer}, Gabriel and {Chisholm}, John and {Eilers}, Anna-Christina and {Goulding}, Andy and {Greene}, Jenny and {Kashino}, Daichi and {Labbe}, Ivo and {Lilly}, Simon J. and {Mackenzie}, Ruari and {Oesch}, Pascal A. and {Weibel}, Andrea and {Wuyts}, Stijn and {Xiao}, Mengyuan and {Bordoloi}, Rongmon and {Bouwens}, Rychard and {van Dokkum}, Pieter and {Illingworth}, Garth and {Kramarenko}, Ivan and {Maseda}, Michael V. and {Mason}, Charlotte and {Meyer}, Romain A. and {Nelson}, Erica J. and {Reddy}, Naveen A. and {Shivaei}, Irene and {Simcoe}, Robert A. and {Yue}, Minghao},
        title = "{Little Red Dots: An Abundant Population of Faint Active Galactic Nuclei at z {\ensuremath{\sim}} 5 Revealed by the EIGER and FRESCO JWST Surveys}",
      journal = {\apj},
         year = 2024,
        month = mar,
       volume = {963},
       number = {2},
          eid = {129},
        pages = {129},
          doi = {10.3847/1538-4357/ad2345},
archivePrefix = {arXiv},
       eprint = {2306.05448},
 primaryClass = {astro-ph.GA},
       adsurl = {https://ui.adsabs.harvard.edu/abs/2024ApJ...963..129M}
}

@ARTICLE{Lambrides26,
       author = {{Lambrides}, Erini and {Larson}, Rebecca L. and {Garofali}, Kristen and {Ptak}, Andrew and {Chiaberge}, Marco and {Long}, Arianna S. and {Hutchison}, Taylor A. and {Norman}, Colin and {McKinney}, Jed and {Akins}, Hollis B. and {Berg}, Danielle A. and {Chisholm}, John and {Civano}, Francesca and {Cloonan}, Aidan P. and {Endsley}, Ryan and {Faisst}, Andreas L. and {Gilli}, Roberto and {Gillman}, Steven and {Hirschmann}, Michaela and {Kartaltepe}, Jeyhan S. and {Kocevski}, Dale D. and {Kokorev}, Vasily and {Pacucci}, Fabio and {Richardson}, Chris T. and {Stiavelli}, Massimo and {Whalen}, Kelly E.},
        title = "{The case for super-Eddington accretion in JWST broad-line active galactic nuclei during the first billion years}",
      journal = {Nature Astronomy},
         year = 2026,
        month = apr,
          doi = {10.1038/s41550-026-02813-w},
archivePrefix = {arXiv},
       eprint = {2409.13047},
 primaryClass = {astro-ph.HE},
       adsurl = {https://ui.adsabs.harvard.edu/abs/2026NatAs.tmp...66L}
}

@ARTICLE{Weibel24,
       author = {{Weibel}, Andrea and {Oesch}, Pascal A. and {Barrufet}, Laia and {Gottumukkala}, Rashmi and {Ellis}, Richard S. and {Santini}, Paola and {Weaver}, John R. and {Allen}, Natalie and {Bouwens}, Rychard and {Bowler}, Rebecca A.~A. and {Brammer}, Gabe and {Carnall}, Adam C. and {Cullen}, Fergus and {Dayal}, Pratika and {Dickinson}, Mark and {Donnan}, Callum T. and {Dunlop}, James S. and {Giavalisco}, Mauro and {Grogin}, Norman A. and {Illingworth}, Garth D. and {Koekemoer}, Anton M. and {Labbe}, Ivo and {Marchesini}, Danilo and {McLeod}, Derek J. and {McLure}, Ross J. and {Naidu}, Rohan P. and {P{\'e}rez-Gonz{\'a}lez}, Pablo G. and {Shuntov}, Marko and {Stefanon}, Mauro and {Toft}, Sune and {Xiao}, Mengyuan},
        title = "{Galaxy build-up in the first 1.5 Gyr of cosmic history: insights from the stellar mass function at z   4-9 from JWST NIRCam observations}",
      journal = {\mnras},
         year = 2024,
        month = sep,
       volume = {533},
       number = {2},
        pages = {1808-1838},
          doi = {10.1093/mnras/stae1891},
archivePrefix = {arXiv},
       eprint = {2403.08872},
 primaryClass = {astro-ph.GA},
       adsurl = {https://ui.adsabs.harvard.edu/abs/2024MNRAS.533.1808W}
}

@ARTICLE{Suess24,
       author = {{Suess}, Katherine A. and {Weaver}, John R. and {Price}, Sedona H. and {Pan}, Richard and {Wang}, Bingjie and {Bezanson}, Rachel and {Brammer}, Gabriel and {Cutler}, Sam E. and {Labb{\'e}}, Ivo and {Leja}, Joel and {Williams}, Christina C. and {Whitaker}, Katherine E. and {Atek}, Hakim and {Dayal}, Pratika and {de Graaff}, Anna and {Feldmann}, Robert and {Franx}, Marijn and {Fudamoto}, Yoshinobu and {Fujimoto}, Seiji and {Furtak}, Lukas J. and {Goulding}, Andy D. and {Greene}, Jenny E. and {Khullar}, Gourav and {Kokorev}, Vasily and {Kriek}, Mariska and {Lorenz}, Brian and {Marchesini}, Danilo and {Maseda}, Michael V. and {Matthee}, Jorryt and {Miller}, Tim B. and {Mitsuhashi}, Ikki and {Mowla}, Lamiya A. and {Muzzin}, Adam and {Naidu}, Rohan P. and {Nanayakkara}, Themiya and {Nelson}, Erica J. and {Oesch}, Pascal A. and {Setton}, David J. and {Shipley}, Heath and {Smit}, Renske and {Spilker}, Justin S. and {van Dokkum}, Pieter and {Zitrin}, Adi},
        title = "{Medium Bands, Mega Science: A JWST/NIRCam Medium-band Imaging Survey of A2744}",
      journal = {\apj},
         year = 2024,
        month = nov,
       volume = {976},
       number = {1},
          eid = {101},
        pages = {101},
          doi = {10.3847/1538-4357/ad75fe},
archivePrefix = {arXiv},
       eprint = {2404.13132},
 primaryClass = {astro-ph.GA},
       adsurl = {https://ui.adsabs.harvard.edu/abs/2024ApJ...976..101S}
}

@ARTICLE{Mukherjee24,
       author = {{Mukherjee}, Sagnick and {Fortney}, Jonathan J. and {Morley}, Caroline V. and {Batalha}, Natasha E. and {Marley}, Mark S. and {Karalidi}, Theodora and {Visscher}, Channon and {Lupu}, Roxana and {Freedman}, Richard and {Gharib-Nezhad}, Ehsan},
        title = "{The Sonora Substellar Atmosphere Models. IV. Elf Owl: Atmospheric Mixing and Chemical Disequilibrium with Varying Metallicity and C/O Ratios}",
      journal = {\apj},
         year = 2024,
        month = mar,
       volume = {963},
       number = {1},
          eid = {73},
        pages = {73},
          doi = {10.3847/1538-4357/ad18c2},
archivePrefix = {arXiv},
       eprint = {2402.00756},
 primaryClass = {astro-ph.EP},
       adsurl = {https://ui.adsabs.harvard.edu/abs/2024ApJ...963...73M}
}

@ARTICLE{Meisner21,
       author = {{Meisner}, Aaron M. and {Schneider}, Adam C. and {Burgasser}, Adam J. and {Marocco}, Federico and {Line}, Michael R. and {Faherty}, Jacqueline K. and {Kirkpatrick}, J. Davy and {Caselden}, Dan and {Kuchner}, Marc J. and {Gelino}, Christopher R. and {Gagn{\'e}}, Jonathan and {Theissen}, Christopher and {Gerasimov}, Roman and {Aganze}, Christian and {Hsu}, Chih-chun and {Wisniewski}, John P. and {Casewell}, Sarah L. and {Bardalez Gagliuffi}, Daniella C. and {Logsdon}, Sarah E. and {Eisenhardt}, Peter R.~M. and {Allers}, Katelyn and {Debes}, John H. and {Allen}, Michaela B. and {Stevnbak Andersen}, Nikolaj and {Goodman}, Sam and {Gramaize}, L{\'e}opold and {Martin}, David W. and {Sainio}, Arttu and {Cushing}, Michael C. and {Backyard Worlds: Planet 9 Collaboration}},
        title = "{New Candidate Extreme T Subdwarfs from the Backyard Worlds: Planet 9 Citizen Science Project}",
      journal = {\apj},
         year = 2021,
        month = jul,
       volume = {915},
       number = {2},
          eid = {120},
        pages = {120},
          doi = {10.3847/1538-4357/ac013c},
archivePrefix = {arXiv},
       eprint = {2106.01387},
 primaryClass = {astro-ph.SR},
       adsurl = {https://ui.adsabs.harvard.edu/abs/2021ApJ...915..120M}
}

@ARTICLE{GAIA2016,
       author = {{Gaia Collaboration} and {Prusti}, T. and {de Bruijne}, J.~H.~J. and {Brown}, A.~G.~A. and {Vallenari}, A. and {Babusiaux}, C. and {Bailer-Jones}, C.~A.~L. and {Bastian}, U. and {Biermann}, M. and {Evans}, D.~W. and {Eyer}, L. and {Jansen}, F. and {Jordi}, C. and {Klioner}, S.~A. and {Lammers}, U. and {Lindegren}, L. and {Luri}, X. and {Mignard}, F. and {Milligan}, D.~J. and {Panem}, C. and {Poinsignon}, V. and {Pourbaix}, D. and {Randich}, S. and {Sarri}, G. and {Sartoretti}, P. and {Siddiqui}, H.~I. and {Soubiran}, C. and {Valette}, V. and {van Leeuwen}, F. and {Walton}, N.~A. and {Aerts}, C. and {Arenou}, F. and {Cropper}, M. and {Drimmel}, R. and {H{\o}g}, E. and {Katz}, D. and {Lattanzi}, M.~G. and {O'Mullane}, W. and {Grebel}, E.~K. and {Holland}, A.~D. and {Huc}, C. and {Passot}, X. and {Bramante}, L. and {Cacciari}, C. and {Casta{\~n}eda}, J. and {Chaoul}, L. and {Cheek}, N. and {De Angeli}, F. and {Fabricius}, C. and {Guerra}, R. and {Hern{\'a}ndez}, J. and {Jean-Antoine-Piccolo}, A. and {Masana}, E. and {Messineo}, R. and {Mowlavi}, N. and {Nienartowicz}, K. and {Ord{\'o}{\~n}ez-Blanco}, D. and {Panuzzo}, P. and {Portell}, J. and {Richards}, P.~J. and {Riello}, M. and {Seabroke}, G.~M. and {Tanga}, P. and {Th{\'e}venin}, F. and {Torra}, J. and {Els}, S.~G. and {Gracia-Abril}, G. and {Comoretto}, G. and {Garcia-Reinaldos}, M. and {Lock}, T. and {Mercier}, E. and {Altmann}, M. and {Andrae}, R. and {Astraatmadja}, T.~L. and {Bellas-Velidis}, I. and {Benson}, K. and {Berthier}, J. and {Blomme}, R. and {Busso}, G. and {Carry}, B. and {Cellino}, A. and {Clementini}, G. and {Cowell}, S. and {Creevey}, O. and {Cuypers}, J. and {Davidson}, M. and {De Ridder}, J. and {de Torres}, A. and {Delchambre}, L. and {Dell'Oro}, A. and {Ducourant}, C. and {Fr{\'e}mat}, Y. and {Garc{\'\i}a-Torres}, M. and {Gosset}, E. and {Halbwachs}, J. -L. and {Hambly}, N.~C. and {Harrison}, D.~L. and {Hauser}, M. and {Hestroffer}, D. and {Hodgkin}, S.~T. and {Huckle}, H.~E. and {Hutton}, A. and {Jasniewicz}, G. and {Jordan}, S. and {Kontizas}, M. and {Korn}, A.~J. and {Lanzafame}, A.~C. and {Manteiga}, M. and {Moitinho}, A. and {Muinonen}, K. and {Osinde}, J. and {Pancino}, E. and {Pauwels}, T. and {Petit}, J. -M. and {Recio-Blanco}, A. and {Robin}, A.~C. and {Sarro}, L.~M. and {Siopis}, C. and {Smith}, M. and {Smith}, K.~W. and {Sozzetti}, A. and {Thuillot}, W. and {van Reeven}, W. and {Viala}, Y. and {Abbas}, U. and {Abreu Aramburu}, A. and {Accart}, S. and {Aguado}, J.~J. and {Allan}, P.~M. and {Allasia}, W. and {Altavilla}, G. and {{\'A}lvarez}, M.~A. and {Alves}, J. and {Anderson}, R.~I. and {Andrei}, A.~H. and {Anglada Varela}, E. and {Antiche}, E. and {Antoja}, T. and {Ant{\'o}n}, S. and {Arcay}, B. and {Atzei}, A. and {Ayache}, L. and {Bach}, N. and {Baker}, S.~G. and {Balaguer-N{\'u}{\~n}ez}, L. and {Barache}, C. and {Barata}, C. and {Barbier}, A. and {Barblan}, F. and {Baroni}, M. and {Barrado y Navascu{\'e}s}, D. and {Barros}, M. and {Barstow}, M.~A. and {Becciani}, U. and {Bellazzini}, M. and {Bellei}, G. and {Bello Garc{\'\i}a}, A. and {Belokurov}, V. and {Bendjoya}, P. and {Berihuete}, A. and {Bianchi}, L. and {Bienaym{\'e}}, O. and {Billebaud}, F. and {Blagorodnova}, N. and {Blanco-Cuaresma}, S. and {Boch}, T. and {Bombrun}, A. and {Borrachero}, R. and {Bouquillon}, S. and {Bourda}, G. and {Bouy}, H. and {Bragaglia}, A. and {Breddels}, M.~A. and {Brouillet}, N. and {Br{\"u}semeister}, T. and {Bucciarelli}, B. and {Budnik}, F. and {Burgess}, P. and {Burgon}, R. and {Burlacu}, A. and {Busonero}, D. and {Buzzi}, R. and {Caffau}, E. and {Cambras}, J. and {Campbell}, H. and {Cancelliere}, R. and {Cantat-Gaudin}, T. and {Carlucci}, T. and {Carrasco}, J.~M. and {Castellani}, M. and {Charlot}, P. and {Charnas}, J. and {Charvet}, P. and {Chassat}, F. and {Chiavassa}, A. and {Clotet}, M. and {Cocozza}, G. and {Collins}, R.~S. and {Collins}, P. and {Costigan}, G. and {Crifo}, F. and {Cross}, N.~J.~G. and {Crosta}, M. and {Crowley}, C. and {Dafonte}, C. and {Damerdji}, Y. and {Dapergolas}, A. and {David}, P. and {David}, M. and {De Cat}, P. and {de Felice}, F. and {de Laverny}, P. and {De Luise}, F. and {De March}, R. and {de Martino}, D. and {de Souza}, R. and {Debosscher}, J. and {del Pozo}, E. and {Delbo}, M. and {Delgado}, A. and {Delgado}, H.~E. and {di Marco}, F. and {Di Matteo}, P. and {Diakite}, S. and {Distefano}, E. and {Dolding}, C. and {Dos Anjos}, S. and {Drazinos}, P. and {Dur{\'a}n}, J. and {Dzigan}, Y. and {Ecale}, E. and {Edvardsson}, B. and {Enke}, H. and {Erdmann}, M. and {Escolar}, D. and {Espina}, M. and {Evans}, N.~W. and {Eynard Bontemps}, G. and {Fabre}, C. and {Fabrizio}, M. and {Faigler}, S. and {Falc{\~a}o}, A.~J. and {Farr{\`a}s Casas}, M. and {Faye}, F. and {Federici}, L. and {Fedorets}, G. and {Fern{\'a}ndez-Hern{\'a}ndez}, J. and {Fernique}, P. and {Fienga}, A. and {Figueras}, F. and {Filippi}, F. and {Findeisen}, K. and {Fonti}, A. and {Fouesneau}, M. and {Fraile}, E. and {Fraser}, M. and {Fuchs}, J. and {Furnell}, R. and {Gai}, M. and {Galleti}, S. and {Galluccio}, L. and {Garabato}, D. and {Garc{\'\i}a-Sedano}, F. and {Gar{\'e}}, P. and {Garofalo}, A. and {Garralda}, N. and {Gavras}, P. and {Gerssen}, J. and {Geyer}, R. and {Gilmore}, G. and {Girona}, S. and {Giuffrida}, G. and {Gomes}, M. and {Gonz{\'a}lez-Marcos}, A. and {Gonz{\'a}lez-N{\'u}{\~n}ez}, J. and {Gonz{\'a}lez-Vidal}, J.~J. and {Granvik}, M. and {Guerrier}, A. and {Guillout}, P. and {Guiraud}, J. and {G{\'u}rpide}, A. and {Guti{\'e}rrez-S{\'a}nchez}, R. and {Guy}, L.~P. and {Haigron}, R. and {Hatzidimitriou}, D. and {Haywood}, M. and {Heiter}, U. and {Helmi}, A. and {Hobbs}, D. and {Hofmann}, W. and {Holl}, B. and {Holland}, G. and {Hunt}, J.~A.~S. and {Hypki}, A. and {Icardi}, V. and {Irwin}, M. and {Jevardat de Fombelle}, G. and {Jofr{\'e}}, P. and {Jonker}, P.~G. and {Jorissen}, A. and {Julbe}, F. and {Karampelas}, A. and {Kochoska}, A. and {Kohley}, R. and {Kolenberg}, K. and {Kontizas}, E. and {Koposov}, S.~E. and {Kordopatis}, G. and {Koubsky}, P. and {Kowalczyk}, A. and {Krone-Martins}, A. and {Kudryashova}, M. and {Kull}, I. and {Bachchan}, R.~K. and {Lacoste-Seris}, F. and {Lanza}, A.~F. and {Lavigne}, J. -B. and {Le Poncin-Lafitte}, C. and {Lebreton}, Y. and {Lebzelter}, T. and {Leccia}, S. and {Leclerc}, N. and {Lecoeur-Taibi}, I. and {Lemaitre}, V. and {Lenhardt}, H. and {Leroux}, F. and {Liao}, S. and {Licata}, E. and {Lindstr{\o}m}, H.~E.~P. and {Lister}, T.~A. and {Livanou}, E. and {Lobel}, A. and {L{\"o}ffler}, W. and {L{\'o}pez}, M. and {Lopez-Lozano}, A. and {Lorenz}, D. and {Loureiro}, T. and {MacDonald}, I. and {Magalh{\~a}es Fernandes}, T. and {Managau}, S. and {Mann}, R.~G. and {Mantelet}, G. and {Marchal}, O. and {Marchant}, J.~M. and {Marconi}, M. and {Marie}, J. and {Marinoni}, S. and {Marrese}, P.~M. and {Marschalk{\'o}}, G. and {Marshall}, D.~J. and {Mart{\'\i}n-Fleitas}, J.~M. and {Martino}, M. and {Mary}, N. and {Matijevi{\v{c}}}, G. and {Mazeh}, T. and {McMillan}, P.~J. and {Messina}, S. and {Mestre}, A. and {Michalik}, D. and {Millar}, N.~R. and {Miranda}, B.~M.~H. and {Molina}, D. and {Molinaro}, R. and {Molinaro}, M. and {Moln{\'a}r}, L. and {Moniez}, M. and {Montegriffo}, P. and {Monteiro}, D. and {Mor}, R. and {Mora}, A. and {Morbidelli}, R. and {Morel}, T. and {Morgenthaler}, S. and {Morley}, T. and {Morris}, D. and {Mulone}, A.~F. and {Muraveva}, T. and {Musella}, I. and {Narbonne}, J. and {Nelemans}, G. and {Nicastro}, L. and {Noval}, L. and {Ord{\'e}novic}, C. and {Ordieres-Mer{\'e}}, J. and {Osborne}, P. and {Pagani}, C. and {Pagano}, I. and {Pailler}, F. and {Palacin}, H. and {Palaversa}, L. and {Parsons}, P. and {Paulsen}, T. and {Pecoraro}, M. and {Pedrosa}, R. and {Pentik{\"a}inen}, H. and {Pereira}, J. and {Pichon}, B. and {Piersimoni}, A.~M. and {Pineau}, F. -X. and {Plachy}, E. and {Plum}, G. and {Poujoulet}, E. and {Pr{\v{s}}a}, A. and {Pulone}, L. and {Ragaini}, S. and {Rago}, S. and {Rambaux}, N. and {Ramos-Lerate}, M. and {Ranalli}, P. and {Rauw}, G. and {Read}, A. and {Regibo}, S. and {Renk}, F. and {Reyl{\'e}}, C. and {Ribeiro}, R.~A. and {Rimoldini}, L. and {Ripepi}, V. and {Riva}, A. and {Rixon}, G. and {Roelens}, M. and {Romero-G{\'o}mez}, M. and {Rowell}, N. and {Royer}, F. and {Rudolph}, A. and {Ruiz-Dern}, L. and {Sadowski}, G. and {Sagrist{\`a} Sell{\'e}s}, T. and {Sahlmann}, J. and {Salgado}, J. and {Salguero}, E. and {Sarasso}, M. and {Savietto}, H. and {Schnorhk}, A. and {Schultheis}, M. and {Sciacca}, E. and {Segol}, M. and {Segovia}, J.~C. and {Segransan}, D. and {Serpell}, E. and {Shih}, I. -C. and {Smareglia}, R. and {Smart}, R.~L. and {Smith}, C. and {Solano}, E. and {Solitro}, F. and {Sordo}, R. and {Soria Nieto}, S. and {Souchay}, J. and {Spagna}, A. and {Spoto}, F. and {Stampa}, U. and {Steele}, I.~A. and {Steidelm{\"u}ller}, H. and {Stephenson}, C.~A. and {Stoev}, H. and {Suess}, F.~F. and {S{\"u}veges}, M. and {Surdej}, J. and {Szabados}, L. and {Szegedi-Elek}, E. and {Tapiador}, D. and {Taris}, F. and {Tauran}, G. and {Taylor}, M.~B. and {Teixeira}, R. and {Terrett}, D. and {Tingley}, B. and {Trager}, S.~C. and {Turon}, C. and {Ulla}, A. and {Utrilla}, E. and {Valentini}, G. and {van Elteren}, A. and {Van Hemelryck}, E. and {van Leeuwen}, M. and {Varadi}, M. and {Vecchiato}, A. and {Veljanoski}, J. and {Via}, T. and {Vicente}, D. and {Vogt}, S. and {Voss}, H. and {Votruba}, V. and {Voutsinas}, S. and {Walmsley}, G. and {Weiler}, M. and {Weingrill}, K. and {Werner}, D. and {Wevers}, T. and {Whitehead}, G. and {Wyrzykowski}, {\L}. and {Yoldas}, A. and {{\v{Z}}erjal}, M. and {Zucker}, S. and {Zurbach}, C. and {Zwitter}, T. and {Alecu}, A. and {Allen}, M. and {Allende Prieto}, C. and {Amorim}, A. and {Anglada-Escud{\'e}}, G. and {Arsenijevic}, V. and {Azaz}, S. and {Balm}, P. and {Beck}, M. and {Bernstein}, H. -H. and {Bigot}, L. and {Bijaoui}, A. and {Blasco}, C. and {Bonfigli}, M. and {Bono}, G. and {Boudreault}, S. and {Bressan}, A. and {Brown}, S. and {Brunet}, P. -M. and {Bunclark}, P. and {Buonanno}, R. and {Butkevich}, A.~G. and {Carret}, C. and {Carrion}, C. and {Chemin}, L. and {Ch{\'e}reau}, F. and {Corcione}, L. and {Darmigny}, E. and {de Boer}, K.~S. and {de Teodoro}, P. and {de Zeeuw}, P.~T. and {Delle Luche}, C. and {Domingues}, C.~D. and {Dubath}, P. and {Fodor}, F. and {Fr{\'e}zouls}, B. and {Fries}, A. and {Fustes}, D. and {Fyfe}, D. and {Gallardo}, E. and {Gallegos}, J. and {Gardiol}, D. and {Gebran}, M. and {Gomboc}, A. and {G{\'o}mez}, A. and {Grux}, E. and {Gueguen}, A. and {Heyrovsky}, A. and {Hoar}, J. and {Iannicola}, G. and {Isasi Parache}, Y. and {Janotto}, A. -M. and {Joliet}, E. and {Jonckheere}, A. and {Keil}, R. and {Kim}, D. -W. and {Klagyivik}, P. and {Klar}, J. and {Knude}, J. and {Kochukhov}, O. and {Kolka}, I. and {Kos}, J. and {Kutka}, A. and {Lainey}, V. and {LeBouquin}, D. and {Liu}, C. and {Loreggia}, D. and {Makarov}, V.~V. and {Marseille}, M.~G. and {Martayan}, C. and {Martinez-Rubi}, O. and {Massart}, B. and {Meynadier}, F. and {Mignot}, S. and {Munari}, U. and {Nguyen}, A. -T. and {Nordlander}, T. and {Ocvirk}, P. and {O'Flaherty}, K.~S. and {Olias Sanz}, A. and {Ortiz}, P. and {Osorio}, J. and {Oszkiewicz}, D. and {Ouzounis}, A. and {Palmer}, M. and {Park}, P. and {Pasquato}, E. and {Peltzer}, C. and {Peralta}, J. and {P{\'e}turaud}, F. and {Pieniluoma}, T. and {Pigozzi}, E. and {Poels}, J. and {Prat}, G. and {Prod'homme}, T. and {Raison}, F. and {Rebordao}, J.~M. and {Risquez}, D. and {Rocca-Volmerange}, B. and {Rosen}, S. and {Ruiz-Fuertes}, M.~I. and {Russo}, F. and {Sembay}, S. and {Serraller Vizcaino}, I. and {Short}, A. and {Siebert}, A. and {Silva}, H. and {Sinachopoulos}, D. and {Slezak}, E. and {Soffel}, M. and {Sosnowska}, D. and {Strai{\v{z}}ys}, V. and {ter Linden}, M. and {Terrell}, D. and {Theil}, S. and {Tiede}, C. and {Troisi}, L. and {Tsalmantza}, P. and {Tur}, D. and {Vaccari}, M. and {Vachier}, F. and {Valles}, P. and {Van Hamme}, W. and {Veltz}, L. and {Virtanen}, J. and {Wallut}, J. -M. and {Wichmann}, R. and {Wilkinson}, M.~I. and {Ziaeepour}, H. and {Zschocke}, S.},
        title = "{The Gaia mission}",
      journal = {\aap},
         year = 2016,
        month = nov,
       volume = {595},
          eid = {A1},
        pages = {A1},
          doi = {10.1051/0004-6361/201629272},
archivePrefix = {arXiv},
       eprint = {1609.04153},
 primaryClass = {astro-ph.IM},
       adsurl = {https://ui.adsabs.harvard.edu/abs/2016A&A...595A...1G}
}

@ARTICLE{GAIA2023,
       author = {{Gaia Collaboration} and {Vallenari}, A. and {Brown}, A.~G.~A. and {Prusti}, T. and {de Bruijne}, J.~H.~J. and {Arenou}, F. and {Babusiaux}, C. and {Biermann}, M. and {Creevey}, O.~L. and {Ducourant}, C. and {Evans}, D.~W. and {Eyer}, L. and {Guerra}, R. and {Hutton}, A. and {Jordi}, C. and {Klioner}, S.~A. and {Lammers}, U.~L. and {Lindegren}, L. and {Luri}, X. and {Mignard}, F. and {Panem}, C. and {Pourbaix}, D. and {Randich}, S. and {Sartoretti}, P. and {Soubiran}, C. and {Tanga}, P. and {Walton}, N.~A. and {Bailer-Jones}, C.~A.~L. and {Bastian}, U. and {Drimmel}, R. and {Jansen}, F. and {Katz}, D. and {Lattanzi}, M.~G. and {van Leeuwen}, F. and {Bakker}, J. and {Cacciari}, C. and {Casta{\~n}eda}, J. and {De Angeli}, F. and {Fabricius}, C. and {Fouesneau}, M. and {Fr{\'e}mat}, Y. and {Galluccio}, L. and {Guerrier}, A. and {Heiter}, U. and {Masana}, E. and {Messineo}, R. and {Mowlavi}, N. and {Nicolas}, C. and {Nienartowicz}, K. and {Pailler}, F. and {Panuzzo}, P. and {Riclet}, F. and {Roux}, W. and {Seabroke}, G.~M. and {Sordo}, R. and {Th{\'e}venin}, F. and {Gracia-Abril}, G. and {Portell}, J. and {Teyssier}, D. and {Altmann}, M. and {Andrae}, R. and {Audard}, M. and {Bellas-Velidis}, I. and {Benson}, K. and {Berthier}, J. and {Blomme}, R. and {Burgess}, P.~W. and {Busonero}, D. and {Busso}, G. and {C{\'a}novas}, H. and {Carry}, B. and {Cellino}, A. and {Cheek}, N. and {Clementini}, G. and {Damerdji}, Y. and {Davidson}, M. and {de Teodoro}, P. and {Nu{\~n}ez Campos}, M. and {Delchambre}, L. and {Dell'Oro}, A. and {Esquej}, P. and {Fern{\'a}ndez-Hern{\'a}ndez}, J. and {Fraile}, E. and {Garabato}, D. and {Garc{\'\i}a-Lario}, P. and {Gosset}, E. and {Haigron}, R. and {Halbwachs}, J. -L. and {Hambly}, N.~C. and {Harrison}, D.~L. and {Hern{\'a}ndez}, J. and {Hestroffer}, D. and {Hodgkin}, S.~T. and {Holl}, B. and {Jan{\ss}en}, K. and {Jevardat de Fombelle}, G. and {Jordan}, S. and {Krone-Martins}, A. and {Lanzafame}, A.~C. and {L{\"o}ffler}, W. and {Marchal}, O. and {Marrese}, P.~M. and {Moitinho}, A. and {Muinonen}, K. and {Osborne}, P. and {Pancino}, E. and {Pauwels}, T. and {Recio-Blanco}, A. and {Reyl{\'e}}, C. and {Riello}, M. and {Rimoldini}, L. and {Roegiers}, T. and {Rybizki}, J. and {Sarro}, L.~M. and {Siopis}, C. and {Smith}, M. and {Sozzetti}, A. and {Utrilla}, E. and {van Leeuwen}, M. and {Abbas}, U. and {{\'A}brah{\'a}m}, P. and {Abreu Aramburu}, A. and {Aerts}, C. and {Aguado}, J.~J. and {Ajaj}, M. and {Aldea-Montero}, F. and {Altavilla}, G. and {{\'A}lvarez}, M.~A. and {Alves}, J. and {Anders}, F. and {Anderson}, R.~I. and {Anglada Varela}, E. and {Antoja}, T. and {Baines}, D. and {Baker}, S.~G. and {Balaguer-N{\'u}{\~n}ez}, L. and {Balbinot}, E. and {Balog}, Z. and {Barache}, C. and {Barbato}, D. and {Barros}, M. and {Barstow}, M.~A. and {Bartolom{\'e}}, S. and {Bassilana}, J. -L. and {Bauchet}, N. and {Becciani}, U. and {Bellazzini}, M. and {Berihuete}, A. and {Bernet}, M. and {Bertone}, S. and {Bianchi}, L. and {Binnenfeld}, A. and {Blanco-Cuaresma}, S. and {Blazere}, A. and {Boch}, T. and {Bombrun}, A. and {Bossini}, D. and {Bouquillon}, S. and {Bragaglia}, A. and {Bramante}, L. and {Breedt}, E. and {Bressan}, A. and {Brouillet}, N. and {Brugaletta}, E. and {Bucciarelli}, B. and {Burlacu}, A. and {Butkevich}, A.~G. and {Buzzi}, R. and {Caffau}, E. and {Cancelliere}, R. and {Cantat-Gaudin}, T. and {Carballo}, R. and {Carlucci}, T. and {Carnerero}, M.~I. and {Carrasco}, J.~M. and {Casamiquela}, L. and {Castellani}, M. and {Castro-Ginard}, A. and {Chaoul}, L. and {Charlot}, P. and {Chemin}, L. and {Chiaramida}, V. and {Chiavassa}, A. and {Chornay}, N. and {Comoretto}, G. and {Contursi}, G. and {Cooper}, W.~J. and {Cornez}, T. and {Cowell}, S. and {Crifo}, F. and {Cropper}, M. and {Crosta}, M. and {Crowley}, C. and {Dafonte}, C. and {Dapergolas}, A. and {David}, M. and {David}, P. and {de Laverny}, P. and {De Luise}, F. and {De March}, R. and {De Ridder}, J. and {de Souza}, R. and {de Torres}, A. and {del Peloso}, E.~F. and {del Pozo}, E. and {Delbo}, M. and {Delgado}, A. and {Delisle}, J. -B. and {Demouchy}, C. and {Dharmawardena}, T.~E. and {Di Matteo}, P. and {Diakite}, S. and {Diener}, C. and {Distefano}, E. and {Dolding}, C. and {Edvardsson}, B. and {Enke}, H. and {Fabre}, C. and {Fabrizio}, M. and {Faigler}, S. and {Fedorets}, G. and {Fernique}, P. and {Fienga}, A. and {Figueras}, F. and {Fournier}, Y. and {Fouron}, C. and {Fragkoudi}, F. and {Gai}, M. and {Garcia-Gutierrez}, A. and {Garcia-Reinaldos}, M. and {Garc{\'\i}a-Torres}, M. and {Garofalo}, A. and {Gavel}, A. and {Gavras}, P. and {Gerlach}, E. and {Geyer}, R. and {Giacobbe}, P. and {Gilmore}, G. and {Girona}, S. and {Giuffrida}, G. and {Gomel}, R. and {Gomez}, A. and {Gonz{\'a}lez-N{\'u}{\~n}ez}, J. and {Gonz{\'a}lez-Santamar{\'\i}a}, I. and {Gonz{\'a}lez-Vidal}, J.~J. and {Granvik}, M. and {Guillout}, P. and {Guiraud}, J. and {Guti{\'e}rrez-S{\'a}nchez}, R. and {Guy}, L.~P. and {Hatzidimitriou}, D. and {Hauser}, M. and {Haywood}, M. and {Helmer}, A. and {Helmi}, A. and {Sarmiento}, M.~H. and {Hidalgo}, S.~L. and {Hilger}, T. and {H{\l}adczuk}, N. and {Hobbs}, D. and {Holland}, G. and {Huckle}, H.~E. and {Jardine}, K. and {Jasniewicz}, G. and {Jean-Antoine Piccolo}, A. and {Jim{\'e}nez-Arranz}, {\'O}. and {Jorissen}, A. and {Juaristi Campillo}, J. and {Julbe}, F. and {Karbevska}, L. and {Kervella}, P. and {Khanna}, S. and {Kontizas}, M. and {Kordopatis}, G. and {Korn}, A.~J. and {K{\'o}sp{\'a}l}, {\'A}. and {Kostrzewa-Rutkowska}, Z. and {Kruszy{\'n}ska}, K. and {Kun}, M. and {Laizeau}, P. and {Lambert}, S. and {Lanza}, A.~F. and {Lasne}, Y. and {Le Campion}, J. -F. and {Lebreton}, Y. and {Lebzelter}, T. and {Leccia}, S. and {Leclerc}, N. and {Lecoeur-Taibi}, I. and {Liao}, S. and {Licata}, E.~L. and {Lindstr{\o}m}, H.~E.~P. and {Lister}, T.~A. and {Livanou}, E. and {Lobel}, A. and {Lorca}, A. and {Loup}, C. and {Madrero Pardo}, P. and {Magdaleno Romeo}, A. and {Managau}, S. and {Mann}, R.~G. and {Manteiga}, M. and {Marchant}, J.~M. and {Marconi}, M. and {Marcos}, J. and {Marcos Santos}, M.~M.~S. and {Mar{\'\i}n Pina}, D. and {Marinoni}, S. and {Marocco}, F. and {Marshall}, D.~J. and {Martin Polo}, L. and {Mart{\'\i}n-Fleitas}, J.~M. and {Marton}, G. and {Mary}, N. and {Masip}, A. and {Massari}, D. and {Mastrobuono-Battisti}, A. and {Mazeh}, T. and {McMillan}, P.~J. and {Messina}, S. and {Michalik}, D. and {Millar}, N.~R. and {Mints}, A. and {Molina}, D. and {Molinaro}, R. and {Moln{\'a}r}, L. and {Monari}, G. and {Mongui{\'o}}, M. and {Montegriffo}, P. and {Montero}, A. and {Mor}, R. and {Mora}, A. and {Morbidelli}, R. and {Morel}, T. and {Morris}, D. and {Muraveva}, T. and {Murphy}, C.~P. and {Musella}, I. and {Nagy}, Z. and {Noval}, L. and {Oca{\~n}a}, F. and {Ogden}, A. and {Ordenovic}, C. and {Osinde}, J.~O. and {Pagani}, C. and {Pagano}, I. and {Palaversa}, L. and {Palicio}, P.~A. and {Pallas-Quintela}, L. and {Panahi}, A. and {Payne-Wardenaar}, S. and {Pe{\~n}alosa Esteller}, X. and {Penttil{\"a}}, A. and {Pichon}, B. and {Piersimoni}, A.~M. and {Pineau}, F. -X. and {Plachy}, E. and {Plum}, G. and {Poggio}, E. and {Pr{\v{s}}a}, A. and {Pulone}, L. and {Racero}, E. and {Ragaini}, S. and {Rainer}, M. and {Raiteri}, C.~M. and {Rambaux}, N. and {Ramos}, P. and {Ramos-Lerate}, M. and {Re Fiorentin}, P. and {Regibo}, S. and {Richards}, P.~J. and {Rios Diaz}, C. and {Ripepi}, V. and {Riva}, A. and {Rix}, H. -W. and {Rixon}, G. and {Robichon}, N. and {Robin}, A.~C. and {Robin}, C. and {Roelens}, M. and {Rogues}, H.~R.~O. and {Rohrbasser}, L. and {Romero-G{\'o}mez}, M. and {Rowell}, N. and {Royer}, F. and {Ruz Mieres}, D. and {Rybicki}, K.~A. and {Sadowski}, G. and {S{\'a}ez N{\'u}{\~n}ez}, A. and {Sagrist{\`a} Sell{\'e}s}, A. and {Sahlmann}, J. and {Salguero}, E. and {Samaras}, N. and {Sanchez Gimenez}, V. and {Sanna}, N. and {Santove{\~n}a}, R. and {Sarasso}, M. and {Schultheis}, M. and {Sciacca}, E. and {Segol}, M. and {Segovia}, J.~C. and {S{\'e}gransan}, D. and {Semeux}, D. and {Shahaf}, S. and {Siddiqui}, H.~I. and {Siebert}, A. and {Siltala}, L. and {Silvelo}, A. and {Slezak}, E. and {Slezak}, I. and {Smart}, R.~L. and {Snaith}, O.~N. and {Solano}, E. and {Solitro}, F. and {Souami}, D. and {Souchay}, J. and {Spagna}, A. and {Spina}, L. and {Spoto}, F. and {Steele}, I.~A. and {Steidelm{\"u}ller}, H. and {Stephenson}, C.~A. and {S{\"u}veges}, M. and {Surdej}, J. and {Szabados}, L. and {Szegedi-Elek}, E. and {Taris}, F. and {Taylor}, M.~B. and {Teixeira}, R. and {Tolomei}, L. and {Tonello}, N. and {Torra}, F. and {Torra}, J. and {Torralba Elipe}, G. and {Trabucchi}, M. and {Tsounis}, A.~T. and {Turon}, C. and {Ulla}, A. and {Unger}, N. and {Vaillant}, M.~V. and {van Dillen}, E. and {van Reeven}, W. and {Vanel}, O. and {Vecchiato}, A. and {Viala}, Y. and {Vicente}, D. and {Voutsinas}, S. and {Weiler}, M. and {Wevers}, T. and {Wyrzykowski}, {\L}. and {Yoldas}, A. and {Yvard}, P. and {Zhao}, H. and {Zorec}, J. and {Zucker}, S. and {Zwitter}, T.},
        title = "{Gaia Data Release 3. Summary of the content and survey properties}",
      journal = {\aap},
         year = 2023,
        month = jun,
       volume = {674},
          eid = {A1},
        pages = {A1},
          doi = {10.1051/0004-6361/202243940},
archivePrefix = {arXiv},
       eprint = {2208.00211},
 primaryClass = {astro-ph.GA},
       adsurl = {https://ui.adsabs.harvard.edu/abs/2023A&A...674A...1G}
}

@ARTICLE{Ando26,
       author = {{Ando}, Makoto and {Harikane}, Yuichi and {Katz}, Harley and {Inayoshi}, Kohei and {Tanaka}, Takumi S.},
        title = "{The UV Side of Little Red Dots: Red, Compact, and Iron-Enhanced Rest-UV Emission with a Strong Downturn around Ly$α$}",
      journal = {arXiv e-prints},
         year = 2026,
        month = jun,
          eid = {arXiv:2606.03522},
        pages = {arXiv:2606.03522},
          doi = {10.48550/arXiv.2606.03522},
archivePrefix = {arXiv},
       eprint = {2606.03522},
 primaryClass = {astro-ph.GA},
       adsurl = {https://ui.adsabs.harvard.edu/abs/2026arXiv260603522A}
}

@ARTICLE{Sneppen26,
       author = {{Sneppen}, A. and {Watson}, D. and {Matthews}, J.~H. and {Nikopoulos}, G. and {Allen}, N. and {Brammer}, G. and {Damgaard}, R. and {Heintz}, K.~E. and {Knigge}, C. and {Long}, K.~S. and {Rusakov}, V. and {Sim}, S.~A. and {Witstok}, J.},
        title = "{Inside the cocoon: a comprehensive explanation of the spectra of Little Red Dots}",
      journal = {arXiv e-prints},
         year = 2026,
        month = jan,
          eid = {arXiv:2601.18864},
        pages = {arXiv:2601.18864},
          doi = {10.48550/arXiv.2601.18864},
archivePrefix = {arXiv},
       eprint = {2601.18864},
 primaryClass = {astro-ph.GA},
       adsurl = {https://ui.adsabs.harvard.edu/abs/2026arXiv260118864S}
}

@article{Oke83,
	Adsurl = {http://adsabs.harvard.edu/abs/1983ApJ...266..713O},
	Author = {{Oke}, J.~B. and {Gunn}, J.~E.},
	Doi = {10.1086/160817},
	Journal = {\apj},
	Month = mar,
	Pages = {713-717},
	Title = {{Secondary standard stars for absolute spectrophotometry}},
	Volume = 266,
	Year = 1983}

@ARTICLE{Chisholm26,
       author = {{Chisholm}, John and {Berg}, Danielle A. and {Boylan-Kolchin}, Michael and {de Graaff}, Anna and {Furtak}, Lukas J. and {Kokorev}, Vasily and {Matthee}, Jorryt and {Mu{\~n}oz}, Julian B. and {Naidu}, Rohan P. and {Sander}, Andreas A.~C.},
        title = "{Little Red Dots as Globular Clusters in Formation}",
      journal = {arXiv e-prints},
         year = 2026,
        month = feb,
          eid = {arXiv:2602.15935},
        pages = {arXiv:2602.15935},
          doi = {10.48550/arXiv.2602.15935},
archivePrefix = {arXiv},
       eprint = {2602.15935},
 primaryClass = {astro-ph.GA},
       adsurl = {https://ui.adsabs.harvard.edu/abs/2026arXiv260215935C}
}

@article{Grogin11,
	Adsurl = {http://adsabs.harvard.edu/abs/2011ApJS..197...35G},
	Archiveprefix = {arXiv},
	Author = {{Grogin}, N.~A. and {Kocevski}, D.~D. and {Faber}, S.~M. and {Ferguson}, H.~C. and {Koekemoer}, A.~M. and {Riess}, A.~G. and {Acquaviva}, V. and {Alexander}, D.~M. and {Almaini}, O. and {Ashby}, M.~L.~N. and {Barden}, M. and {Bell}, E.~F. and {Bournaud}, F. and {Brown}, T.~M. and {Caputi}, K.~I. and {Casertano}, S. and {Cassata}, P. and {Castellano}, M. and {Challis}, P. and {Chary}, R.-R. and {Cheung}, E. and {Cirasuolo}, M. and {Conselice}, C.~J. and {Roshan Cooray}, A. and {Croton}, D.~J. and {Daddi}, E. and {Dahlen}, T. and {Dav{\'e}}, R. and {de Mello}, D.~F. and {Dekel}, A. and {Dickinson}, M. and {Dolch}, T. and {Donley}, J.~L. and {Dunlop}, J.~S. and {Dutton}, A.~A. and {Elbaz}, D. and {Fazio}, G.~G. and {Filippenko}, A.~V. and {Finkelstein}, S.~L. and {Fontana}, A. and {Gardner}, J.~P. and {Garnavich}, P.~M. and {Gawiser}, E. and {Giavalisco}, M. and {Grazian}, A. and {Guo}, Y. and {Hathi}, N.~P. and {H{\"a}ussler}, B. and {Hopkins}, P.~F. and {Huang}, J.-S. and {Huang}, K.-H. and {Jha}, S.~W. and {Kartaltepe}, J.~S. and {Kirshner}, R.~P. and {Koo}, D.~C. and {Lai}, K. and {Lee}, K.-S. and {Li}, W. and {Lotz}, J.~M. and {Lucas}, R.~A. and {Madau}, P. and {McCarthy}, P.~J. and {McGrath}, E.~J. and {McIntosh}, D.~H. and {McLure}, R.~J. and {Mobasher}, B. and {Moustakas}, L.~A. and {Mozena}, M. and {Nandra}, K. and {Newman}, J.~A. and {Niemi}, S.-M. and {Noeske}, K.~G. and {Papovich}, C.~J. and {Pentericci}, L. and {Pope}, A. and {Primack}, J.~R. and {Rajan}, A. and {Ravindranath}, S. and {Reddy}, N.~A. and {Renzini}, A. and {Rix}, H.-W. and {Robaina}, A.~R. and {Rodney}, S.~A. and {Rosario}, D.~J. and {Rosati}, P. and {Salimbeni}, S. and {Scarlata}, C. and {Siana}, B. and {Simard}, L. and {Smidt}, J. and {Somerville}, R.~S. and {Spinrad}, H. and {Straughn}, A.~N. and {Strolger}, L.-G. and {Telford}, O. and {Teplitz}, H.~I. and {Trump}, J.~R. and {van der Wel}, A. and {Villforth}, C. and {Wechsler}, R.~H. and {Weiner}, B.~J. and {Wiklind}, T. and {Wild}, V. and {Wilson}, G. and {Wuyts}, S. and {Yan}, H.-J. and {Yun}, M.~S.},
	Doi = {10.1088/0067-0049/197/2/35},
	Eid = {35},
	Eprint = {1105.3753},
	Journal = {\apjs},
	Month = dec,
	Pages = {35},
	Primaryclass = {astro-ph.CO},
	Title = {{CANDELS: The Cosmic Assembly Near-infrared Deep Extragalactic Legacy Survey}},
	Volume = 197,
	Year = 2011}

@ARTICLE{Rusakov26,
       author = {{Rusakov}, V. and {Watson}, D. and {Nikopoulos}, G.~P. and {Brammer}, G. and {Gottumukkala}, R. and {Harvey}, T. and {Heintz}, K.~E. and {Damgaard}, R. and {Sim}, S.~A. and {Sneppen}, A. and {Vijayan}, A.~P. and {Adams}, N. and {Austin}, D. and {Conselice}, C.~J. and {Goolsby}, C.~M. and {Toft}, S. and {Witstok}, J.},
        title = "{Little red dots as young supermassive black holes in dense ionized cocoons}",
      journal = {\nat},
         year = 2026,
        month = jan,
       volume = {649},
       number = {8097},
        pages = {574-579},
          doi = {10.1038/s41586-025-09900-4},
archivePrefix = {arXiv},
       eprint = {2503.16595},
 primaryClass = {astro-ph.GA},
       adsurl = {https://ui.adsabs.harvard.edu/abs/2026Natur.649..574R}
}

@ARTICLE{Weaver24,
       author = {{Weaver}, John R. and {Cutler}, Sam E. and {Pan}, Richard and {Whitaker}, Katherine E. and {Labb{\'e}}, Ivo and {Price}, Sedona H. and {Bezanson}, Rachel and {Brammer}, Gabriel and {Marchesini}, Danilo and {Leja}, Joel and {Wang}, Bingjie and {Furtak}, Lukas J. and {Zitrin}, Adi and {Atek}, Hakim and {Chemerynska}, Iryna and {Coe}, Dan and {Dayal}, Pratika and {van Dokkum}, Pieter and {Feldmann}, Robert and {F{\"o}rster Schreiber}, Natascha M. and {Franx}, Marijn and {Fujimoto}, Seiji and {Fudamoto}, Yoshinobu and {Glazebrook}, Karl and {de Graaff}, Anna and {Greene}, Jenny E. and {Juneau}, St{\'e}phanie and {Kassin}, Susan and {Kriek}, Mariska and {Khullar}, Gourav and {Maseda}, Michael V. and {Mowla}, Lamiya A. and {Muzzin}, Adam and {Nanayakkara}, Themiya and {Nelson}, Erica J. and {Oesch}, Pascal A. and {Pacifici}, Camilla and {Papovich}, Casey and {Setton}, David J. and {Shapley}, Alice E. and {Shipley}, Heath V. and {Smit}, Renske and {Stefanon}, Mauro and {Taylor}, Edward N. and {Weibel}, Andrea and {Williams}, Christina C.},
        title = "{The UNCOVER Survey: A First-look HST + JWST Catalog of 60,000 Galaxies near A2744 and beyond}",
      journal = {\apjs},
         year = 2024,
        month = jan,
       volume = {270},
       number = {1},
          eid = {7},
        pages = {7},
          doi = {10.3847/1538-4365/ad07e0},
archivePrefix = {arXiv},
       eprint = {2301.02671},
 primaryClass = {astro-ph.GA},
       adsurl = {https://ui.adsabs.harvard.edu/abs/2024ApJS..270....7W}
}

@ARTICLE{Wang25Outflow,
       author = {{Wang}, Bingjie and {de Graaff}, Anna and {Davies}, Rebecca L. and {Greene}, Jenny E. and {Leja}, Joel and {Brammer}, Gabriel B. and {Goulding}, Andy D. and {Miller}, Tim B. and {Suess}, Katherine A. and {Weibel}, Andrea and {Williams}, Christina C. and {Bezanson}, Rachel and {Boogaard}, Leindert A. and {Cleri}, Nikko J. and {Hirschmann}, Michaela and {Katz}, Harley and {Labb{\'e}}, Ivo and {Maseda}, Michael V. and {Matthee}, Jorryt and {McConachie}, Ian and {Naidu}, Rohan P. and {Oesch}, Pascal A. and {Rix}, Hans-Walter and {Setton}, David J. and {Whitaker}, Katherine E.},
        title = "{RUBIES: JWST/NIRSpec Confirmation of an Infrared-luminous, Broad-line Little Red Dot with an Ionized Outflow}",
      journal = {\apj},
         year = 2025,
        month = may,
       volume = {984},
       number = {2},
          eid = {121},
        pages = {121},
          doi = {10.3847/1538-4357/adc1ca},
archivePrefix = {arXiv},
       eprint = {2403.02304},
 primaryClass = {astro-ph.GA},
       adsurl = {https://ui.adsabs.harvard.edu/abs/2025ApJ...984..121W}
}

@ARTICLE{Wang26,
       author = {{Wang}, Bingjie and {Leja}, Joel and {Labbe}, Ivo and {Greene}, Jenny E. and {Liu}, Hanpu and {de Graaff}, Anna and {Hviding}, Raphael E. and {Matthee}, Jorryt and {Quataert}, Eliot and {Bezanson}, Rachel and {Boogaard}, Leindert A. and {Brammer}, Gabriel and {Burgasser}, Adam J. and {Chen}, Yi-Xian and {Cleri}, Nikko J. and {Cutler}, Sam E. and {Dayal}, Pratika and {Furtak}, Lukas J. and {Fujimoto}, Seiji and {Glazebrook}, Karl and {Goulding}, Andy D. and {Helton}, Jakob M. and {Hirschmann}, Michaela and {Jiang}, Yan-Fei and {Kokorev}, Vasily and {Ma}, Yilun and {Miller}, Tim B. and {Naidu}, Rohan P. and {Oesch}, Pascal and {Pan}, Richard and {Papovich}, Casey and {Price}, Sedona H. and {Rix}, Hans-Walter and {Setton}, David J. and {Sun}, Wendy Q. and {Weaver}, John R. and {Whitaker}, Katherine E. and {Zitrin}, Adi},
        title = "{Water absorption confirms cool atmospheres in two little red dots}",
      journal = {arXiv e-prints},
         year = 2026,
        month = feb,
          eid = {arXiv:2602.06024},
        pages = {arXiv:2602.06024},
          doi = {10.48550/arXiv.2602.06024},
archivePrefix = {arXiv},
       eprint = {2602.06024},
 primaryClass = {astro-ph.GA},
       adsurl = {https://ui.adsabs.harvard.edu/abs/2026arXiv260206024W}
}

@ARTICLE{Cloonan2026,
       author = {{Cloonan}, Aidan P. and {Whitaker}, Katherine E. and {Manning}, Sinclaire M. and {Williams}, Christina C. and {Greene}, Jenny E. and {Oesch}, Pascal A. and {Weibel}, Andrea and {Brammer}, Gabriel and {de Graaff}, Anna and {Hviding}, Raphael E. and {Dayal}, Pratika and {Jespersen}, Christian Kragh and {Ji}, Zhiyuan and {Labbe}, Ivo and {Xiao}, Mengyuan and {Zhang}, Yunchong},
        title = "{A PANORAMIC of UV-optical morphologies of ``Little Red Dots'': Two groups of LRDs distinguished by UV half-light radius}",
      journal = {arXiv e-prints},
         year = 2026,
        month = mar,
          eid = {arXiv:2603.24700},
        pages = {arXiv:2603.24700},
          doi = {10.48550/arXiv.2603.24700},
archivePrefix = {arXiv},
       eprint = {2603.24700},
 primaryClass = {astro-ph.GA},
       adsurl = {https://ui.adsabs.harvard.edu/abs/2026arXiv260324700C}
}

@ARTICLE{Weibel26,
       author = {{Weibel}, Andrea and {Oesch}, Pascal A. and {Williams}, Christina C. and {Jespersen}, Christian Kragh and {Shuntov}, Marko and {Whitaker}, Katherine E. and {Atek}, Hakim and {Bezanson}, Rachel and {Brammer}, Gabriel and {Chemerynska}, Iryna and {Cloonan}, Aidan P. and {Dayal}, Pratika and {Furtak}, Lukas J. and {Hutter}, Anne and {Ji}, Zhiyuan and {Maseda}, Michael V. and {Xiao}, Mengyuan},
        title = "{Exploring Cosmic Dawn with PANORAMIC. I. The Bright End of the UV Luminosity Function at z {\ensuremath{\sim}} 9─17}",
      journal = {\apj},
         year = 2026,
        month = may,
       volume = {1002},
       number = {2},
          eid = {136},
        pages = {136},
          doi = {10.3847/1538-4357/ae5a9c},
archivePrefix = {arXiv},
       eprint = {2507.06292},
 primaryClass = {astro-ph.GA},
       adsurl = {https://ui.adsabs.harvard.edu/abs/2026ApJ..1002..136W}
}

@ARTICLE{Ferland17,
       author = {{Ferland}, G.~J. and {Chatzikos}, M. and {Guzm{\'a}n}, F. and {Lykins}, M.~L. and {van Hoof}, P.~A.~M. and {Williams}, R.~J.~R. and {Abel}, N.~P. and {Badnell}, N.~R. and {Keenan}, F.~P. and {Porter}, R.~L. and {Stancil}, P.~C.},
        title = "{The 2017 Release Cloudy}",
      journal = {\rmxaa},
         year = 2017,
        month = oct,
       volume = {53},
        pages = {385-438},
          doi = {10.48550/arXiv.1705.10877},
archivePrefix = {arXiv},
       eprint = {1705.10877},
 primaryClass = {astro-ph.GA},
       adsurl = {https://ui.adsabs.harvard.edu/abs/2017RMxAA..53..385F}
}

@ARTICLE{Brammer08,
   author = {{Brammer}, G.~B. and {van Dokkum}, P.~G. and {Coppi}, P.},
    title = "{EAZY: A Fast, Public Photometric Redshift Code}",
  journal = {\apj},
archivePrefix = "arXiv",
   eprint = {0807.1533},
     year = 2008,
    month = oct,
   volume = 686,
      eid = {1503-1513},
    pages = {1503-1513},
      doi = {10.1086/591786},
   adsurl = {http://adsabs.harvard.edu/abs/2008ApJ...686.1503B}
}

@ARTICLE{astropy,
   author = {{Astropy Collaboration} and {Robitaille}, T.~P. and {Tollerud}, E.~J. and 
	{Greenfield}, P. and {Droettboom}, M. and {Bray}, E. and {Aldcroft}, T. and 
	{Davis}, M. and {Ginsburg}, A. and {Price-Whelan}, A.~M. and 
	{Kerzendorf}, W.~E. and {Conley}, A. and {Crighton}, N. and 
	{Barbary}, K. and {Muna}, D. and {Ferguson}, H. and {Grollier}, F. and 
	{Parikh}, M.~M. and {Nair}, P.~H. and {Unther}, H.~M. and {Deil}, C. and 
	{Woillez}, J. and {Conseil}, S. and {Kramer}, R. and {Turner}, J.~E.~H. and 
	{Singer}, L. and {Fox}, R. and {Weaver}, B.~A. and {Zabalza}, V. and 
	{Edwards}, Z.~I. and {Azalee Bostroem}, K. and {Burke}, D.~J. and 
	{Casey}, A.~R. and {Crawford}, S.~M. and {Dencheva}, N. and 
	{Ely}, J. and {Jenness}, T. and {Labrie}, K. and {Lim}, P.~L. and 
	{Pierfederici}, F. and {Pontzen}, A. and {Ptak}, A. and {Refsdal}, B. and 
	{Servillat}, M. and {Streicher}, O.},
    title = "{Astropy: A community Python package for astronomy}",
  journal = {\aap},
archivePrefix = "arXiv",
   eprint = {1307.6212},
 primaryClass = "astro-ph.IM",
     year = 2013,
    month = oct,
   volume = 558,
      eid = {A33},
    pages = {A33},
      doi = {10.1051/0004-6361/201322068},
   adsurl = {http://adsabs.harvard.edu/abs/2013A%26A...558A..33A}
}

@ARTICLE{Carnall23,
       author = {{Carnall}, A.~C. and {Begley}, R. and {McLeod}, D.~J. and {Hamadouche}, M.~L. and {Donnan}, C.~T. and {McLure}, R.~J. and {Dunlop}, J.~S. and {Milvang-Jensen}, B. and {Bondestam}, C.~L. and {Cullen}, F. and {Jewell}, S.~M. and {Pollock}, C.~L.},
        title = "{A first look at the SMACS0723 JWST ERO: spectroscopic redshifts, stellar masses, and star-formation histories}",
      journal = {\mnras},
         year = 2023,
        month = jan,
       volume = {518},
       number = {1},
        pages = {L45-L50},
          doi = {10.1093/mnrasl/slac136},
archivePrefix = {arXiv},
       eprint = {2207.08778},
 primaryClass = {astro-ph.GA},
       adsurl = {https://ui.adsabs.harvard.edu/abs/2023MNRAS.518L..45C}
}

@Article{ipython,
  Author    = {P\'erez, Fernando and Granger, Brian E.},
  Title     = {{IP}ython: a System for Interactive Scientific Computing},
  Journal   = {Computing in Science and Engineering},
  Volume    = {9},
  Number    = {3},
  Pages     = {21--29},
  month     = may,
  year      = 2007,
  url       = "https://ipython.org",
  ISSN      = "1521-9615",
  doi       = {10.1109/MCSE.2007.53},
  publisher = {IEEE Computer Society},
}

@ARTICLE{scipy,
       author = {{Virtanen}, Pauli and {Gommers}, Ralf and {Oliphant}, Travis E. and {Haberland}, Matt and {Reddy}, Tyler and {Cournapeau}, David and {Burovski}, Evgeni and {Peterson}, Pearu and {Weckesser}, Warren and {Bright}, Jonathan and {van der Walt}, St{\'e}fan J. and {Brett}, Matthew and {Wilson}, Joshua and {Millman}, K. Jarrod and {Mayorov}, Nikolay and {Nelson}, Andrew R.~J. and {Jones}, Eric and {Kern}, Robert and {Larson}, Eric and {Carey}, C.~J. and {Polat}, {\.I}lhan and {Feng}, Yu and {Moore}, Eric W. and {VanderPlas}, Jake and {Laxalde}, Denis and {Perktold}, Josef and {Cimrman}, Robert and {Henriksen}, Ian and {Quintero}, E.~A. and {Harris}, Charles R. and {Archibald}, Anne M. and {Ribeiro}, Ant{\^o}nio H. and {Pedregosa}, Fabian and {van Mulbregt}, Paul and {SciPy 1. 0 Contributors}},
        title = "{SciPy 1.0: fundamental algorithms for scientific computing in Python}",
      journal = {Nature Methods},
         year = 2020,
        month = feb,
       volume = {17},
        pages = {261-272},
          doi = {10.1038/s41592-019-0686-2},
archivePrefix = {arXiv},
       eprint = {1907.10121},
 primaryClass = {cs.MS},
       adsurl = {https://ui.adsabs.harvard.edu/abs/2020NatMe..17..261V}
}

@book{numpy, place={Austin, TX}, title={Guide to NumPy}, publisher={Continuum Press}, author={Oliphant, Travis E.}, year={2015}}

@Article{matplotlib,
  Author    = {Hunter, J. D.},
  Title     = {Matplotlib: A 2D graphics environment},
  Journal   = {Computing In Science \& Engineering},
  Volume    = {9},
  Number    = {3},
  Pages     = {90--95},
  publisher = {IEEE COMPUTER SOC},
  doi       = {10.1109/MCSE.2007.55},
  year      = 2007
}

@conference{jupyter,
	Author = {Thomas Kluyver and Benjamin Ragan-Kelley and Fernando P{\'e}rez and Brian Granger and Matthias Bussonnier and Jonathan Frederic and Kyle Kelley and Jessica Hamrick and Jason Grout and Sylvain Corlay and Paul Ivanov and Dami{\'a}n Avila and Safia Abdalla and Carol Willing},
	Booktitle = {Positioning and Power in Academic Publishing: Players, Agents and Agendas},
	Editor = {F. Loizides and B. Schmidt},
	Organization = {IOS Press},
	Pages = {87 - 90},
	Title = {Jupyter Notebooks -- a publishing format for reproducible computational workflows},
	Year = {2016}}

@ARTICLE{Jiang26,
       author = {{Jiang}, Fangzhou and {Jia}, Zixiang and {Zheng}, Haonan and {Ho}, Luis C. and {Inayoshi}, Kohei and {Shen}, Xuejian and {Vogelsberger}, Mark and {Feng}, Wei-Xiang},
        title = "{Formation of the Little Red Dots from the Core Collapse of Self-interacting Dark Matter Halos}",
      journal = {\apjl},
         year = 2026,
        month = jan,
       volume = {996},
       number = {1},
          eid = {L19},
        pages = {L19},
          doi = {10.3847/2041-8213/ae247a},
archivePrefix = {arXiv},
       eprint = {2503.23710},
 primaryClass = {astro-ph.GA},
       adsurl = {https://ui.adsabs.harvard.edu/abs/2026ApJ...996L..19J}
}

@ARTICLE{Begelman08,
       author = {{Begelman}, Mitchell C. and {Rossi}, Elena M. and {Armitage}, Philip J.},
        title = "{Quasi-stars: accreting black holes inside massive envelopes}",
      journal = {\mnras},
         year = 2008,
        month = jul,
       volume = {387},
       number = {4},
        pages = {1649-1659},
          doi = {10.1111/j.1365-2966.2008.13344.x},
archivePrefix = {arXiv},
       eprint = {0711.4078},
 primaryClass = {astro-ph},
       adsurl = {https://ui.adsabs.harvard.edu/abs/2008MNRAS.387.1649B}
}

@ARTICLE{Ma26,
       author = {{Ma}, Yilun and {Greene}, Jenny E. and {Setton}, David J. and {Goulding}, Andy D. and {Annunziatella}, Marianna and {Fan}, Xiaohui and {Kokorev}, Vasily and {Labbe}, Ivo and {Li}, Jiaxuan and {Lin}, Xiaojing and {Marchesini}, Danilo and {Matthee}, Jorryt and {Robbins}, Luke and {Sajina}, Anna and {Sawicki}, Marcin and {Telford}, O. Grace},
        title = "{Counting Little Red Dots at z < 4 with Ground-based Surveys and Spectroscopic Follow-up}",
      journal = {\apj},
         year = 2026,
        month = mar,
       volume = {1000},
       number = {1},
          eid = {59},
        pages = {59},
          doi = {10.3847/1538-4357/ae4596},
archivePrefix = {arXiv},
       eprint = {2504.08032},
 primaryClass = {astro-ph.GA},
       adsurl = {https://ui.adsabs.harvard.edu/abs/2026ApJ..1000...59M}
}

@ARTICLE{Ma25cutoff,
       author = {{Ma}, Yilun and {Greene}, Jenny E. and {Volonteri}, Marta and {Goulding}, Andy D. and {Setton}, David J. and {Annunziatella}, Marianna and {Egami}, Eiichi and {Fan}, Xiaohui and {Kokorev}, Vasily and {Labbe}, Ivo and {Lin}, Xiaojing and {Marchesini}, Danilo and {Matthee}, Jorryt and {Nanayakkara}, Themiya and {Robbins}, Luke and {Sajina}, Anna and {Sawicki}, Marcin},
        title = "{No Luminous Little Red Dots: A Sharp Cutoff in Their Luminosity Function}",
      journal = {arXiv e-prints},
         year = 2025,
        month = sep,
          eid = {arXiv:2509.02662},
        pages = {arXiv:2509.02662},
          doi = {10.48550/arXiv.2509.02662},
archivePrefix = {arXiv},
       eprint = {2509.02662},
 primaryClass = {astro-ph.GA},
       adsurl = {https://ui.adsabs.harvard.edu/abs/2025arXiv250902662M}
}

@ARTICLE{Xiao25,
       author = {{Xiao}, Mengyuan and {Oesch}, Pascal A. and {Bing}, Longji and {Elbaz}, David and {Matthee}, Jorryt and {Fudamoto}, Yoshinobu and {Fujimoto}, Seiji and {Marques-Chaves}, Rui and {Williams}, Christina C. and {Dessauges-Zavadsky}, Miroslava and {Valentino}, Francesco and {Brammer}, Gabriel and {Covelo-Paz}, Alba and {Daddi}, Emanuele and {Fynbo}, Johan P.~U. and {Gillman}, Steven and {Ginolfi}, Michele and {Giovinazzo}, Emma and {Greene}, Jenny E. and {Gu}, Qiusheng and {Illingworth}, Garth and {Inayoshi}, Kohei and {Kokorev}, Vasily and {Meyer}, Romain A. and {Naidu}, Rohan P. and {Reddy}, Naveen A. and {Schaerer}, Daniel and {Shapley}, Alice and {Stefanon}, Mauro and {Steinhardt}, Charles L. and {Setton}, David J. and {Vestergaard}, Marianne and {Wang}, Tao},
        title = "{No [C II] or dust detection in two Little Red Dots at z$_{spec}$> 7}",
      journal = {\aap},
         year = 2025,
        month = aug,
       volume = {700},
          eid = {A231},
        pages = {A231},
          doi = {10.1051/0004-6361/202554361},
archivePrefix = {arXiv},
       eprint = {2503.01945},
 primaryClass = {astro-ph.GA},
       adsurl = {https://ui.adsabs.harvard.edu/abs/2025A&A...700A.231X}
}

@ARTICLE{Setton25,
       author = {{Setton}, David J. and {Greene}, Jenny E. and {Spilker}, Justin S. and {Williams}, Christina C. and {Labb{\'e}}, Ivo and {Ma}, Yilun and {Wang}, Bingjie and {Whitaker}, Katherine E. and {Leja}, Joel and {de Graaff}, Anna and {Alberts}, Stacey and {Bezanson}, Rachel and {Boogaard}, Leindert A. and {Brammer}, Gabriel and {Cutler}, Sam E. and {Cleri}, Nikko J. and {Cooper}, Olivia R. and {Dayal}, Pratika and {Fujimoto}, Seiji and {Furtak}, Lukas J. and {Goulding}, Andy D. and {Hirschmann}, Michaela and {Kokorev}, Vasily and {Maseda}, Michael V. and {McConachie}, Ian and {Matthee}, Jorryt and {Miller}, Tim B. and {Naidu}, Rohan P. and {Oesch}, Pascal A. and {Pan}, Richard and {Price}, Sedona H. and {Suess}, Katherine A. and {Weaver}, John R. and {Xiao}, Mengyuan and {Zhang}, Yunchong and {Zitrin}, Adi},
        title = "{A Confirmed Deficit of Hot and Cold Dust Emission in the Most Luminous Little Red Dots}",
      journal = {\apjl},
         year = 2025,
        month = sep,
       volume = {991},
       number = {1},
          eid = {L10},
        pages = {L10},
          doi = {10.3847/2041-8213/ade78b},
archivePrefix = {arXiv},
       eprint = {2503.02059},
 primaryClass = {astro-ph.GA},
       adsurl = {https://ui.adsabs.harvard.edu/abs/2025ApJ...991L..10S}
}

@ARTICLE{Setton25inflection,
       author = {{Setton}, David J. and {Greene}, Jenny E. and {de Graaff}, Anna and {Ma}, Yilun and {Leja}, Joel and {Matthee}, Jorryt and {Bezanson}, Rachel and {Boogaard}, Leindert A. and {Cleri}, Nikko J. and {Katz}, Harley and {Labbe}, Ivo and {Maseda}, Michael V. and {McConachie}, Ian and {Miller}, Tim B. and {Price}, Sedona H. and {Suess}, Katherine A. and {van Dokkum}, Pieter and {Wang}, Bingjie and {Weibel}, Andrea and {Whitaker}, Katherine E. and {Williams}, Christina C.},
        title = "{Little Red Dots at an Inflection Point: Ubiquitous V-shaped Turnover Consistently Occurs at the Balmer Limit}",
      journal = {\apj},
         year = 2025,
        month = dec,
       volume = {995},
       number = {1},
          eid = {118},
        pages = {118},
          doi = {10.3847/1538-4357/ae1500},
archivePrefix = {arXiv},
       eprint = {2411.03424},
 primaryClass = {astro-ph.GA},
       adsurl = {https://ui.adsabs.harvard.edu/abs/2025ApJ...995..118S}
}

@ARTICLE{Greene26,
       author = {{Greene}, Jenny E. and {Setton}, David J. and {Furtak}, Lukas J. and {Naidu}, Rohan P. and {Volonteri}, Marta and {Dayal}, Pratika and {Labbe}, Ivo and {van Dokkum}, Pieter and {Bezanson}, Rachel and {Brammer}, Gabriel and {Cutler}, Sam E. and {Glazebrook}, Karl and {de Graaff}, Anna and {Hirschmann}, Michaela and {Hviding}, Raphael E. and {Kokorev}, Vasily and {Leja}, Joel and {Liu}, Hanpu and {Ma}, Yilun and {Matthee}, Jorryt and {Nanayakkara}, Themiya and {Oesch}, Pascal A. and {Pan}, Richard and {Price}, Sedona H. and {Spilker}, Justin S. and {Wang}, Bingjie and {Weaver}, John R. and {Whitaker}, Katherine E. and {Williams}, Christina C. and {Zitrin}, Adi},
        title = "{What You See Is What You Get: Empirically Measured Bolometric Luminosities of Little Red Dots}",
      journal = {\apj},
         year = 2026,
        month = jan,
       volume = {996},
       number = {2},
          eid = {129},
        pages = {129},
          doi = {10.3847/1538-4357/ae1836},
archivePrefix = {arXiv},
       eprint = {2509.05434},
 primaryClass = {astro-ph.GA},
       adsurl = {https://ui.adsabs.harvard.edu/abs/2026ApJ...996..129G}
}

@ARTICLE{Akins25,
       author = {{Akins}, Hollis B. and {Casey}, Caitlin M. and {Lambrides}, Erini and {Allen}, Natalie and {Andika}, Irham T. and {Brinch}, Malte and {Champagne}, Jaclyn B. and {Cooper}, Olivia and {Ding}, Xuheng and {Drakos}, Nicole E. and {Faisst}, Andreas and {Finkelstein}, Steven L. and {Franco}, Maximilien and {Fujimoto}, Seiji and {Gentile}, Fabrizio and {Gillman}, Steven and {Gozaliasl}, Ghassem and {Harish}, Santosh and {Hayward}, Christopher C. and {Hirschmann}, Michaela and {Ilbert}, Olivier and {Kartaltepe}, Jeyhan S. and {Kocevski}, Dale D. and {Koekemoer}, Anton M. and {Kokorev}, Vasily and {Liu}, Daizhong and {Long}, Arianna S. and {McCracken}, Henry Joy and {McKinney}, Jed and {Onoue}, Masafusa and {Paquereau}, Louise and {Renzini}, Alvio and {Rhodes}, Jason and {Robertson}, Brant E. and {Shuntov}, Marko and {Silverman}, John D. and {Tanaka}, Takumi S. and {Toft}, Sune and {Trakhtenbrot}, Benny and {Valentino}, Francesco and {Zavala}, Jorge},
        title = "{COSMOS-Web: The Overabundance and Physical Nature of ``Little Red Dots''{\textemdash}Implications for Early Galaxy and SMBH Assembly}",
      journal = {\apj},
         year = 2025,
        month = sep,
       volume = {991},
       number = {1},
          eid = {37},
        pages = {37},
          doi = {10.3847/1538-4357/ade984},
archivePrefix = {arXiv},
       eprint = {2406.10341},
 primaryClass = {astro-ph.GA},
       adsurl = {https://ui.adsabs.harvard.edu/abs/2025ApJ...991...37A}
}

@ARTICLE{Labbe25LRD,
       author = {{Labbe}, Ivo and {Greene}, Jenny E. and {Bezanson}, Rachel and {Fujimoto}, Seiji and {Furtak}, Lukas J. and {Goulding}, Andy D. and {Matthee}, Jorryt and {Naidu}, Rohan P. and {Oesch}, Pascal A. and {Atek}, Hakim and {Brammer}, Gabriel and {Chemerynska}, Iryna and {Coe}, Dan and {Cutler}, Sam E. and {Dayal}, Pratika and {Feldmann}, Robert and {Franx}, Marijn and {Glazebrook}, Karl and {Leja}, Joel and {Maseda}, Michael and {Marchesini}, Danilo and {Nanayakkara}, Themiya and {Nelson}, Erica J. and {Pan}, Richard and {Papovich}, Casey and {Price}, Sedona H. and {Suess}, Katherine A. and {Wang}, Bingjie and {Weaver}, John R. and {Whitaker}, Katherine E. and {Williams}, Christina C. and {Zitrin}, Adi},
        title = "{UNCOVER: Candidate Red Active Galactic Nuclei at 3 < z < 7 with JWST and ALMA}",
      journal = {\apj},
         year = 2025,
        month = jan,
       volume = {978},
       number = {1},
          eid = {92},
        pages = {92},
          doi = {10.3847/1538-4357/ad3551},
archivePrefix = {arXiv},
       eprint = {2306.07320},
 primaryClass = {astro-ph.GA},
       adsurl = {https://ui.adsabs.harvard.edu/abs/2025ApJ...978...92L}
}

@ARTICLE{Hviding25,
       author = {{Hviding}, Raphael E. and {de Graaff}, Anna and {Miller}, Tim B. and {Setton}, David J. and {Greene}, Jenny E. and {Labb{\'e}}, Ivo and {Brammer}, Gabriel and {Bezanson}, Rachel and {Boogaard}, Leindert A. and {Cleri}, Nikko J. and {Leja}, Joel and {Maseda}, Michael V. and {McConachie}, Ian and {Matthee}, Jorryt and {Naidu}, Rohan P. and {Oesch}, Pascal A. and {Wang}, Bingjie and {Whitaker}, Katherine E. and {Williams}, Christina C.},
        title = "{RUBIES: A spectroscopic census of little red dots: All point sources with v-shaped continua have broad lines}",
      journal = {\aap},
         year = 2025,
        month = oct,
       volume = {702},
          eid = {A57},
        pages = {A57},
          doi = {10.1051/0004-6361/202555816},
archivePrefix = {arXiv},
       eprint = {2506.05459},
 primaryClass = {astro-ph.GA},
       adsurl = {https://ui.adsabs.harvard.edu/abs/2025A&A...702A..57H}
}

@ARTICLE{Nandal26,
       author = {{Nandal}, Devesh and {Loeb}, Abraham},
        title = "{Supermassive Stars Match the Spectral Signatures of JWST's Little Red Dots}",
      journal = {\apj},
         year = 2026,
        month = feb,
       volume = {998},
       number = {1},
          eid = {124},
        pages = {124},
          doi = {10.3847/1538-4357/ae32f3},
archivePrefix = {arXiv},
       eprint = {2507.12618},
 primaryClass = {astro-ph.GA},
       adsurl = {https://ui.adsabs.harvard.edu/abs/2026ApJ...998..124N}
}

@ARTICLE{Yue24,
       author = {{Yue}, Minghao and {Eilers}, Anna-Christina and {Ananna}, Tonima Tasnim and {Panagiotou}, Christos and {Kara}, Erin and {Miyaji}, Takamitsu},
        title = "{Stacking X-Ray Observations of ``Little Red Dots'': Implications for Their Active Galactic Nucleus Properties}",
      journal = {\apjl},
         year = 2024,
        month = oct,
       volume = {974},
       number = {2},
          eid = {L26},
        pages = {L26},
          doi = {10.3847/2041-8213/ad7eba},
archivePrefix = {arXiv},
       eprint = {2404.13290},
 primaryClass = {astro-ph.GA},
       adsurl = {https://ui.adsabs.harvard.edu/abs/2024ApJ...974L..26Y}
}

@ARTICLE{degraaff25pop,
       author = {{de Graaff}, Anna and {Hviding}, Raphael E. and {Naidu}, Rohan P. and {Greene}, Jenny E. and {Miller}, Tim B. and {Leja}, Joel and {Matthee}, Jorryt and {Brammer}, Gabriel and {Katz}, Harley and {Bezanson}, Rachel and {Boogaard}, Leindert A. and {Bose}, Sownak and {Chisholm}, John and {Cleri}, Nikko J. and {Dayal}, Pratika and {Feldmann}, Robert and {Fudamoto}, Yoshinobu and {Fujimoto}, Seiji and {Furtak}, Lukas J. and {Glazebrook}, Karl and {Gottumukkala}, Rashmi and {Heintz}, Kasper E. and {Kokorev}, Vasily and {Labbe}, Ivo and {Maseda}, Michael V. and {McConachie}, Ian and {Nanayakkara}, Themiya and {Nelson}, Erica and {Nowaczyk}, Przemys{\l}aw and {Oesch}, Pascal A. and {Rix}, Hans-Walter and {Setton}, David J. and {Torralba}, Alberto and {Walter}, Fabian and {Wang}, Bingjie and {Weibel}, Andrea and {van der Wel}, Arjen},
        title = "{Little Red Dots host Black Hole Stars: A unified family of gas-reddened AGN revealed by JWST/NIRSpec spectroscopy}",
      journal = {arXiv e-prints},
         year = 2025,
        month = nov,
          eid = {arXiv:2511.21820},
        pages = {arXiv:2511.21820},
          doi = {10.48550/arXiv.2511.21820},
archivePrefix = {arXiv},
       eprint = {2511.21820},
 primaryClass = {astro-ph.GA},
       adsurl = {https://ui.adsabs.harvard.edu/abs/2025arXiv251121820D}
}

@ARTICLE{degraaff25,
       author = {{de Graaff}, Anna and {Rix}, Hans-Walter and {Naidu}, Rohan P. and {Labb{\'e}}, Ivo and {Wang}, Bingjie and {Leja}, Joel and {Matthee}, Jorryt and {Katz}, Harley and {Greene}, Jenny E. and {Hviding}, Raphael E. and {Baggen}, Josephine and {Bezanson}, Rachel and {Boogaard}, Leindert A. and {Brammer}, Gabriel and {Dayal}, Pratika and {van Dokkum}, Pieter and {Goulding}, Andy D. and {Hirschmann}, Michaela and {Maseda}, Michael V. and {McConachie}, Ian and {Miller}, Tim B. and {Nelson}, Erica and {Oesch}, Pascal A. and {Setton}, David J. and {Shivaei}, Irene and {Weibel}, Andrea and {Whitaker}, Katherine E. and {Williams}, Christina C.},
        title = "{A remarkable ruby: Absorption in dense gas, rather than evolved stars, drives the extreme Balmer break of a little red dot at z = 3.5}",
      journal = {\aap},
         year = 2025,
        month = sep,
       volume = {701},
          eid = {A168},
        pages = {A168},
          doi = {10.1051/0004-6361/202554681},
archivePrefix = {arXiv},
       eprint = {2503.16600},
 primaryClass = {astro-ph.GA},
       adsurl = {https://ui.adsabs.harvard.edu/abs/2025A&A...701A.168D}
}

@ARTICLE{Labbe24,
       author = {{Labbe}, Ivo and {Greene}, Jenny E. and {Matthee}, Jorryt and {Treiber}, Helena and {Kokorev}, Vasily and {Miller}, Tim B. and {Kramarenko}, Ivan and {Setton}, David J. and {Ma}, Yilun and {Goulding}, Andy D. and {Bezanson}, Rachel and {Naidu}, Rohan P. and {Williams}, Christina C. and {Atek}, Hakim and {Brammer}, Gabriel and {Cutler}, Sam E. and {Chemerynska}, Iryna and {Cloonan}, Aidan P. and {Dayal}, Pratika and {de Graaff}, Anna and {Fudamoto}, Yoshinobu and {Fujimoto}, Seiji and {Furtak}, Lukas J. and {Glazebrook}, Karl and {Heintz}, Kasper E. and {Leja}, Joel and {Marchesini}, Danilo and {Nanayakkara}, Themiya and {Nelson}, Erica J. and {Oesch}, Pascal A. and {Pan}, Richard and {Price}, Sedona H. and {Shivaei}, Irene and {Sobral}, David and {Suess}, Katherine A. and {van Dokkum}, Pieter and {Wang}, Bingjie and {Weaver}, John R. and {Whitaker}, Katherine E. and {Zitrin}, Adi},
        title = "{An unambiguous AGN and a Balmer break in an Ultraluminous Little Red Dot at z=4.47 from Ultradeep UNCOVER and All the Little Things Spectroscopy}",
      journal = {arXiv e-prints},
         year = 2024,
        month = dec,
          eid = {arXiv:2412.04557},
        pages = {arXiv:2412.04557},
          doi = {10.48550/arXiv.2412.04557},
archivePrefix = {arXiv},
       eprint = {2412.04557},
 primaryClass = {astro-ph.GA},
       adsurl = {https://ui.adsabs.harvard.edu/abs/2024arXiv241204557L}
}

@ARTICLE{CoveloPaz25,
       author = {{Covelo-Paz}, Alba and {Giovinazzo}, Emma and {Oesch}, Pascal A. and {Meyer}, Romain A. and {Weibel}, Andrea and {Brammer}, Gabriel and {Fudamoto}, Yoshinobu and {Kerutt}, Josephine and {Lin}, Jamie and {Matharu}, Jasleen and {Naidu}, Rohan P. and {Velichko}, Anna and {Bollo}, Victoria and {Bouwens}, Rychard and {Chisholm}, John and {Illingworth}, Garth D. and {Kramarenko}, Ivan and {Magee}, Daniel and {Maseda}, Michael and {Matthee}, Jorryt and {Nelson}, Erica and {Reddy}, Naveen and {Schaerer}, Daniel and {Stefanon}, Mauro and {Xiao}, Mengyuan},
        title = "{An H{\ensuremath{\alpha}} view of galaxy buildup in the first 2 Gyr: Luminosity functions at z {\ensuremath{\sim}} 4‑6.5 from NIRCam/grism spectroscopy}",
      journal = {\aap},
         year = 2025,
        month = feb,
       volume = {694},
          eid = {A178},
        pages = {A178},
          doi = {10.1051/0004-6361/202452363},
archivePrefix = {arXiv},
       eprint = {2409.17241},
 primaryClass = {astro-ph.GA},
       adsurl = {https://ui.adsabs.harvard.edu/abs/2025A&A...694A.178C}
}

@ARTICLE{CoveloPaz26,
       author = {{Covelo-Paz}, Alba and {Meuwly}, Corentin and {Oesch}, Pascal A. and {Witten}, Callum and {Weibel}, Andrea and {Carvajal-Bohorquez}, Cristian and {Ciesla}, Laure and {Giovinazzo}, Emma and {Brammer}, Gabriel},
        title = "{A systematic search for dormant galaxies at z {\ensuremath{\sim}} 5{\ensuremath{-}}7 from the JWST NIRSpec archive}",
      journal = {\aap},
         year = 2026,
        month = jan,
       volume = {705},
          eid = {A155},
        pages = {A155},
          doi = {10.1051/0004-6361/202556119},
archivePrefix = {arXiv},
       eprint = {2506.22540},
 primaryClass = {astro-ph.GA},
       adsurl = {https://ui.adsabs.harvard.edu/abs/2026A&A...705A.155C}
}

@ARTICLE{Naidu24,
       author = {{Naidu}, Rohan P. and {Matthee}, Jorryt and {Kramarenko}, Ivan and {Weibel}, Andrea and {Brammer}, Gabriel and {Oesch}, Pascal A. and {Lechner}, Peter and {Furtak}, Lukas J. and {Di Cesare}, Claudia and {Torralba}, Alberto and {Kotiwale}, Gauri and {Bezanson}, Rachel and {Bouwens}, Rychard J. and {Chandra}, Vedant and {Claeyssens}, Ad{\'e}la{\"\i}de and {Danhaive}, A. Lola and {Frebel}, Anna and {de Graaff}, Anna and {Greene}, Jenny E. and {Heintz}, Kasper E. and {Ji}, Alexander P. and {Kashino}, Daichi and {Katz}, Harley and {Labbe}, Ivo and {Leja}, Joel and {Li}, Yijia and {Maseda}, Michael V. and {Richard}, Johan and {Shivaei}, Irene and {Simcoe}, Robert A. and {Sobral}, David and {Suess}, Katherine A. and {Tacchella}, Sandro and {Williams}, Christina C.},
        title = "{All the Little Things in Abell 2744: $>$1000 Gravitationally Lensed Dwarf Galaxies at $z=0-9$ from JWST NIRCam Grism Spectroscopy}",
      journal = {arXiv e-prints},
         year = 2024,
        month = oct,
          eid = {arXiv:2410.01874},
        pages = {arXiv:2410.01874},
          doi = {10.48550/arXiv.2410.01874},
archivePrefix = {arXiv},
       eprint = {2410.01874},
 primaryClass = {astro-ph.GA},
       adsurl = {https://ui.adsabs.harvard.edu/abs/2024arXiv241001874N}
}

@ARTICLE{Torralba26IFU,
       author = {{Torralba}, Alberto and {Matthee}, Jorryt and {Pezzulli}, Gabriele and {Naidu}, Rohan P. and {Ishikawa}, Yuzo and {Brammer}, Gabriel B. and {Chang}, Seok-Jun and {Chisholm}, John and {de Graaff}, Anna and {D'Eugenio}, Francesco and {Di Cesare}, Claudia and {Eilers}, Anna-Christina and {Greene}, Jenny E. and {Gronke}, Max and {Iani}, Edoardo and {Kokorev}, Vasily and {Kotiwale}, Gauri and {Kramarenko}, Ivan and {Ma}, Yilun and {Mascia}, Sara and {Navarrete}, Benjam{\'\i}n and {Nelson}, Erica and {Oesch}, Pascal and {Simcoe}, Robert A. and {Wuyts}, Stijn},
        title = "{The warm outer layer of a little red dot as the source of [Fe II] and collisional Balmer lines with scattering wings}",
      journal = {\aap},
         year = 2026,
        month = feb,
       volume = {707},
          eid = {A75},
        pages = {A75},
          doi = {10.1051/0004-6361/202557537},
archivePrefix = {arXiv},
       eprint = {2510.00103},
 primaryClass = {astro-ph.GA},
       adsurl = {https://ui.adsabs.harvard.edu/abs/2026A&A...707A..75T}
}

@ARTICLE{Naidu25BHstar,
       author = {{Naidu}, Rohan P. and {Matthee}, Jorryt and {Katz}, Harley and {de Graaff}, Anna and {Oesch}, Pascal and {Smith}, Aaron and {Greene}, Jenny E. and {Brammer}, Gabriel and {Weibel}, Andrea and {Hviding}, Raphael and {Chisholm}, John and {Labb\textbackslash'e}, Ivo and {Simcoe}, Robert A. and {Witten}, Callum and {Atek}, Hakim and {Baggen}, Josephine F.~W. and {Belli}, Sirio and {Bezanson}, Rachel and {Boogaard}, Leindert A. and {Bose}, Sownak and {Covelo-Paz}, Alba and {Dayal}, Pratika and {Fudamoto}, Yoshinobu and {Furtak}, Lukas J. and {Giovinazzo}, Emma and {Goulding}, Andy and {Gronke}, Max and {Heintz}, Kasper E. and {Hirschmann}, Michaela and {Illingworth}, Garth and {Inoue}, Akio K. and {Johnson}, Benjamin D. and {Leja}, Joel and {Leonova}, Ecaterina and {McConachie}, Ian and {Maseda}, Michael V. and {Natarajan}, Priyamvada and {Nelson}, Erica and {Setton}, David J. and {Shivaei}, Irene and {Sobral}, David and {Stefanon}, Mauro and {Tacchella}, Sandro and {Toft}, Sune and {Torralba}, Alberto and {van Dokkum}, Pieter and {van der Wel}, Arjen and {Volonteri}, Marta and {Walter}, Fabian and {Wang}, Bingjie and {Watson}, Darach},
        title = "{A ``Black Hole Star'' Reveals the Remarkable Gas-Enshrouded Hearts of the Little Red Dots}",
      journal = {arXiv e-prints},
         year = 2025,
        month = mar,
          eid = {arXiv:2503.16596},
        pages = {arXiv:2503.16596},
          doi = {10.48550/arXiv.2503.16596},
archivePrefix = {arXiv},
       eprint = {2503.16596},
 primaryClass = {astro-ph.GA},
       adsurl = {https://ui.adsabs.harvard.edu/abs/2025arXiv250316596N}
}

@ARTICLE{Donnan25CAPERS,
       author = {{Donnan}, Callum T. and {Dickinson}, Mark and {Taylor}, Anthony J. and {Arrabal Haro}, Pablo and {Finkelstein}, Steven L. and {Stanton}, Thomas M. and {Jung}, Intae and {Papovich}, Casey and {Akins}, Hollis B. and {Koekemoer}, Anton M. and {McLeod}, Derek J. and {Napolitano}, Lorenzo and {Amor{\'\i}n}, Ricardo O. and {Begley}, Ryan and {Burgarella}, Denis and {Carnall}, Adam C. and {Casey}, Caitlin M. and {Calabr{\`o}}, Antonello and {Cullen}, Fergus and {Dunlop}, James S. and {Ellis}, Richard S. and {Fern{\'a}ndez}, Vital and {Giavalisco}, Mauro and {Hirschmann}, Michaela and {Hu}, Weida and {Illingworth}, Garth and {Kartaltepe}, Jeyhan S. and {Kocevski}, Dale D. and {Kokorev}, Vasily and {Leung}, Ho-Hin and {Lucas}, Ray A. and {Morales}, Alexa M. and {McLure}, Ross and {Pentericci}, Laura and {P{\'e}rez-Gonz{\'a}lez}, Pablo G. and {Somerville}, Rachel S. and {Stevenson}, Struan and {Trump}, Jonathan R. and {Yung}, L.~Y. Aaron and {Zavala}, Jorge A.},
        title = "{Very Bright, Very Blue, and Very Red: JWST CAPERS Analysis of Highly Luminous Galaxies with Extreme Ultraviolet Slopes at z = 10}",
      journal = {\apj},
         year = 2025,
        month = nov,
       volume = {993},
       number = {2},
          eid = {224},
        pages = {224},
          doi = {10.3847/1538-4357/ae0a1f},
archivePrefix = {arXiv},
       eprint = {2507.10518},
 primaryClass = {astro-ph.GA},
       adsurl = {https://ui.adsabs.harvard.edu/abs/2025ApJ...993..224D}
}

@ARTICLE{Planck20,
       author = {{Planck Collaboration} and {Aghanim}, N. and {Akrami}, Y. and {Ashdown}, M. and {Aumont}, J. and {Baccigalupi}, C. and {Ballardini}, M. and {Banday}, A.~J. and {Barreiro}, R.~B. and {Bartolo}, N. and {Basak}, S. and {Battye}, R. and {Benabed}, K. and {Bernard}, J.-P. and {Bersanelli}, M. and {Bielewicz}, P. and {Bock}, J.~J. and {Bond}, J.~R. and {Borrill}, J. and {Bouchet}, F.~R. and {Boulanger}, F. and {Bucher}, M. and {Burigana}, C. and {Butler}, R.~C. and {Calabrese}, E. and {Cardoso}, J.-F. and {Carron}, J. and {Challinor}, A. and {Chiang}, H.~C. and {Chluba}, J. and {Colombo}, L.~P.~L. and {Combet}, C. and {Contreras}, D. and {Crill}, B.~P. and {Cuttaia}, F. and {de Bernardis}, P. and {de Zotti}, G. and {Delabrouille}, J. and {Delouis}, J.-M. and {Di Valentino}, E. and {Diego}, J.~M. and {Dor{\'e}}, O. and {Douspis}, M. and {Ducout}, A. and {Dupac}, X. and {Dusini}, S. and {Efstathiou}, G. and {Elsner}, F. and {En{\ss}lin}, T.~A. and {Eriksen}, H.~K. and {Fantaye}, Y. and {Farhang}, M. and {Fergusson}, J. and {Fernandez-Cobos}, R. and {Finelli}, F. and {Forastieri}, F. and {Frailis}, M. and {Fraisse}, A.~A. and {Franceschi}, E. and {Frolov}, A. and {Galeotta}, S. and {Galli}, S. and {Ganga}, K. and {G{\'e}nova-Santos}, R.~T. and {Gerbino}, M. and {Ghosh}, T. and {Gonz{\'a}lez-Nuevo}, J. and {G{\'o}rski}, K.~M. and {Gratton}, S. and {Gruppuso}, A. and {Gudmundsson}, J.~E. and {Hamann}, J. and {Handley}, W. and {Hansen}, F.~K. and {Herranz}, D. and {Hildebrandt}, S.~R. and {Hivon}, E. and {Huang}, Z. and {Jaffe}, A.~H. and {Jones}, W.~C. and {Karakci}, A. and {Keih{\"a}nen}, E. and {Keskitalo}, R. and {Kiiveri}, K. and {Kim}, J. and {Kisner}, T.~S. and {Knox}, L. and {Krachmalnicoff}, N. and {Kunz}, M. and {Kurki-Suonio}, H. and {Lagache}, G. and {Lamarre}, J.-M. and {Lasenby}, A. and {Lattanzi}, M. and {Lawrence}, C.~R. and {Le Jeune}, M. and {Lemos}, P. and {Lesgourgues}, J. and {Levrier}, F. and {Lewis}, A. and {Liguori}, M. and {Lilje}, P.~B. and {Lilley}, M. and {Lindholm}, V. and {L{\'o}pez-Caniego}, M. and {Lubin}, P.~M. and {Ma}, Y.-Z. and {Mac{\'\i}as-P{\'e}rez}, J.~F. and {Maggio}, G. and {Maino}, D. and {Mandolesi}, N. and {Mangilli}, A. and {Marcos-Caballero}, A. and {Maris}, M. and {Martin}, P.~G. and {Martinelli}, M. and {Mart{\'\i}nez-Gonz{\'a}lez}, E. and {Matarrese}, S. and {Mauri}, N. and {McEwen}, J.~D. and {Meinhold}, P.~R. and {Melchiorri}, A. and {Mennella}, A. and {Migliaccio}, M. and {Millea}, M. and {Mitra}, S. and {Miville-Desch{\^e}nes}, M.-A. and {Molinari}, D. and {Montier}, L. and {Morgante}, G. and {Moss}, A. and {Natoli}, P. and {N{\o}rgaard-Nielsen}, H.~U. and {Pagano}, L. and {Paoletti}, D. and {Partridge}, B. and {Patanchon}, G. and {Peiris}, H.~V. and {Perrotta}, F. and {Pettorino}, V. and {Piacentini}, F. and {Polastri}, L. and {Polenta}, G. and {Puget}, J.-L. and {Rachen}, J.~P. and {Reinecke}, M. and {Remazeilles}, M. and {Renzi}, A. and {Rocha}, G. and {Rosset}, C. and {Roudier}, G. and {Rubi{\~n}o-Mart{\'\i}n}, J.~A. and {Ruiz-Granados}, B. and {Salvati}, L. and {Sandri}, M. and {Savelainen}, M. and {Scott}, D. and {Shellard}, E.~P.~S. and {Sirignano}, C. and {Sirri}, G. and {Spencer}, L.~D. and {Sunyaev}, R. and {Suur-Uski}, A.-S. and {Tauber}, J.~A. and {Tavagnacco}, D. and {Tenti}, M. and {Toffolatti}, L. and {Tomasi}, M. and {Trombetti}, T. and {Valenziano}, L. and {Valiviita}, J. and {Van Tent}, B. and {Vibert}, L. and {Vielva}, P. and {Villa}, F. and {Vittorio}, N. and {Wandelt}, B.~D. and {Wehus}, I.~K. and {White}, M. and {White}, S.~D.~M. and {Zacchei}, A. and {Zonca}, A.},
        title = "{Planck 2018 results. VI. Cosmological parameters}",
      journal = {\aap},
         year = 2020,
        month = sep,
       volume = {641},
          eid = {A6},
        pages = {A6},
          doi = {10.1051/0004-6361/201833910},
archivePrefix = {arXiv},
       eprint = {1807.06209},
 primaryClass = {astro-ph.CO},
       adsurl = {https://ui.adsabs.harvard.edu/abs/2020A&A...641A...6P}
}

@ARTICLE{Nikopoulos25,
       author = {{Nikopoulos}, G.~P. and {Watson}, D. and {Sneppen}, A. and {Rusakov}, V. and {Heintz}, K.~E. and {Witstok}, J. and {Brammer}, G.},
        title = "{Evidence of violation of Case B recombination in Little Red Dots}",
      journal = {arXiv e-prints},
         year = 2025,
        month = oct,
          eid = {arXiv:2510.06362},
        pages = {arXiv:2510.06362},
          doi = {10.48550/arXiv.2510.06362},
archivePrefix = {arXiv},
       eprint = {2510.06362},
 primaryClass = {astro-ph.GA},
       adsurl = {https://ui.adsabs.harvard.edu/abs/2025arXiv251006362N}
}

@ARTICLE{Williams24,
       author = {{Williams}, Christina C. and {Alberts}, Stacey and {Ji}, Zhiyuan and {Hainline}, Kevin N. and {Lyu}, Jianwei and {Rieke}, George and {Endsley}, Ryan and {Suess}, Katherine A. and {Sun}, Fengwu and {Johnson}, Benjamin D. and {Florian}, Michael and {Shivaei}, Irene and {Rujopakarn}, Wiphu and {Baker}, William M. and {Bhatawdekar}, Rachana and {Boyett}, Kristan and {Bunker}, Andrew J. and {Cameron}, Alex J. and {Carniani}, Stefano and {Charlot}, Stephane and {Curtis-Lake}, Emma and {DeCoursey}, Christa and {de Graaff}, Anna and {Egami}, Eiichi and {Eisenstein}, Daniel J. and {Gibson}, Justus L. and {Hausen}, Ryan and {Helton}, Jakob M. and {Maiolino}, Roberto and {Maseda}, Michael V. and {Nelson}, Erica J. and {P{\'e}rez-Gonz{\'a}lez}, Pablo G. and {Rieke}, Marcia J. and {Robertson}, Brant E. and {Saxena}, Aayush and {Tacchella}, Sandro and {Willmer}, Christopher N.~A. and {Willott}, Chris J.},
        title = "{The Galaxies Missed by Hubble and ALMA: The Contribution of Extremely Red Galaxies to the Cosmic Census at 3 < z < 8}",
      journal = {\apj},
         year = 2024,
        month = jun,
       volume = {968},
       number = {1},
          eid = {34},
        pages = {34},
          doi = {10.3847/1538-4357/ad3f17},
archivePrefix = {arXiv},
       eprint = {2311.07483},
 primaryClass = {astro-ph.GA},
       adsurl = {https://ui.adsabs.harvard.edu/abs/2024ApJ...968...34W}
}

@ARTICLE{Williams25,
       author = {{Williams}, Christina C. and {Oesch}, Pascal A. and {Weibel}, Andrea and {Brammer}, Gabriel and {Cloonan}, Aidan P. and {Whitaker}, Katherine E. and {Barrufet}, Laia and {Bezanson}, Rachel and {Bowler}, Rebecca A.~A. and {Dayal}, Pratika and {Franx}, Marijn and {Greene}, Jenny E. and {Hutter}, Anne and {Ji}, Zhiyuan and {Labb{\'e}}, Ivo and {Manning}, Sinclaire M. and {Maseda}, Michael V. and {Xiao}, Mengyuan},
        title = "{The PANORAMIC Survey: Pure Parallel Wide Area Legacy Imaging with JWST/NIRCam}",
      journal = {\apj},
         year = 2025,
        month = feb,
       volume = {979},
       number = {2},
          eid = {140},
        pages = {140},
          doi = {10.3847/1538-4357/ad97bc},
archivePrefix = {arXiv},
       eprint = {2410.01875},
 primaryClass = {astro-ph.GA},
       adsurl = {https://ui.adsabs.harvard.edu/abs/2025ApJ...979..140W}
}

@ARTICLE{Secunda26,
       author = {{Secunda}, Amy and {Somerville}, Rachel S. and {Jiang}, Yan-Fei and {Greene}, Jenny E. and {Furtak}, Lukas J. and {Zitrin}, Adi},
        title = "{Do Little Red Dots Vary?}",
      journal = {\apj},
         year = 2026,
        month = jan,
       volume = {996},
       number = {1},
          eid = {6},
        pages = {6},
          doi = {10.3847/1538-4357/ae1f08},
archivePrefix = {arXiv},
       eprint = {2509.03571},
 primaryClass = {astro-ph.GA},
       adsurl = {https://ui.adsabs.harvard.edu/abs/2026ApJ...996....6S}
}

@ARTICLE{Ito25,
       author = {{Ito}, K. and {Valentino}, F. and {Brammer}, G. and {Hamadouche}, M.~L. and {Whitaker}, K.~E. and {Kokorev}, V. and {Zhu}, P. and {Kakimoto}, T. and {Wu}, P.-F. and {Antwi-Danso}, J. and {Baker}, W.~M. and {Ceverino}, D. and {Faisst}, A.~L. and {Farcy}, M. and {Fujimoto}, S. and {Gallazzi}, A. and {Gillman}, S. and {Gottumukkala}, R. and {Heintz}, K.~E. and {Hirschmann}, M. and {Jespersen}, C.~K. and {Kubo}, M. and {Lee}, M. and {Magdis}, G. and {Onodera}, M. and {Shimakawa}, R. and {Tanaka}, M. and {Toft}, S. and {Weaver}, J. R},
        title = "{DeepDive: A deep dive into the physics of the first massive quiescent galaxies in the Universe}",
      journal = {arXiv e-prints},
         year = 2025,
        month = jun,
          eid = {arXiv:2506.22642},
        pages = {arXiv:2506.22642},
          doi = {10.48550/arXiv.2506.22642},
archivePrefix = {arXiv},
       eprint = {2506.22642},
 primaryClass = {astro-ph.GA},
       adsurl = {https://ui.adsabs.harvard.edu/abs/2025arXiv250622642I}
}

@ARTICLE{PerezGonzalez24,
       author = {{P{\'e}rez-Gonz{\'a}lez}, Pablo G. and {Barro}, Guillermo and {Rieke}, George H. and {Lyu}, Jianwei and {Rieke}, Marcia and {Alberts}, Stacey and {Williams}, Christina C. and {Hainline}, Kevin and {Sun}, Fengwu and {Pusk{\'a}s}, D{\'a}vid and {Annunziatella}, Marianna and {Baker}, William M. and {Bunker}, Andrew J. and {Egami}, Eiichi and {Ji}, Zhiyuan and {Johnson}, Benjamin D. and {Robertson}, Brant and {Rodr{\'\i}guez Del Pino}, Bruno and {Rujopakarn}, Wiphu and {Shivaei}, Irene and {Tacchella}, Sandro and {Willmer}, Christopher N.~A. and {Willott}, Chris},
        title = "{What Is the Nature of Little Red Dots and what Is Not, MIRI SMILES Edition}",
      journal = {\apj},
         year = 2024,
        month = jun,
       volume = {968},
       number = {1},
          eid = {4},
        pages = {4},
          doi = {10.3847/1538-4357/ad38bb},
archivePrefix = {arXiv},
       eprint = {2401.08782},
 primaryClass = {astro-ph.GA},
       adsurl = {https://ui.adsabs.harvard.edu/abs/2024ApJ...968....4P}
}

@ARTICLE{KormendyHo13,
       author = {{Kormendy}, John and {Ho}, Luis C.},
        title = "{Coevolution (Or Not) of Supermassive Black Holes and Host Galaxies}",
      journal = {\araa},
         year = 2013,
        month = aug,
       volume = {51},
       number = {1},
        pages = {511-653},
          doi = {10.1146/annurev-astro-082708-101811},
archivePrefix = {arXiv},
       eprint = {1304.7762},
 primaryClass = {astro-ph.CO},
       adsurl = {https://ui.adsabs.harvard.edu/abs/2013ARA&A..51..511K}
}

@ARTICLE{Barro24,
       author = {{Barro}, Guillermo and {P{\'e}rez-Gonz{\'a}lez}, Pablo G. and {Kocevski}, Dale D. and {McGrath}, Elizabeth J. and {Trump}, Jonathan R. and {Simons}, Raymond C. and {Somerville}, Rachel S. and {Yung}, L.~Y. Aaron and {Arrabal Haro}, Pablo and {Akins}, Hollis B. and {Bagley}, Michaela B. and {Cleri}, Nikko J. and {Costantin}, Luca and {Davis}, Kelcey and {Dickinson}, Mark and {Finkelstein}, Steve L. and {Giavalisco}, Mauro and {G{\'o}mez-Guijarro}, Carlos and {Hathi}, Nimish P. and {Hirschmann}, Michaela and {Holwerda}, Benne W. and {Huertas-Company}, Marc and {Kartaltepe}, Jeyhan S. and {Koekemoer}, Anton M. and {Lucas}, Ray A. and {Papovich}, Casey and {Pirzkal}, Nor and {Seill{\'e}}, Lise-Marie and {Tacchella}, Sandro and {Wuyts}, Stijn and {Wilkins}, Stephen M. and {de la Vega}, Alexander and {Yang}, Guang and {Zavala}, Jorge A.},
        title = "{Extremely Red Galaxies at z = 5{\textendash}9 with MIRI and NIRSpec: Dusty Galaxies or Obscured Active Galactic Nuclei?}",
      journal = {\apj},
         year = 2024,
        month = mar,
       volume = {963},
       number = {2},
          eid = {128},
        pages = {128},
          doi = {10.3847/1538-4357/ad167e},
archivePrefix = {arXiv},
       eprint = {2305.14418},
 primaryClass = {astro-ph.GA},
       adsurl = {https://ui.adsabs.harvard.edu/abs/2024ApJ...963..128B}
}

@ARTICLE{Barro26,
       author = {{Barro}, Guillermo and {P{\'e}rez-Gonz{\'a}lez}, Pablo G. and {Kocevski}, Dale and {Trump}, Jonathan R. and {Dickinson}, Mark and {Arrabal Haro}, Pablo and {Brooks}, Madisyn and {Donnan}, Callum T. and {Dunlop}, James S. and {Finkelstein}, Steven L. and {Franco}, Maximilien and {Gandolfi}, Giovanni and {Giavalisco}, Mauro and {Grogin}, Norman A. and {Hirschmann}, Michaela and {Kartaltepe}, Jeyhan S. and {Koekemoer}, Anton M. and {Larson}, Rebecca L. and {Leung}, Gene C.~K. and {Lucas}, Ray A. and {McGrath}, Elizabeth J. and {Papovich}, Casey and {P{\'e}rez-D{\'\i}az}, Borja and {Somerville}, Rachel S. and {Taylor}, Elizabeth and {Taylor}, Anthony J. and {Tripodi}, Roberta and {Yung}, L.~Y. Aaron and {Wang}, Xin},
        title = "{From ``The Cliff'' to ``Virgil'': Mapping the Spectral Diversity of Little Red Dots with JWST/NIRSpec}",
      journal = {\apj},
         year = 2026,
        month = may,
       volume = {1003},
       number = {1},
          eid = {96},
        pages = {96},
          doi = {10.3847/1538-4357/ae5d2c},
archivePrefix = {arXiv},
       eprint = {2512.15853},
 primaryClass = {astro-ph.GA},
       adsurl = {https://ui.adsabs.harvard.edu/abs/2026ApJ..1003...96B}
}

@ARTICLE{Merida26,
       author = {{M{\'e}rida}, Rosa M. and {Sawicki}, Marcin and {Gaspar}, Gaia and {Willott}, Chris J. and {Iyer}, Kartheik G.},
        title = "{Testing the BH$^*$ Model: a UV-to-Optical Spectral Fitting of The Cliff}",
      journal = {arXiv e-prints},
         year = 2026,
        month = may,
          eid = {arXiv:2605.07976},
        pages = {arXiv:2605.07976},
          doi = {10.48550/arXiv.2605.07976},
archivePrefix = {arXiv},
       eprint = {2605.07976},
 primaryClass = {astro-ph.GA},
       adsurl = {https://ui.adsabs.harvard.edu/abs/2026arXiv260507976M}
}

@ARTICLE{Pan26,
       author = {{Pan}, Zhiwei and {Zhuang}, Ming-Yang and {Shen}, Yue and {Wang}, Feige and {Greene}, Jenny E. and {Burgasser}, Adam J. and {Li}, Junyao and {Stone}, Zachary and {Venkatraman}, Padmavathi},
        title = "{NEXUS: Abundance, Environments, and Spectral Diversity of Little Red Dots from the NIRSpec MSA Sample}",
      journal = {arXiv e-prints},
         year = 2026,
        month = jun,
          eid = {arXiv:2606.09721},
        pages = {arXiv:2606.09721},
archivePrefix = {arXiv},
       eprint = {2606.09721},
 primaryClass = {astro-ph.GA},
       adsurl = {https://ui.adsabs.harvard.edu/abs/2026arXiv260609721P}
}

@ARTICLE{Furtak23,
       author = {{Furtak}, Lukas J. and {Zitrin}, Adi and {Weaver}, John R. and {Atek}, Hakim and {Bezanson}, Rachel and {Labb{\'e}}, Ivo and {Whitaker}, Katherine E. and {Leja}, Joel and {Price}, Sedona H. and {Brammer}, Gabriel B. and {Wang}, Bingjie and {Marchesini}, Danilo and {Pan}, Richard and {Dayal}, Pratika and {van Dokkum}, Pieter and {Feldmann}, Robert and {Fujimoto}, Seiji and {Franx}, Marijn and {Khullar}, Gourav and {Nelson}, Erica J. and {Mowla}, Lamiya A.},
        title = "{UNCOVERing the extended strong lensing structures of Abell 2744 with the deepest JWST imaging}",
      journal = {\mnras},
         year = 2023,
        month = aug,
       volume = {523},
       number = {3},
        pages = {4568-4582},
          doi = {10.1093/mnras/stad1627},
archivePrefix = {arXiv},
       eprint = {2212.04381},
 primaryClass = {astro-ph.GA},
       adsurl = {https://ui.adsabs.harvard.edu/abs/2023MNRAS.523.4568F}
}

@ARTICLE{Oesch23,
       author = {{Oesch}, P.~A. and {Brammer}, G. and {Naidu}, R.~P. and {Bouwens}, R.~J. and {Chisholm}, J. and {Illingworth}, G.~D. and {Matthee}, J. and {Nelson}, E. and {Qin}, Y. and {Reddy}, N. and {Shapley}, A. and {Shivaei}, I. and {van Dokkum}, P. and {Weibel}, A. and {Whitaker}, K. and {Wuyts}, S. and {Covelo-Paz}, A. and {Endsley}, R. and {Fudamoto}, Y. and {Giovinazzo}, E. and {Herard-Demanche}, T. and {Kerutt}, J. and {Kramarenko}, I. and {Labbe}, I. and {Leonova}, E. and {Lin}, J. and {Magee}, D. and {Marchesini}, D. and {Maseda}, M. and {Mason}, C. and {Matharu}, J. and {Meyer}, R.~A. and {Neufeld}, C. and {Prieto Lyon}, G. and {Schaerer}, D. and {Sharma}, R. and {Shuntov}, M. and {Smit}, R. and {Stefanon}, M. and {Wyithe}, J.~S.~B. and {Xiao}, M.},
        title = "{The JWST FRESCO survey: legacy NIRCam/grism spectroscopy and imaging in the two GOODS fields}",
      journal = {\mnras},
         year = 2023,
        month = oct,
       volume = {525},
       number = {2},
        pages = {2864-2874},
          doi = {10.1093/mnras/stad2411},
archivePrefix = {arXiv},
       eprint = {2304.02026},
 primaryClass = {astro-ph.GA},
       adsurl = {https://ui.adsabs.harvard.edu/abs/2023MNRAS.525.2864O}
}

@ARTICLE{Kashino23,
       author = {{Kashino}, Daichi and {Lilly}, Simon J. and {Matthee}, Jorryt and {Eilers}, Anna-Christina and {Mackenzie}, Ruari and {Bordoloi}, Rongmon and {Simcoe}, Robert A.},
        title = "{EIGER. I. A Large Sample of [O III]-emitting Galaxies at 5.3 < z < 6.9 and Direct Evidence for Local Reionization by Galaxies}",
      journal = {\apj},
         year = 2023,
        month = jun,
       volume = {950},
       number = {1},
          eid = {66},
        pages = {66},
          doi = {10.3847/1538-4357/acc588},
archivePrefix = {arXiv},
       eprint = {2211.08254},
 primaryClass = {astro-ph.GA},
       adsurl = {https://ui.adsabs.harvard.edu/abs/2023ApJ...950...66K}
}

@ARTICLE{Madau26,
       author = {{Madau}, Piero},
        title = "{Chasing the light: Shadowing, collimation, and the super-Eddington growth of infant black holes in JWST broad-line AGNs}",
      journal = {\aap},
         year = 2026,
        month = apr,
       volume = {708},
          eid = {A116},
        pages = {A116},
          doi = {10.1051/0004-6361/202659244},
archivePrefix = {arXiv},
       eprint = {2501.09854},
 primaryClass = {astro-ph.HE},
       adsurl = {https://ui.adsabs.harvard.edu/abs/2026A&A...708A.116M}
}

@ARTICLE{MadauMaiolino26,
       author = {{Madau}, Piero and {Maiolino}, Roberto},
        title = "{Little Red Dots as Obscured Little Blue Dots: A Super-Eddington Unification Model}",
      journal = {arXiv e-prints},
         year = 2026,
        month = feb,
          eid = {arXiv:2602.22386},
        pages = {arXiv:2602.22386},
          doi = {10.48550/arXiv.2602.22386},
archivePrefix = {arXiv},
       eprint = {2602.22386},
 primaryClass = {astro-ph.GA},
       adsurl = {https://ui.adsabs.harvard.edu/abs/2026arXiv260222386M}
}

@ARTICLE{MadauMaiolino26LBDs,
       author = {{Madau}, Piero and {Maiolino}, Roberto},
        title = "{Little red dots as obscured little blue dots: relative abundances, luminosities, and black-hole masses}",
      journal = {arXiv e-prints},
         year = 2026,
        month = may,
          eid = {arXiv:2605.05074},
        pages = {arXiv:2605.05074},
          doi = {10.48550/arXiv.2605.05074},
archivePrefix = {arXiv},
       eprint = {2605.05074},
 primaryClass = {astro-ph.GA},
       adsurl = {https://ui.adsabs.harvard.edu/abs/2026arXiv260505074M}
}

@ARTICLE{DeLucia25,
       author = {{De Lucia}, Gabriella and {Fontanot}, Fabio and {Hirschmann}, Michaela and {Xie}, Lizhi},
        title = "{Cosmic quenching}",
      journal = {arXiv e-prints},
         year = 2025,
        month = feb,
          eid = {arXiv:2502.01724},
        pages = {arXiv:2502.01724},
          doi = {10.48550/arXiv.2502.01724},
archivePrefix = {arXiv},
       eprint = {2502.01724},
 primaryClass = {astro-ph.GA},
       adsurl = {https://ui.adsabs.harvard.edu/abs/2025arXiv250201724D}
}

@ARTICLE{Weinberger18,
       author = {{Weinberger}, Rainer and {Springel}, Volker and {Pakmor}, R{\"u}diger and {Nelson}, Dylan and {Genel}, Shy and {Pillepich}, Annalisa and {Vogelsberger}, Mark and {Marinacci}, Federico and {Naiman}, Jill and {Torrey}, Paul and {Hernquist}, Lars},
        title = "{Supermassive black holes and their feedback effects in the IllustrisTNG simulation}",
      journal = {\mnras},
         year = 2018,
        month = sep,
       volume = {479},
       number = {3},
        pages = {4056-4072},
          doi = {10.1093/mnras/sty1733},
archivePrefix = {arXiv},
       eprint = {1710.04659},
 primaryClass = {astro-ph.GA},
       adsurl = {https://ui.adsabs.harvard.edu/abs/2018MNRAS.479.4056W}
}

@ARTICLE{Brazzini26,
       author = {{Brazzini}, M. and {D'Eugenio}, F. and {Maiolino}, R. and {Lyu}, J. and {DeCoursey}, C. and {{\"U}bler}, H. and {Ji}, X. and {Juod{\v{z}}balis}, I. and {Scholtz}, J. and {Jones}, G.~C. and {Hainline}, K. and {Dalla Bont{\`a}}, E. and {{\'e}rez-Gonz{\'a}lez}, P.~G. P and {Geris}, S. and {Harshan}, A. and {Feruglio}, C. and {Bischetti}, M. and {Mazzolari}, G. and {Rieke}, G. and {Alberts}, S. and {Trefoloni}, B. and {Carniani}, S. and {Parlanti}, E. and {Marconi}, A. and {Risaliti}, G. and {Ramos Almeida}, C. and {Rinaldi}, P. and {Perna}, M. and {Zamora}, S. and {Lamperti}, I. and {Venturi}, G. and {Cresci}, G. and {Bunker}, Andrew J. and {Ivey}, L.~R.},
        title = "{The Little Blue and Red Dots Rosetta Stones: Non-Gaussian broad lines, hot dust, and X-ray weakness}",
      journal = {arXiv e-prints},
         year = 2026,
        month = jan,
          eid = {arXiv:2601.22214},
        pages = {arXiv:2601.22214},
          doi = {10.48550/arXiv.2601.22214},
archivePrefix = {arXiv},
       eprint = {2601.22214},
 primaryClass = {astro-ph.GA},
       adsurl = {https://ui.adsabs.harvard.edu/abs/2026arXiv260122214B}
}

@ARTICLE{Torralba26,
       author = {{Torralba}, Alberto and {Matthee}, Jorryt and {Weibel}, Andrea and {Naidu}, Rohan P. and {Ma}, Yilun and {Cloonan}, Aidan P. and {Desai}, Aayush and {de Graaff}, Anna and {Greene}, Jenny E. and {Jespersen}, Christian Kragh and {Kramarenko}, Ivan G. and {Mascia}, Sara and {Oesch}, Pascal A. and {Sun}, Wendy Q. and {Williams}, Christina C.},
        title = "{A Black Hole Star at Cosmic Noon: Extreme Balmer break, photospheric continuum, and broad absorption by thick winds in a Little Red Dot at z=1.7}",
      journal = {arXiv e-prints},
         year = 2026,
        month = mar,
          eid = {arXiv:2603.28335},
        pages = {arXiv:2603.28335},
archivePrefix = {arXiv},
       eprint = {2603.28335},
 primaryClass = {astro-ph.GA},
       adsurl = {https://ui.adsabs.harvard.edu/abs/2026arXiv260328335T}
}

@ARTICLE{Conroy2009b,
       author = {{Conroy}, Charlie and {Gunn}, James E. and {White}, Martin},
        title = "{The Propagation of Uncertainties in Stellar Population Synthesis Modeling. I. The Relevance of Uncertain Aspects of Stellar Evolution and the Initial Mass Function to the Derived Physical Properties of Galaxies}",
      journal = {\apj},
         year = 2009,
        month = jul,
       volume = {699},
       number = {1},
        pages = {486-506},
          doi = {10.1088/0004-637X/699/1/486},
archivePrefix = {arXiv},
       eprint = {0809.4261},
 primaryClass = {astro-ph},
       adsurl = {https://ui.adsabs.harvard.edu/abs/2009ApJ...699..486C}
}

@ARTICLE{Conroy2010,
       author = {{Conroy}, Charlie and {Gunn}, James E.},
        title = "{The Propagation of Uncertainties in Stellar Population Synthesis Modeling. III. Model Calibration, Comparison, and Evaluation}",
      journal = {\apj},
         year = 2010,
        month = apr,
       volume = {712},
       number = {2},
        pages = {833-857},
          doi = {10.1088/0004-637X/712/2/833},
archivePrefix = {arXiv},
       eprint = {0911.3151},
 primaryClass = {astro-ph.CO},
       adsurl = {https://ui.adsabs.harvard.edu/abs/2010ApJ...712..833C}
}

@ARTICLE{GrantRoberts25,
       author = {{Grant Roberts}, M. and {Braff}, Lila and {Garg}, Aarna and {Profumo}, Stefano and {Jeltema}, Tesla and {O'Donnell}, Jackson},
        title = "{Early formation of supermassive black holes from the collapse of strongly self-interacting dark matter}",
      journal = {\jcap},
         year = 2025,
        month = jan,
       volume = {2025},
       number = {1},
          eid = {060},
        pages = {060},
          doi = {10.1088/1475-7516/2025/01/060},
archivePrefix = {arXiv},
       eprint = {2410.17480},
 primaryClass = {astro-ph.GA},
       adsurl = {https://ui.adsabs.harvard.edu/abs/2025JCAP...01..060G}
}

@ARTICLE{Ananna24,
       author = {{Ananna}, Tonima Tasnim and {Bogd{\'a}n}, {\'A}kos and {Kov{\'a}cs}, Orsolya E. and {Natarajan}, Priyamvada and {Hickox}, Ryan C.},
        title = "{X-Ray View of Little Red Dots: Do They Host Supermassive Black Holes?}",
      journal = {\apjl},
         year = 2024,
        month = jul,
       volume = {969},
       number = {1},
          eid = {L18},
        pages = {L18},
          doi = {10.3847/2041-8213/ad5669},
archivePrefix = {arXiv},
       eprint = {2404.19010},
 primaryClass = {astro-ph.GA},
       adsurl = {https://ui.adsabs.harvard.edu/abs/2024ApJ...969L..18A}
}

@ARTICLE{Fei26,
       author = {{Fei}, Qinyue and {Fujimoto}, Seiji and {Brammer}, Gabriel and {Li}, Ruancun and {Ho}, Luis C. and {Bromm}, Volker and {{\'A}lvarez-M{\'a}rquez}, Javier and {Asada}, Yoshihisa and {Barro}, Guillermo and {Colina}, Luis and {Dayal}, Pratika and {Finkelstein}, Steven L. and {Fynbo}, Johan P.~U. and {Ginolfi}, Michele and {Inayoshi}, Kohei and {Kokorev}, Vasily and {Leung}, Gene C.~K. and {Matthee}, Jorryt and {Meyer}, Romain A. and {Naidu}, Rohan P. and {Onoue}, Masafusa and {P{\'e}rez-Gonz{\'a}lez}, Pablo G. and {Steinhardt}, Charles L. and {Valentino}, Francesco and {Walter}, Fabian and {Xiao}, Mengyuan and {Zhang}, Haowen},
        title = "{Direct pathway to the Early Supermassive Black Holes: A Red Super-Eddington Quasar in a Massive Starburst Host at $z=7.2$}",
      journal = {arXiv e-prints},
         year = 2026,
        month = feb,
          eid = {arXiv:2602.12325},
        pages = {arXiv:2602.12325},
          doi = {10.48550/arXiv.2602.12325},
archivePrefix = {arXiv},
       eprint = {2602.12325},
 primaryClass = {astro-ph.GA},
       adsurl = {https://ui.adsabs.harvard.edu/abs/2026arXiv260212325F}
}

@ARTICLE{Furtak23QSO1,
       author = {{Furtak}, Lukas J. and {Zitrin}, Adi and {Plat}, Ad{\`e}le and {Fujimoto}, Seiji and {Wang}, Bingjie and {Nelson}, Erica J. and {Labb{\'e}}, Ivo and {Bezanson}, Rachel and {Brammer}, Gabriel B. and {van Dokkum}, Pieter and {Endsley}, Ryan and {Glazebrook}, Karl and {Greene}, Jenny E. and {Leja}, Joel and {Price}, Sedona H. and {Smit}, Renske and {Stark}, Daniel P. and {Weaver}, John R. and {Whitaker}, Katherine E. and {Atek}, Hakim and {Chevallard}, Jacopo and {Curtis-Lake}, Emma and {Dayal}, Pratika and {Feltre}, Anna and {Franx}, Marijn and {Fudamoto}, Yoshinobu and {Marchesini}, Danilo and {Mowla}, Lamiya A. and {Pan}, Richard and {Suess}, Katherine A. and {Vidal-Garc{\'\i}a}, Alba and {Williams}, Christina C.},
        title = "{JWST UNCOVER: Extremely Red and Compact Object at z $_{phot}$ ≃ 7.6 Triply Imaged by A2744}",
      journal = {\apj},
         year = 2023,
        month = aug,
       volume = {952},
       number = {2},
          eid = {142},
        pages = {142},
          doi = {10.3847/1538-4357/acdc9d},
archivePrefix = {arXiv},
       eprint = {2212.10531},
 primaryClass = {astro-ph.GA},
       adsurl = {https://ui.adsabs.harvard.edu/abs/2023ApJ...952..142F}
}

@ARTICLE{Kokubo25,
       author = {{Kokubo}, Mitsuru and {Harikane}, Yuichi},
        title = "{Challenging the Active Galactic Nucleus Scenario for JWST/NIRSpec Little Red Dot and Non─Little Red Dot Broad H{\ensuremath{\alpha}} Emitters in Light of Nondetection of NIRCam Photometric Variability and X-Ray}",
      journal = {\apj},
         year = 2025,
        month = dec,
       volume = {995},
       number = {1},
          eid = {24},
        pages = {24},
          doi = {10.3847/1538-4357/ae119e},
archivePrefix = {arXiv},
       eprint = {2407.04777},
 primaryClass = {astro-ph.GA},
       adsurl = {https://ui.adsabs.harvard.edu/abs/2025ApJ...995...24K}
}

@ARTICLE{Ji25BT,
       author = {{Ji}, Xihan and {Maiolino}, Roberto and {{\"U}bler}, Hannah and {Scholtz}, Jan and {D'Eugenio}, Francesco and {Sun}, Fengwu and {Perna}, Michele and {Turner}, Hannah and {Carniani}, Stefano and {Arribas}, Santiago and {Bennett}, Jake S. and {Bunker}, Andrew and {Charlot}, St{\'e}phane and {Cresci}, Giovanni and {Curti}, Mirko and {Egami}, Eiichi and {Fabian}, Andy and {Inayoshi}, Kohei and {Isobe}, Yuki and {Jones}, Gareth and {Juod{\v{z}}balis}, Ignas and {Kumari}, Nimisha and {Lyu}, Jianwei and {Mazzolari}, Giovanni and {Parlanti}, Eleonora and {Robertson}, Brant and {Rodr{\'\i}guez Del Pino}, Bruno and {Schneider}, Raffaella and {Sijacki}, Debora and {Tacchella}, Sandro and {Trinca}, Alessandro and {Valiante}, Rosa and {Venturi}, Giacomo and {Volonteri}, Marta and {Willott}, Chris and {Witten}, Callum and {Witstok}, Joris},
        title = "{BlackTHUNDER ─ A non-stellar Balmer break in a black hole-dominated little red dot at z = 7.04}",
      journal = {\mnras},
         year = 2025,
        month = dec,
       volume = {544},
       number = {4},
        pages = {3900-3935},
          doi = {10.1093/mnras/staf1867},
       adsurl = {https://ui.adsabs.harvard.edu/abs/2025MNRAS.544.3900J}
}

@ARTICLE{Taylor25,
       author = {{Taylor}, Anthony J. and {Kokorev}, Vasily and {Kocevski}, Dale D. and {Akins}, Hollis B. and {Cullen}, Fergus and {Dickinson}, Mark and {Finkelstein}, Steven L. and {Arrabal Haro}, Pablo and {Bromm}, Volker and {Giavalisco}, Mauro and {Inayoshi}, Kohei and {Juneau}, St{\'e}phanie and {Leung}, Gene C.~K. and {P{\'e}rez-Gonz{\'a}lez}, Pablo G. and {Somerville}, Rachel S. and {Trump}, Jonathan R. and {Amor{\'\i}n}, Ricardo O. and {Barro}, Guillermo and {Burgarella}, Denis and {Brooks}, Madisyn and {Carnall}, Adam C. and {Casey}, Caitlin M. and {Cheng}, Yingjie and {Chisholm}, John and {Chworowsky}, Katherine and {Davis}, Kelcey and {Donnan}, Callum T. and {Dunlop}, James S. and {Ellis}, Richard S. and {Fern{\'a}ndez}, Vital and {Fujimoto}, Seiji and {Grogin}, Norman A. and {Gupta}, Ansh R. and {Hathi}, Nimish P. and {Jung}, Intae and {Hirschmann}, Michaela and {Kartaltepe}, Jeyhan S. and {Koekemoer}, Anton M. and {Larson}, Rebecca L. and {Leung}, Ho-Hin and {Llerena}, Mario and {Lucas}, Ray A. and {McLeod}, Derek J. and {McLure}, Ross and {Napolitano}, Lorenzo and {Papovich}, Casey and {Stanton}, Thomas M. and {Tripodi}, Roberta and {Wang}, Xin and {Wilkins}, Stephen M. and {Yung}, L.~Y. Aaron and {Zavala}, Jorge A.},
        title = "{CAPERS-LRD-z9: A Gas-enshrouded Little Red Dot Hosting a Broad-line Active Galactic Nucleus at z = 9.288}",
      journal = {\apjl},
         year = 2025,
        month = aug,
       volume = {989},
       number = {1},
          eid = {L7},
        pages = {L7},
          doi = {10.3847/2041-8213/ade789},
archivePrefix = {arXiv},
       eprint = {2505.04609},
 primaryClass = {astro-ph.GA},
       adsurl = {https://ui.adsabs.harvard.edu/abs/2025ApJ...989L...7T}
}

@ARTICLE{Billand26,
       author = {{Billand}, Jean-Baptiste and {Elbaz}, David and {Franco}, Maximilien and {Gentile}, Fabrizio and {Daddi}, Emanuele and {Giavalisco}, Mauro and {Kocevski}, Dale D. and {Lewis}, Joseph S.~W. and {Magnelli}, Benjamin and {Sangalli}, Valentina and {Tarrasse}, Maxime},
        title = "{Do little red dots really form a distinct class of astronomical objects?}",
      journal = {arXiv e-prints},
         year = 2026,
        month = apr,
          eid = {arXiv:2604.11677},
        pages = {arXiv:2604.11677},
          doi = {10.48550/arXiv.2604.11677},
archivePrefix = {arXiv},
       eprint = {2604.11677},
 primaryClass = {astro-ph.GA},
       adsurl = {https://ui.adsabs.harvard.edu/abs/2026arXiv260411677B}
}

@ARTICLE{Rinaldi26,
       author = {{Rinaldi}, Pierluigi and {Hainline}, Kevin and {D'Eugenio}, Francesco and {P{\'e}rez-Gonz{\'a}lez}, Pablo G. and {Eisenstein}, Daniel J. and {Willmer}, Christopher N.~A. and {Carreira}, Courtney and {Robertson}, Brant and {Johnson}, Benjamin D. and {Alberts}, Stacey and {Baker}, William M. and {Bunker}, Andrew J. and {Carniani}, Stefano and {Egami}, Eiichi and {Helton}, Jakob M. and {Ji}, Zhiyuan and {Juod{\v{z}}balis}, Ignas and {Lin}, Xiaojing and {Lyu}, Jianwei and {Ma}, Zheng and {Maiolino}, Roberto and {Parlanti}, Eleonora and {Scholtz}, Jan and {Sun}, Yang and {Tacchella}, Sandro and {Venturi}, Giacomo and {Williams}, Christina C. and {Willott}, Chris and {Witstok}, Joris and {Wu}, Zihao},
        title = "{The Way We Tally Becomes the Tale: the Impact of Selection Strategies on the Inferred Evolution of Little Red Dots Across Cosmic Time}",
      journal = {arXiv e-prints},
         year = 2026,
        month = apr,
          eid = {arXiv:2604.07138},
        pages = {arXiv:2604.07138},
          doi = {10.48550/arXiv.2604.07138},
archivePrefix = {arXiv},
       eprint = {2604.07138},
 primaryClass = {astro-ph.GA},
       adsurl = {https://ui.adsabs.harvard.edu/abs/2026arXiv260407138R}
}

@article{Kass95,
    author = "Kass, Robert E. and Raftery, Adrian E.",
    title = "{Bayes Factors}",
    doi = "10.1080/01621459.1995.10476572",
    journal = "J. Am. Statist. Assoc.",
    volume = "90",
    number = "430",
    pages = "773--795",
    year = "1995"
}
\bibliographystyle{apj}

\appendix

\section{List of JWST Programs Contributing NIRSpec Spectra}
\label{sec:appendix_list_of_programs}

As described in Section \ref{sec:sample_sel_spec}, we cross-match our catalog of BH*-dominated candidates with all public spectra available on the DJA as of April 29, 2026. Considering only the highest SNR spectrum of each source (PRISM if available, medium resolution grating otherwise), these come from the following programs: 19 PRISM spectra and one G395M spectrum are from RUBIES (GO-4233, PIs de Graaff \& Brammer, \citealt{degraaff25rubies}); CAPERS (GO-6368, PI Dickinson, e.g., \citealt{Donnan25CAPERS}) and Mirage or Miracle (MoM, GO-5224, PIs Oesch \& Naidu) each contribute seven PRISM spectra; another six are from UNCOVER (GO-2561, PI Labb\'e, \citealt{Bezanson24}); JADES \citep[e.g.,][]{Eisenstein26} through GTO-1180, GTO-1181 (PI Eisenstein), and GTO-1287 (PI Isaak), and GO-5545 (PI Barrufet) each contribute three; another two PRISM spectra are from GO-8060 (PI Egami); finally, one spectrum is associated with each of the programs GO-2565 (PI Glazebrook, \citealt{Nanayakkara25}), GO-4106 (PI Nelson), GO-5997 (PI Looser), and DIVER (GO-8018, PI Lin). The remaining three medium grating spectra (one in G395M and two in G235M) are from GO-3567 (PI Valentino, \citealt{Ito25}).

\section{Spectroscopic V-shape Classification}
\label{sec:appendix_vshape_classification}

To provide an estimate of the contamination rate of our photometric selection, we apply the spectroscopic V-shape classification from \citet{Hviding25} to the subset of 57 sources with grade 3 PRISM spectra. One spectrum is immediately identified as a BD (Section \ref{sec:overview_contaminants}), and removed from the sample, another 34 sources are part of the spectroscopic LRD sample of \citet{degraaff25pop} and considered robust LRDs. Therefore, we only apply the V-shape classification to the remaining 22 spectra. Specifically, we fit a power-law, $f_\lambda=a\cdot\lambda_{\rm rest}^\beta$, to both the photometry and the spectrum on either side of ${\rm H}_\infty$. The V-shape is then defined following Section 3.2.2 of \citet{Hviding25} as (1) $\beta_{\rm UV}<-0.2$ at the 2$\sigma$ level, and $a_{\rm UV}>0$ from either spectroscopy or photometry, (2) $\beta_{\rm opt}({\rm Spec.})>0$ at the 2$\sigma$ level and $a_{\rm opt}({\rm Spec.})>0$, and (3) $\beta_{\rm opt}-\beta_{\rm UV}>0.5$ with $\beta_{\rm UV}$ from spectroscopy if it satisfies (1) or from photometry otherwise. According to this definition, 9/22 sources display V-shaped continua. However, for none of the other 13 sources can a V-shaped SED be confidently ruled out. Most sources do show $\beta_{\rm UV}<-0.2$ and $\beta_{\rm opt}>0$ but lack the SNR to confirm these slopes at the $>2\sigma$ level. For example, the source GDN-59854 which is shown in Figure \ref{fig:gallery_spectra} shows a clear red optical slope ($\beta_{\rm opt}=1.39\pm0.09$), a tentatively broadened H$\alpha$ line and a clear break/turnover around the Balmer limit. However, the SNR in both the photometry and the spectrum is insufficient to confirm a blue UV-slope for this source (e.g., $\beta_{\rm UV}=-0.21\pm0.54$ from the spectrum). Further, the spectrum of the source COS-EAST-85239 is affected by the chip gap and only covers $\lambda_{\rm rest}\gtrsim0.7\mu{\rm m}$, which means that neither slope can be measured. However, this spectrum shows tentatively broadened Paschen lines, and the photometry clearly favors a BH*-dominated interpretation. We only find one likely contaminant, UDS-27591, whose spectrum shows blue UV- and optical slopes ($\beta_{\rm UV}=-3.72\pm0.60$ and $\beta_{\rm opt}=-2.99\pm2.04$), albeit with large uncertainties. The spectrum shows strong lines that boost the NIRCam/LW photometry which has likely caused \texttt{eazy} to wrongly prefer a fit with a BH* template. Resulting estimates of the contamination rate of our sample are provided in Section \ref{sec:sample_sel_spec}.

\section{Likely Contaminant in the COSMOS Field}
\label{sec:appendix_contaminant}

Upon visual inspection of all our BH*-dominated candidates, we identify COS-WEST-29852 as an extremely bright source that is fit with the BH* template based on GN-9771 \citep{Torralba26IFU} at $z_{\rm phot}=5.61$. We show imaging cutouts, the measured photometry, and the best-fitting \texttt{eazy} SED of the source in Figure \ref{fig:cosmos_contaminant}. If at $z_{\rm phot}=5.61$, the implied optical luminosity of COS-WEST-29852 would be more than an order of magnitude higher than that of any other candidate in our sample (log($L_{5100}/{\rm erg}\,{\rm s}^{-1}) = 45.68$). While the fit with the BH* template is nominally robust (Bayes factors $\gg100$), the photometry shows no distinctive features such as a (Lyman or Balmer) break, or individual filters boosted by emission lines. Instead, it suggests a smoothly declining red SED which happens to be consistent with the GN-9771 template at the best-fitting redshift. The imaging cutouts further shows a dominant point source component in the NIRCam/LW filters, and a clearly extended morphology in the SW bands, suggesting it could be an obscured AGN or quasar \citep[see, e.g.,][]{Fei26} whose host galaxy dominates the light at $\lesssim2\mu{\rm m}$. Further investigation of COS-WEST-29852 at (R.A., Dec.) = (150.0646281, 2.1909212) is beyond the scope of this work, but we argue that it is likely a peculiar contaminant, and not a BH*-dominated source at $z_{\rm phot}=5.61$, and therefore remove it from our sample.

\begin{figure}
     \centering
     \includegraphics[width=0.47\textwidth]{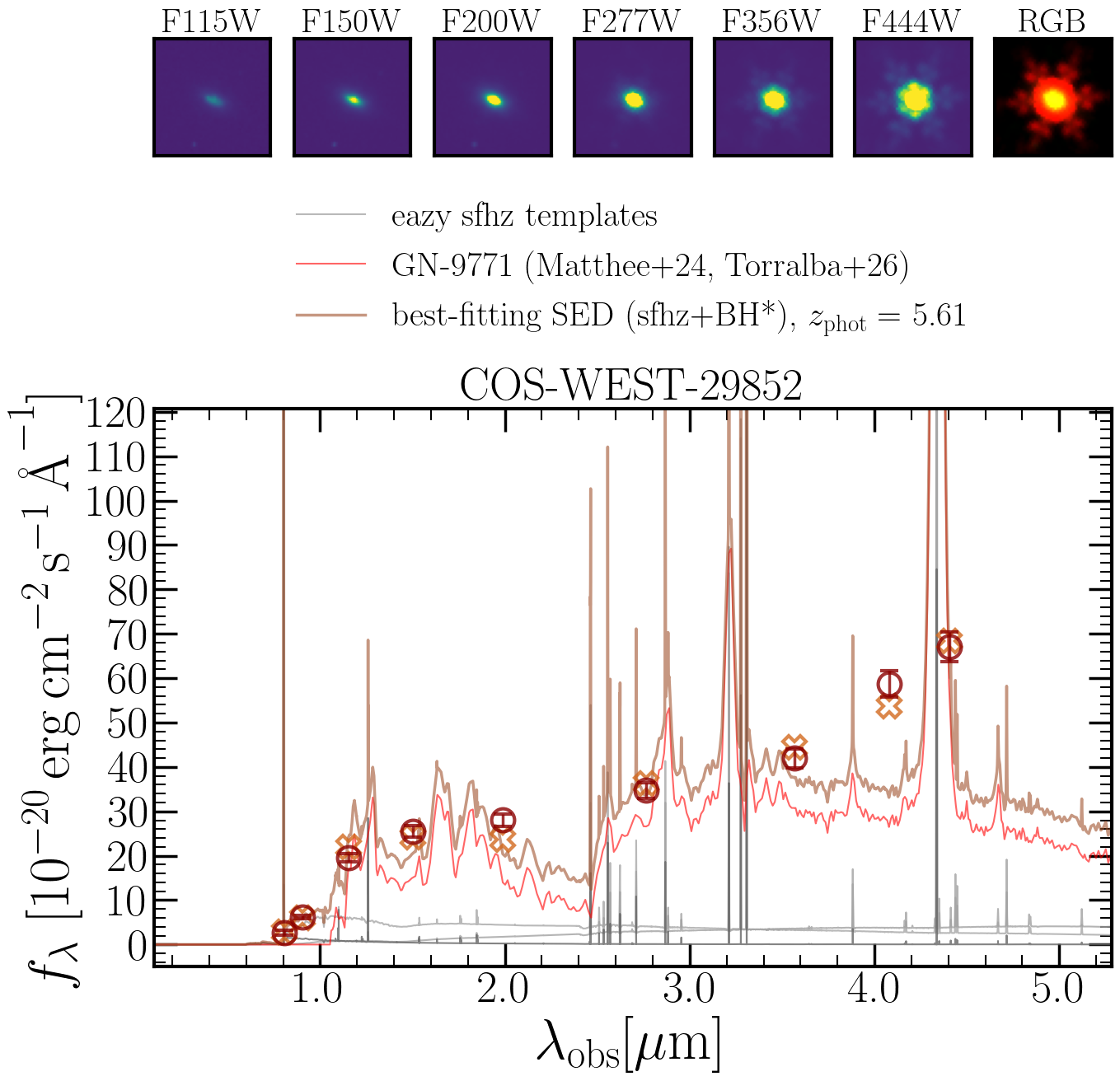}
     \caption{BH*-dominated candidate identified from a fit with the template based on GN-9771 at $z_{\rm phot}=5.61$. Its brightness would imply an optical luminosity that is more than an order of magnitude higher than that of any other source in our sample. We argue that it is likely a contaminant and do not further investigate it here.}
     \label{fig:cosmos_contaminant}
 \end{figure}
 
\section{Gallery of all Candidates in the Gold Sample}
\label{sec:appendix_gallery}

In Figures \ref{fig:gallery_appendix_1} to \ref{fig:gallery_appendix_13}, we show the photometry, best-fitting \texttt{eazy} SEDs, and an RGB imaging stamp for each BH*-dominated candidate that is part of our gold sample (see Section \ref{sec:overview_gold_sample}). The gold candidates are sorted in redshift, and we specify their photometric and - if available - spectroscopic redshift in the title of each panel respectively. The contribution of the respective BH* template to the best-fitting SED, as well as the contributions of \texttt{sfhz} templates are shown in each panel. 

\begin{figure*}
     \centering
     \includegraphics[width=1.9\columnwidth]{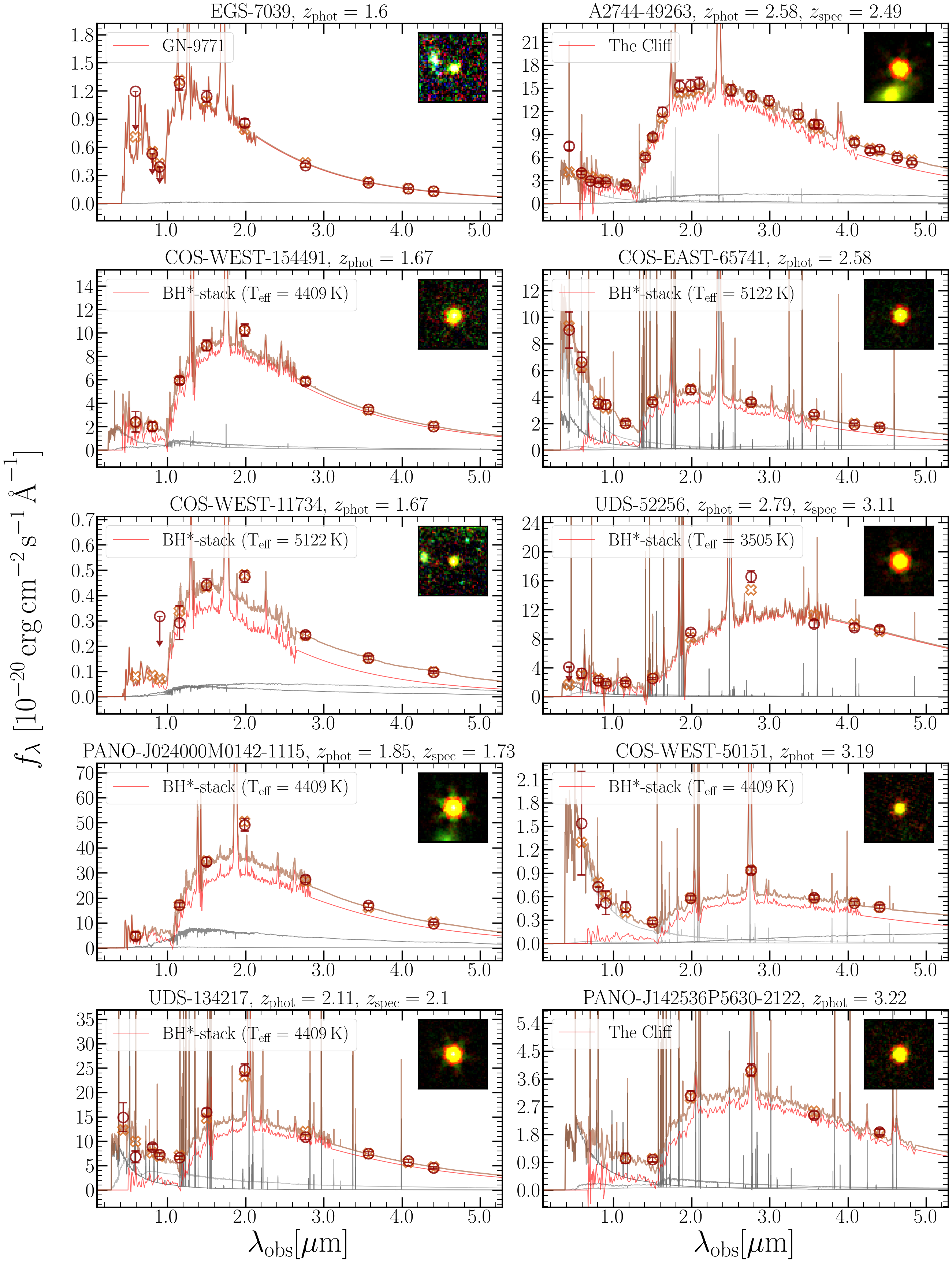}
     \caption{Gallery of BH*-dominated candidates that are part of our gold sample ($B(\texttt{sfhz})>100$ and $B({\rm stars})>100$), in analogy to Figure \ref{fig:overview}. The RGB stamp shown for each source is constructed from F444W (red), F277W (green) and F115W (blue). Gray lines show the contribution of different \texttt{eazy} templates to the best-fitting SED (brown line). The contribution of the respective BH* template that dominates the fit is shown in red.}
     \label{fig:gallery_appendix_1}
 \end{figure*}

 \begin{figure*}
     \centering
     \includegraphics[width=1.9\columnwidth]{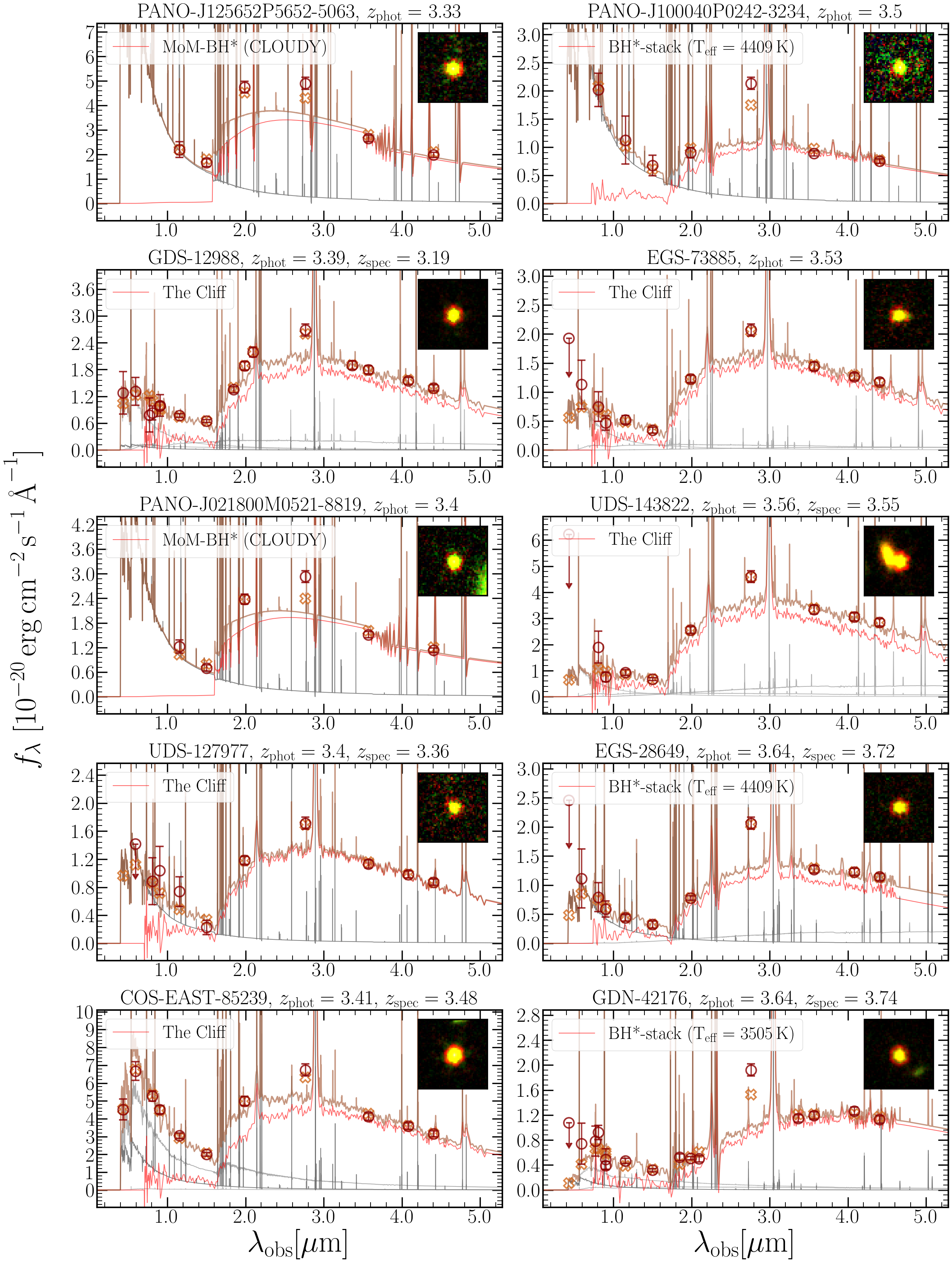}
     \caption{Same as Figure \ref{fig:gallery_appendix_1} (continued).}
     \label{fig:gallery_appendix_2}
 \end{figure*}

  \begin{figure*}
     \centering
     \includegraphics[width=1.9\columnwidth]{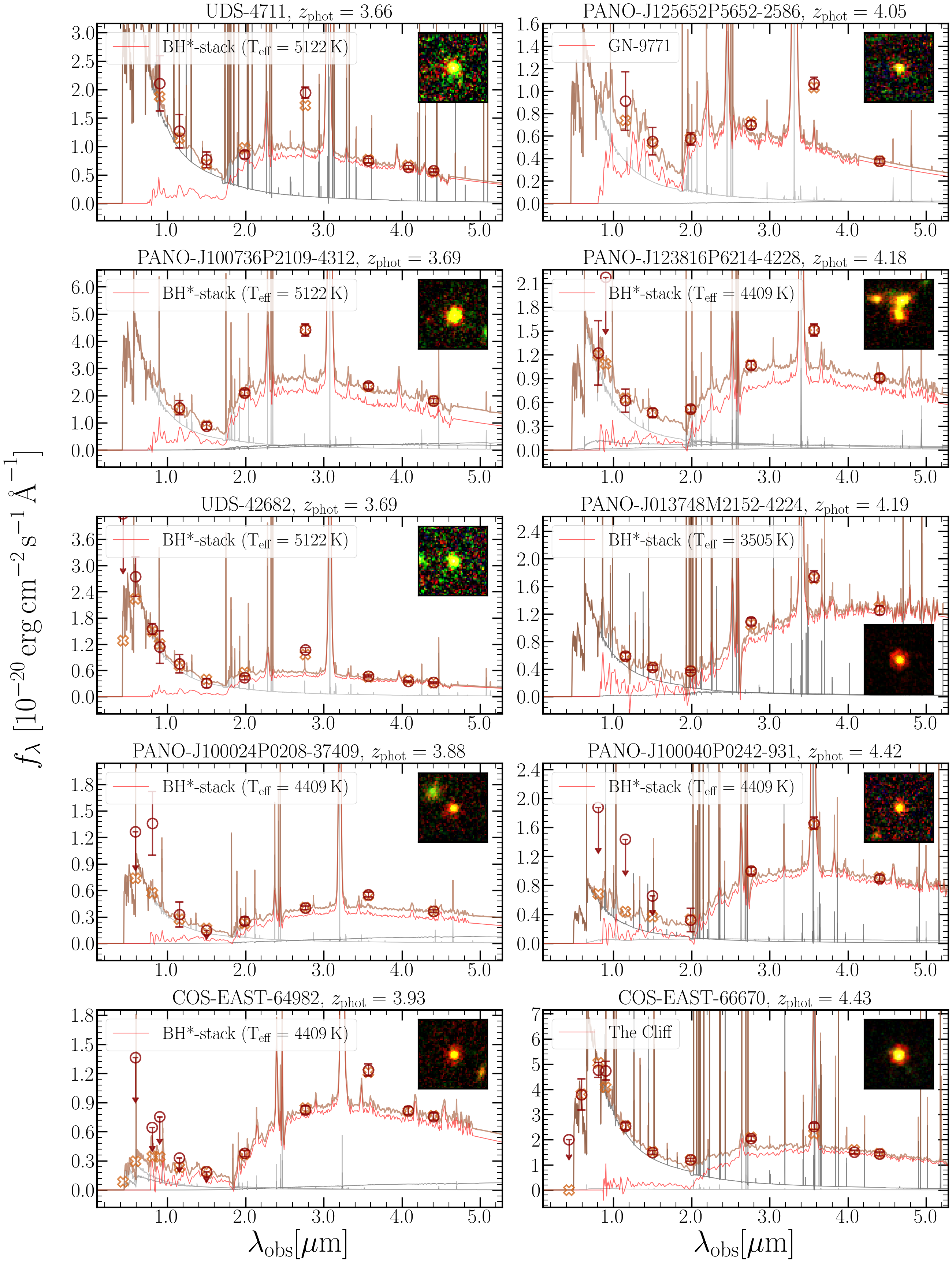}
     \caption{Same as Figure \ref{fig:gallery_appendix_1} (continued).}
     \label{fig:gallery_appendix_3}
 \end{figure*}

  \begin{figure*}
     \centering
     \includegraphics[width=1.9\columnwidth]{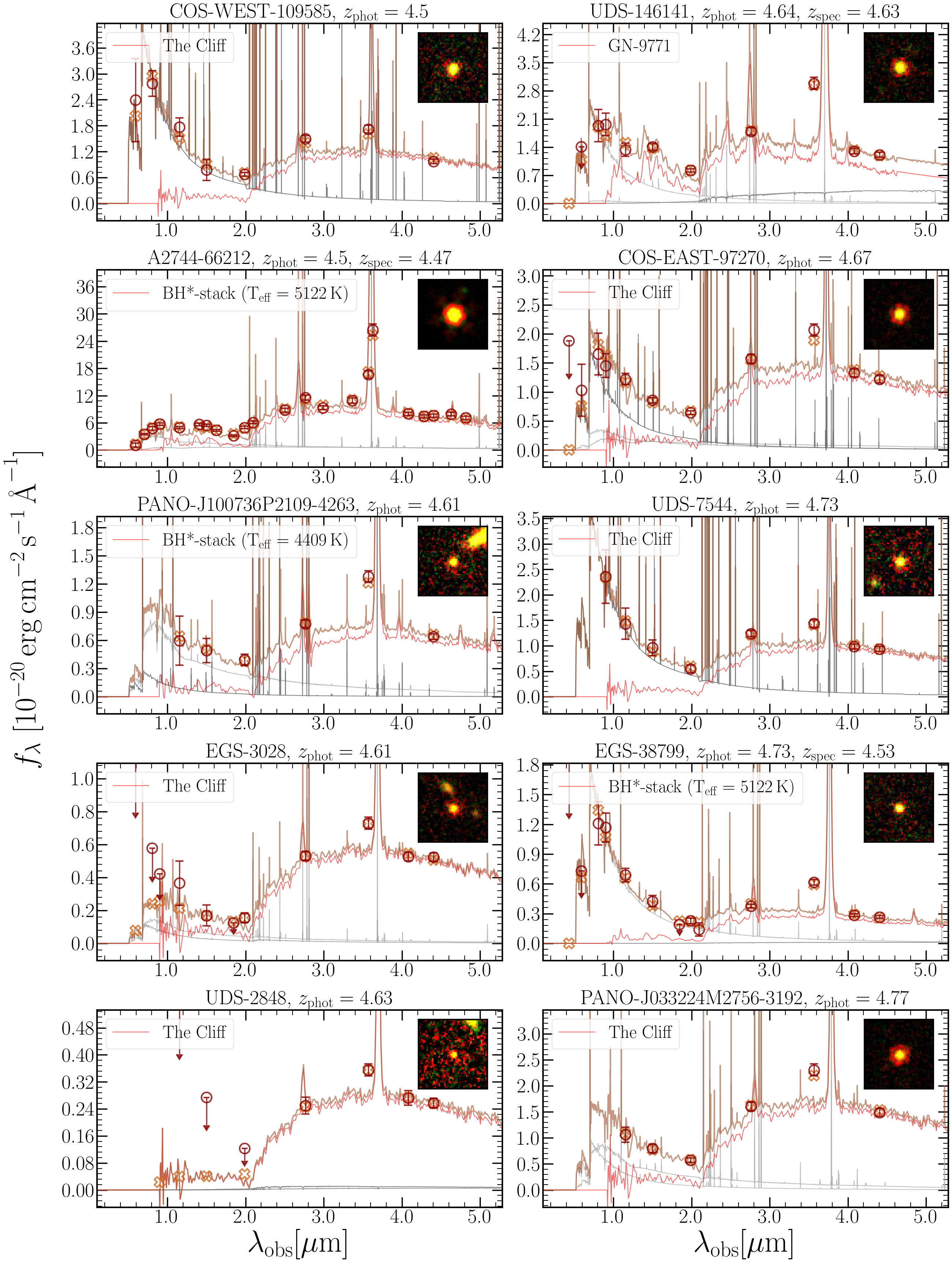}
     \caption{Same as Figure \ref{fig:gallery_appendix_1} (continued).}
     \label{fig:gallery_appendix_4}
 \end{figure*}

  \begin{figure*}
     \centering
     \includegraphics[width=1.9\columnwidth]{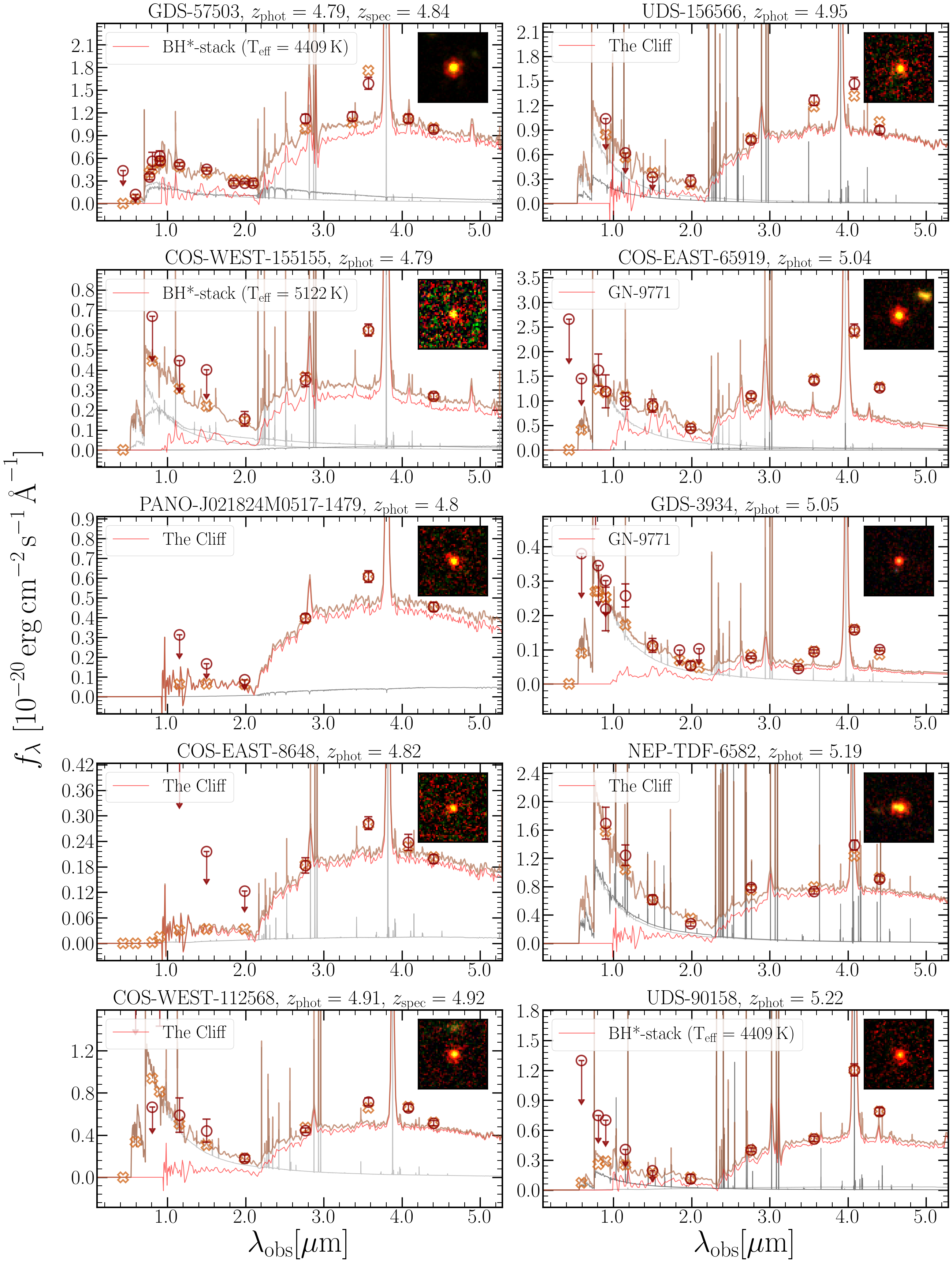}
     \caption{Same as Figure \ref{fig:gallery_appendix_1} (continued).}
     \label{fig:gallery_appendix_5}
 \end{figure*}

  \begin{figure*}
     \centering
     \includegraphics[width=1.9\columnwidth]{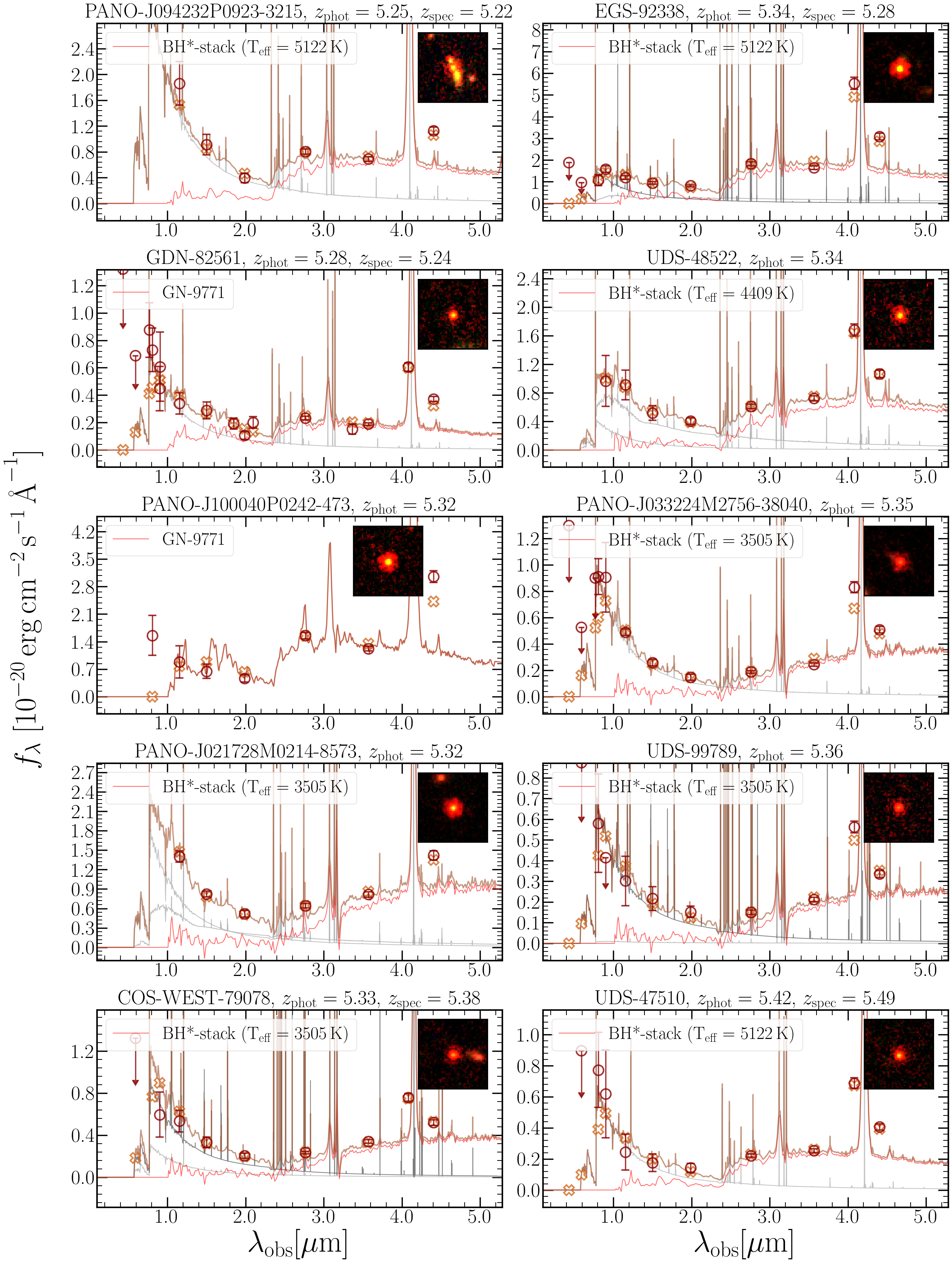}
     \caption{Same as Figure \ref{fig:gallery_appendix_1} (continued).}
     \label{fig:gallery_appendix_6}
 \end{figure*}

  \begin{figure*}
     \centering
     \includegraphics[width=1.9\columnwidth]{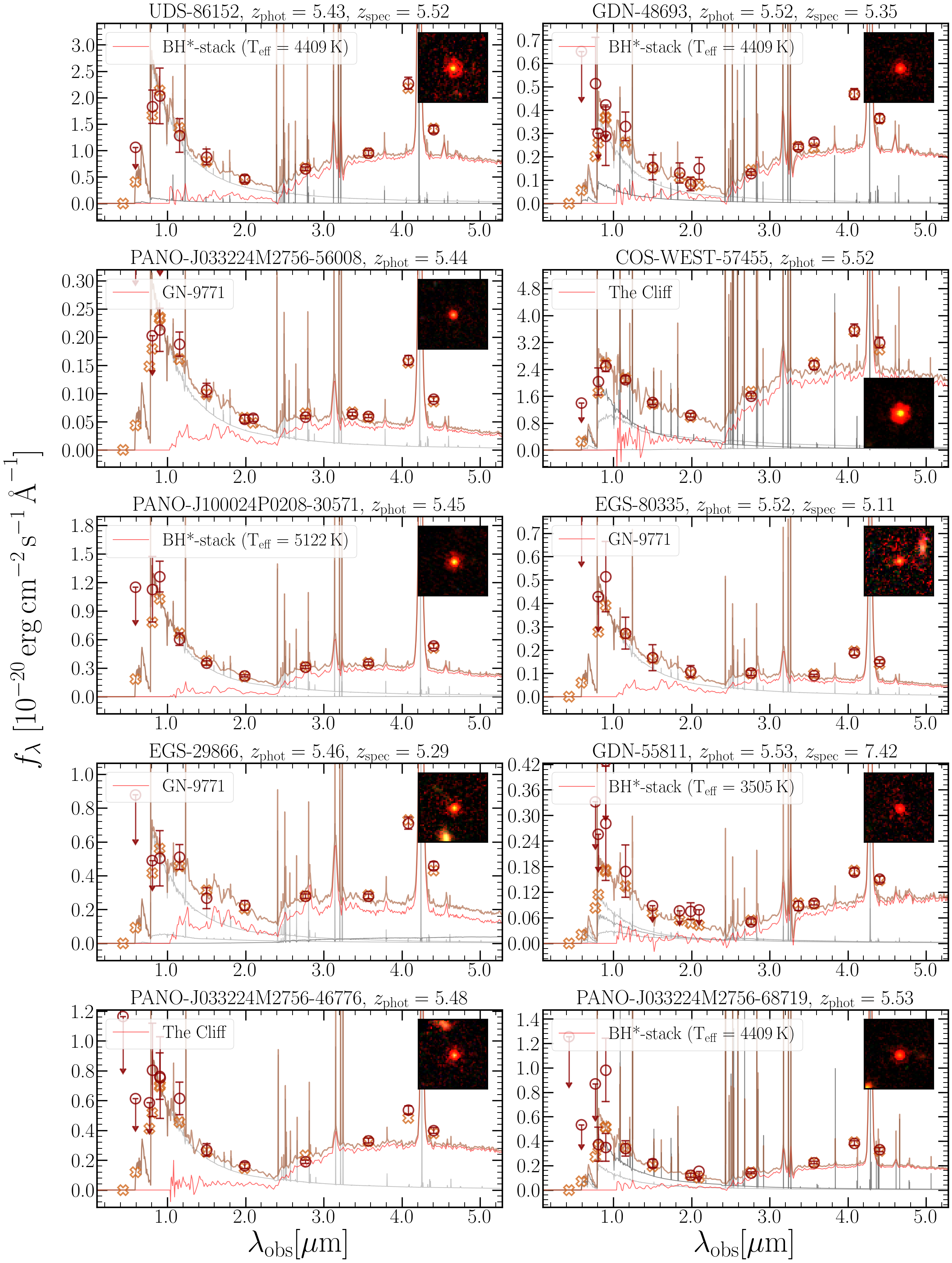}
     \caption{Same as Figure \ref{fig:gallery_appendix_1} (continued).}
     \label{fig:gallery_appendix_7}
 \end{figure*}

  \begin{figure*}
     \centering
     \includegraphics[width=1.9\columnwidth]{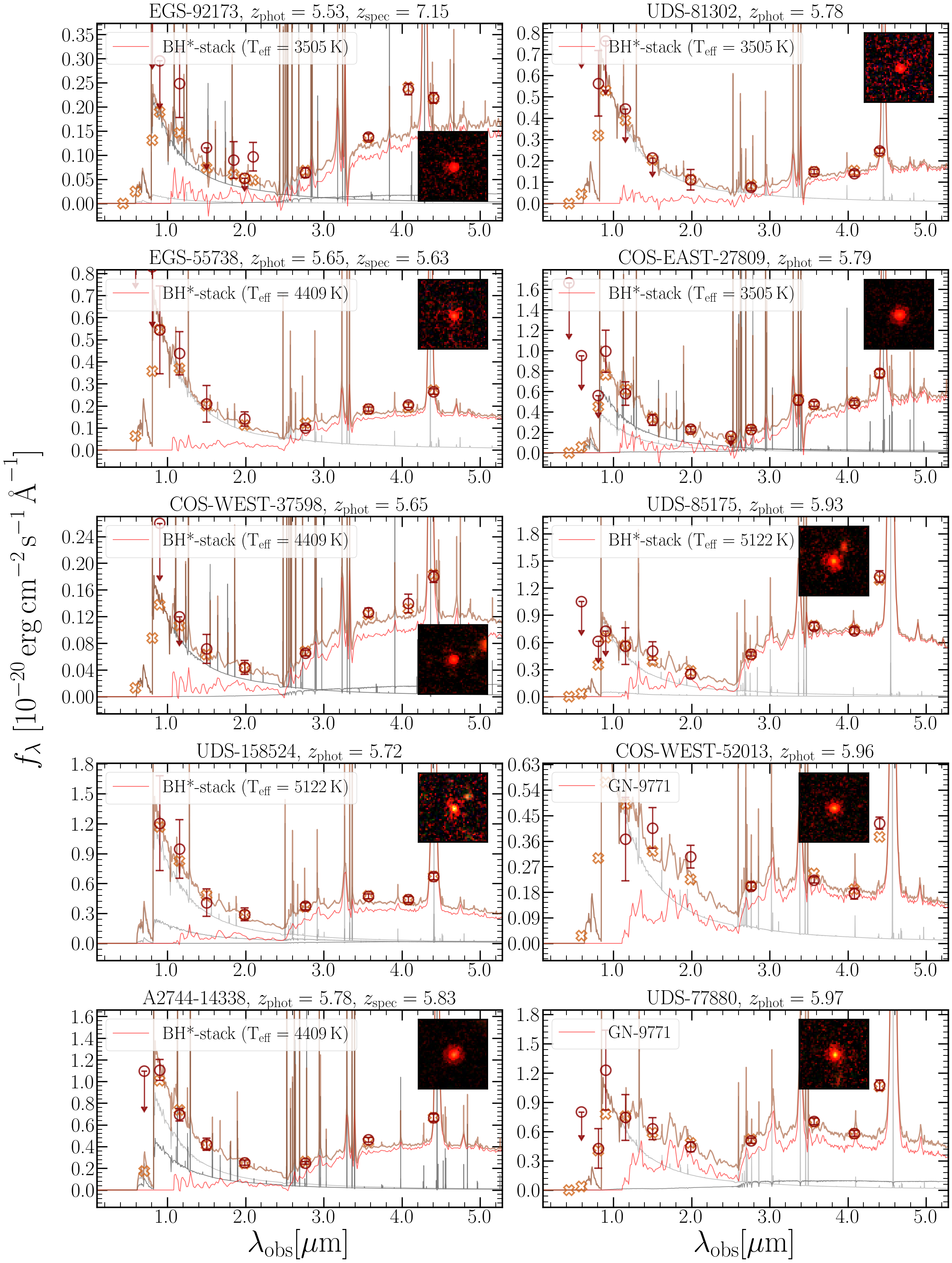}
     \caption{Same as Figure \ref{fig:gallery_appendix_1} (continued).}
     \label{fig:gallery_appendix_8}
 \end{figure*}

  \begin{figure*}
     \centering
     \includegraphics[width=1.9\columnwidth]{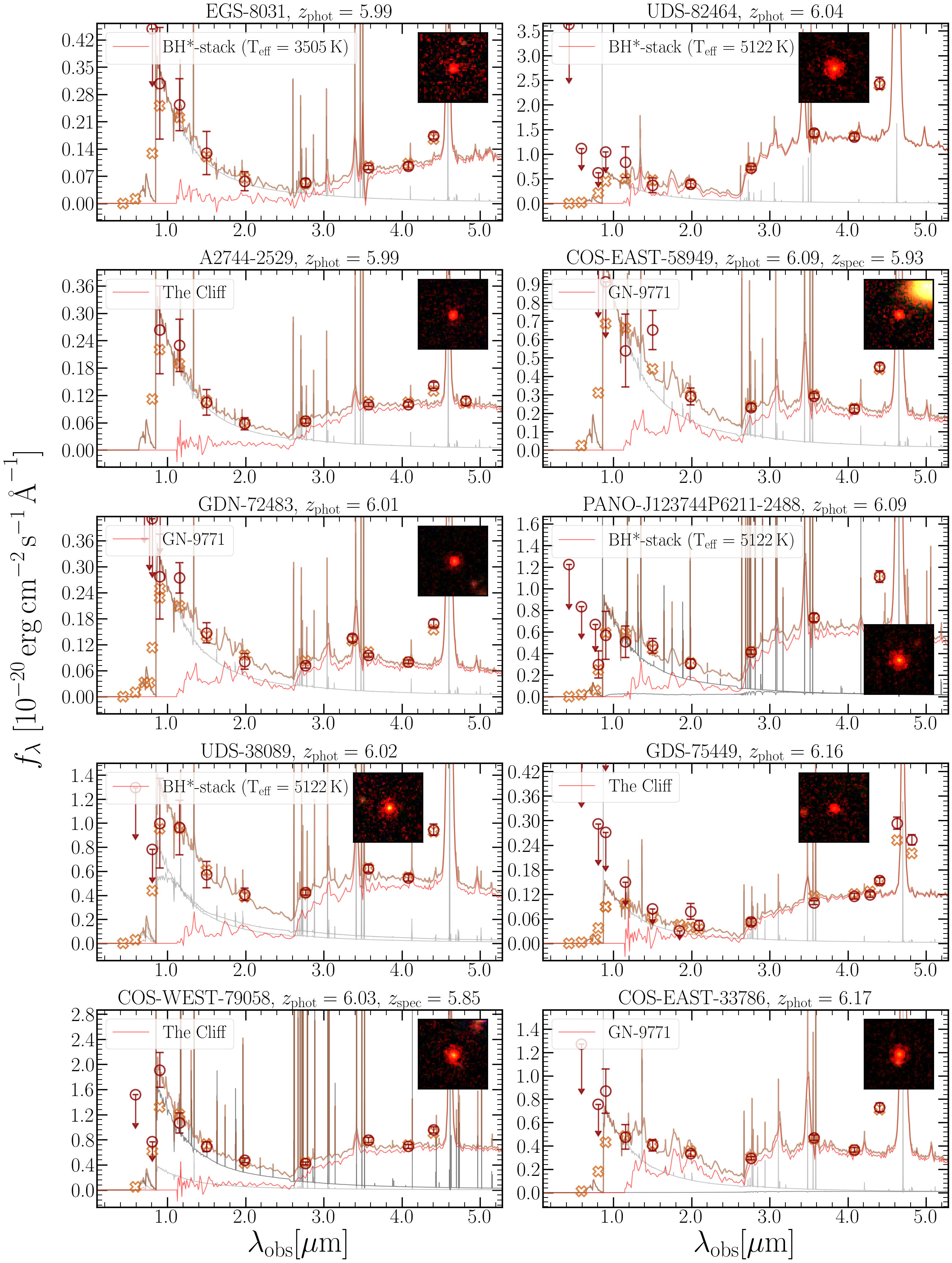}
     \caption{Same as Figure \ref{fig:gallery_appendix_1} (continued).}
     \label{fig:gallery_appendix_9}
 \end{figure*}

  \begin{figure*}
     \centering
     \includegraphics[width=1.9\columnwidth]{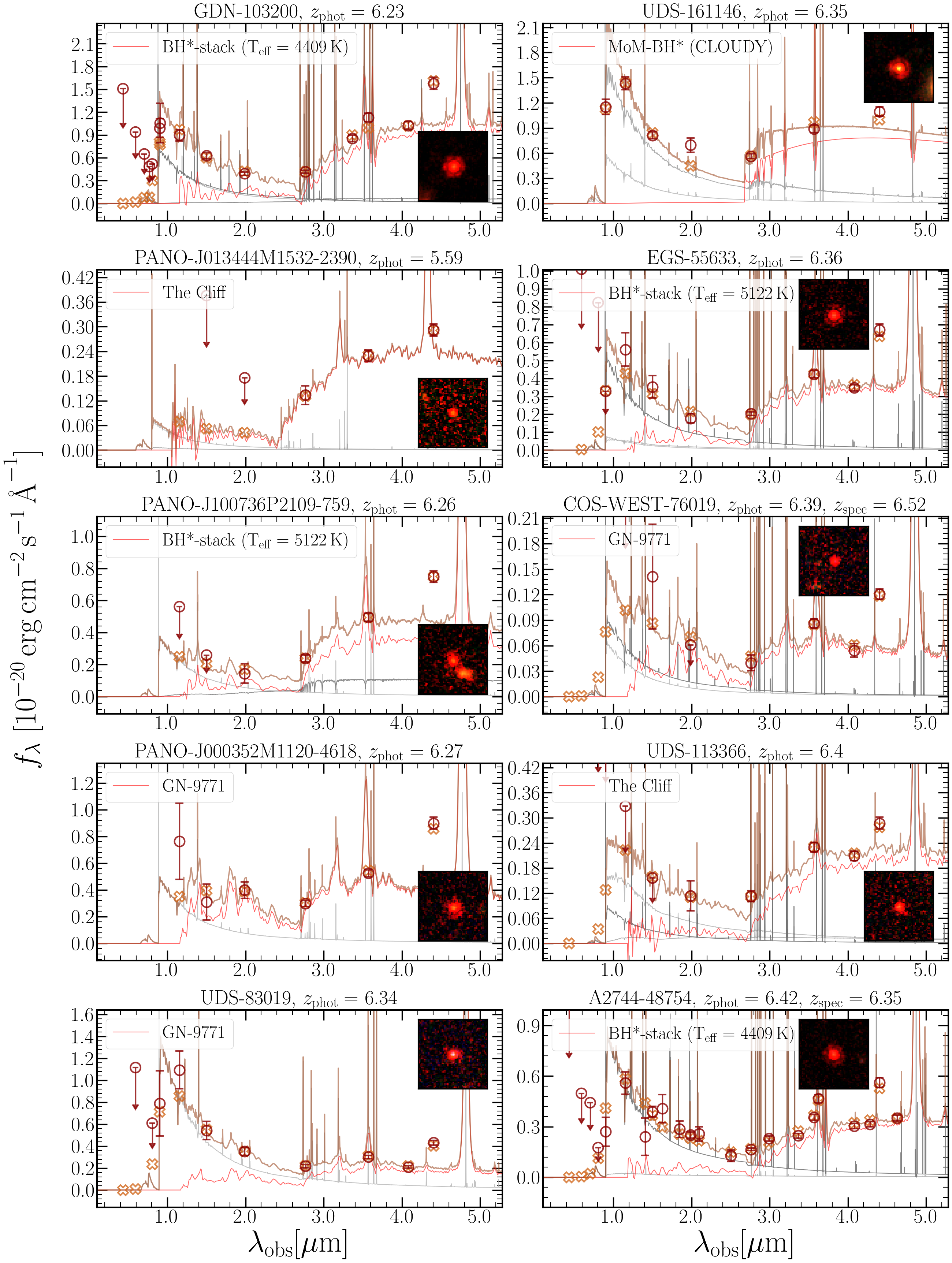}
     \caption{Same as Figure \ref{fig:gallery_appendix_1} (continued).}
     \label{fig:gallery_appendix_10}
 \end{figure*}

  \begin{figure*}
     \centering
     \includegraphics[width=1.9\columnwidth]{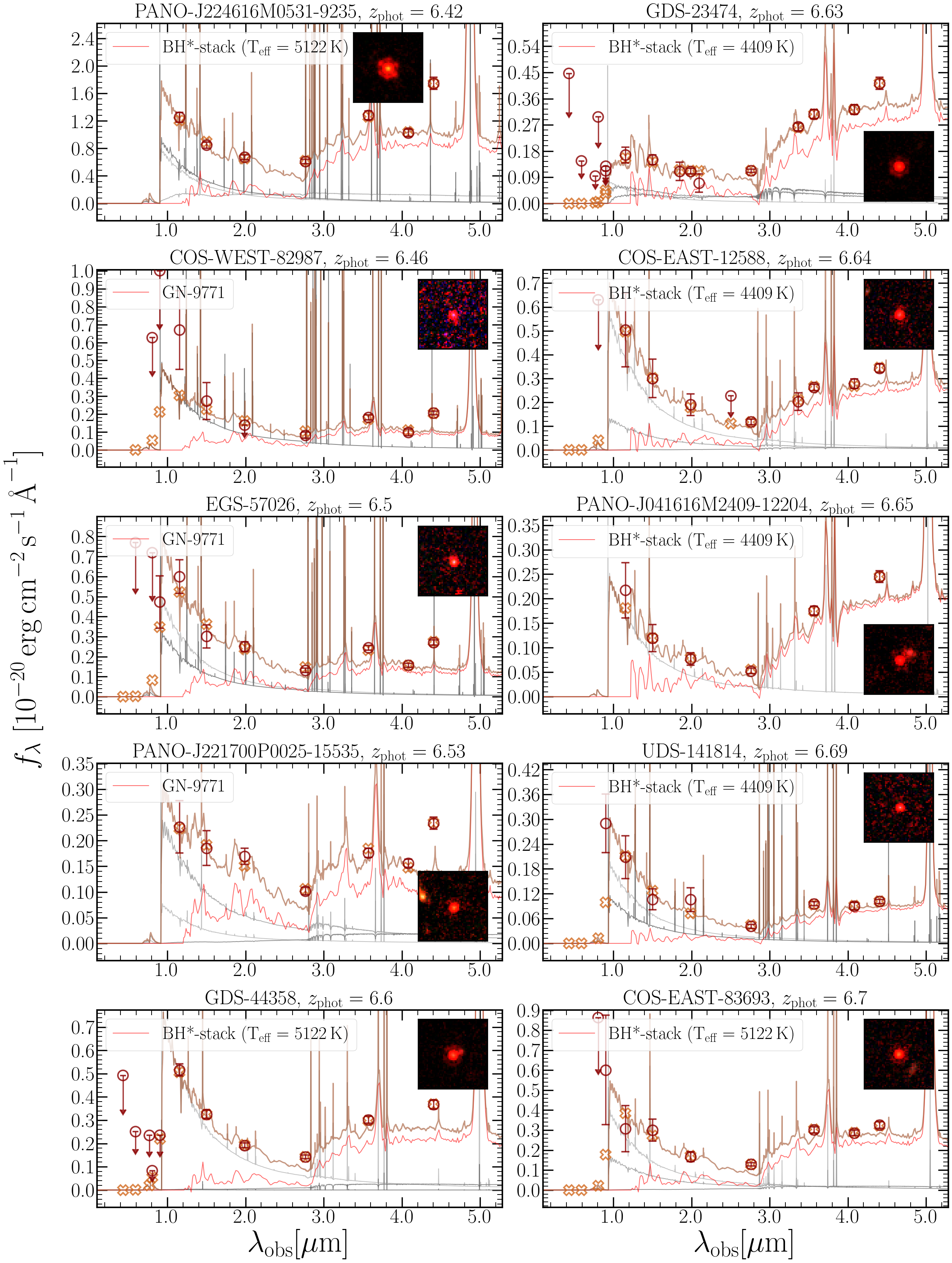}
     \caption{Same as Figure \ref{fig:gallery_appendix_1} (continued).}
     \label{fig:gallery_appendix_11}
 \end{figure*}

  \begin{figure*}
     \centering
     \includegraphics[width=1.9\columnwidth]{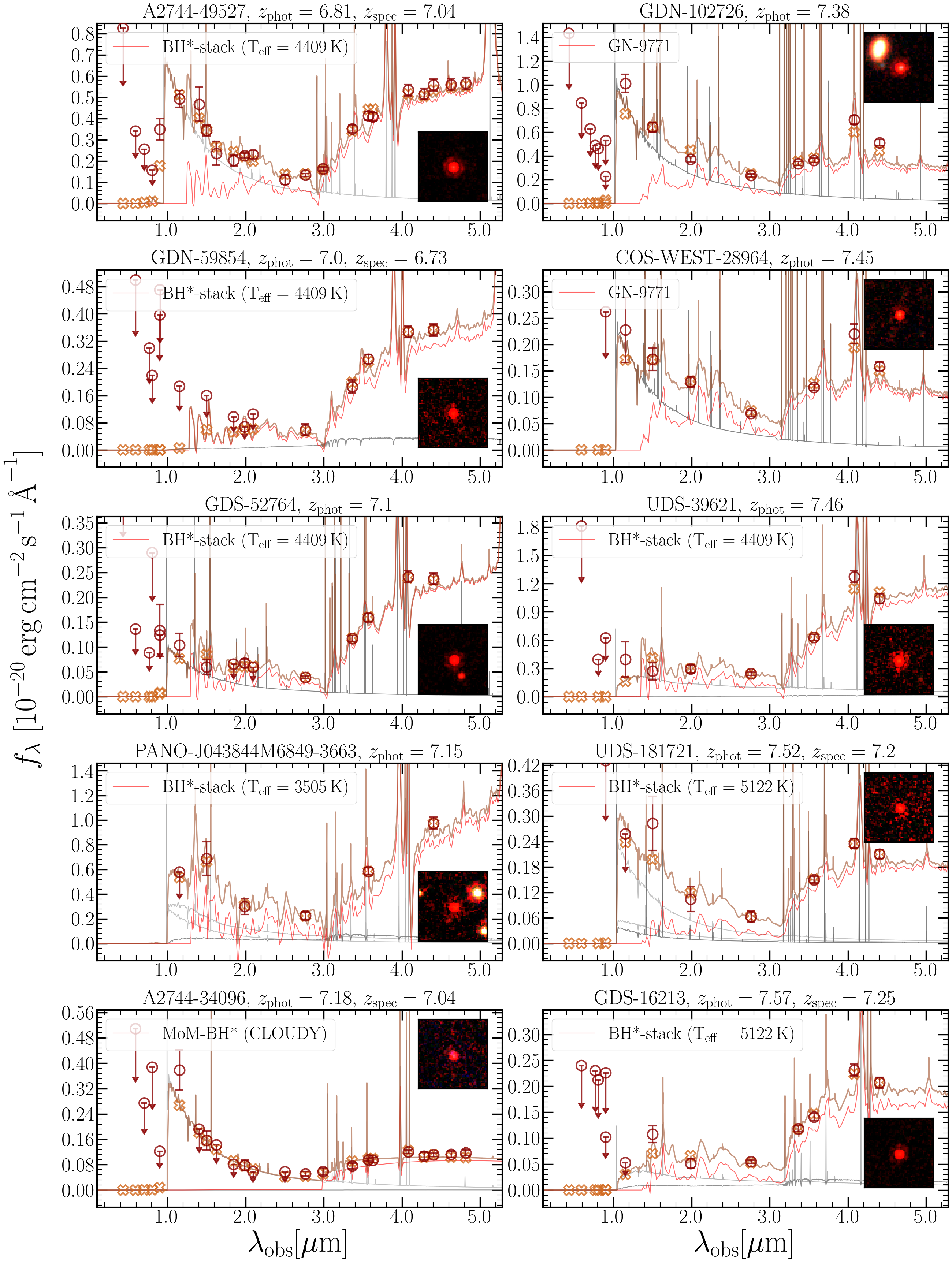}
     \caption{Same as Figure \ref{fig:gallery_appendix_1} (continued).}
     \label{fig:gallery_appendix_12}
 \end{figure*}

  \begin{figure*}
     \centering
     \includegraphics[width=1.9\columnwidth]{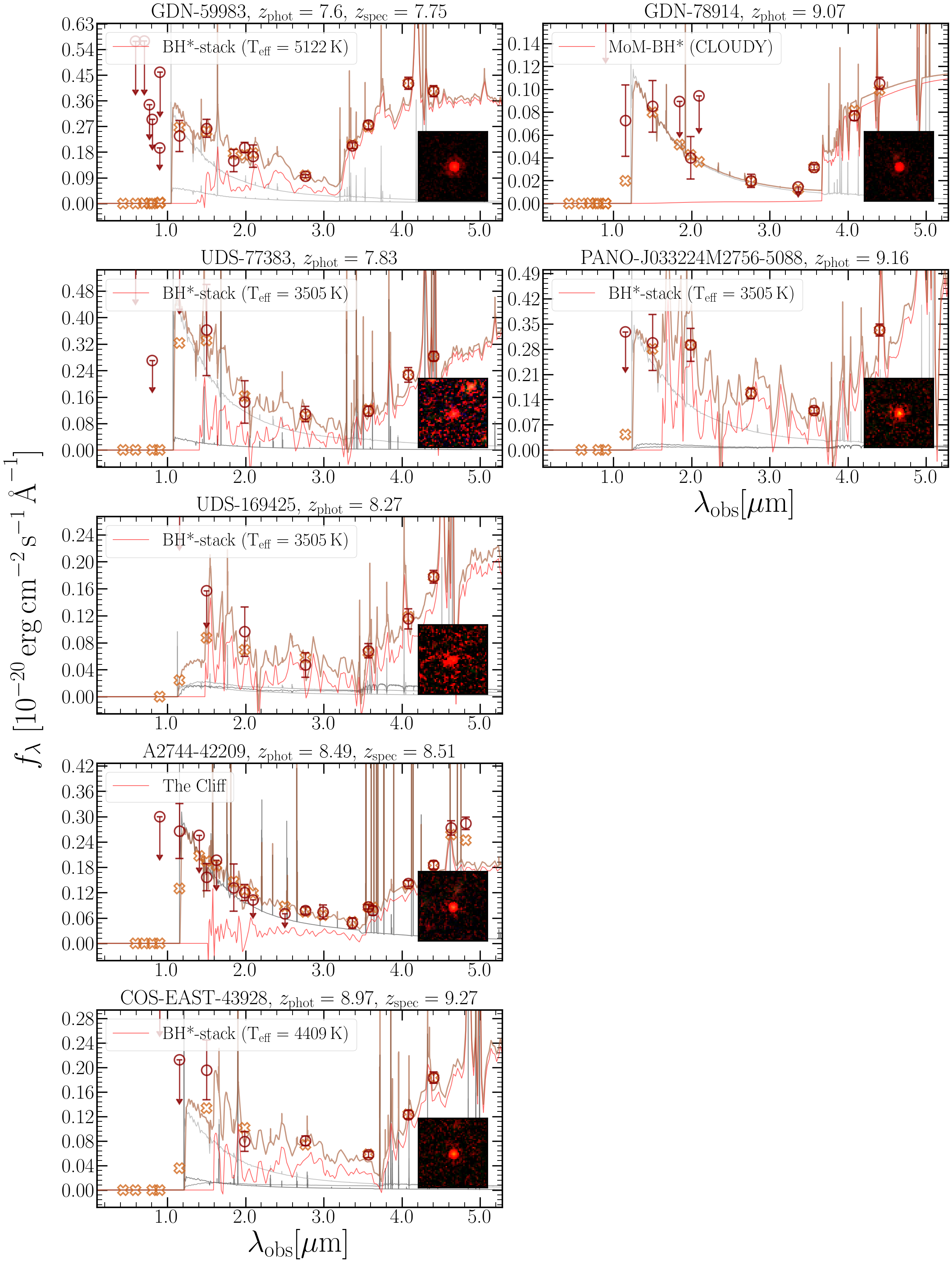}
     \caption{Same as Figure \ref{fig:gallery_appendix_1} (continued).}
     \label{fig:gallery_appendix_13}
 \end{figure*}

\end{document}